\documentclass[%
 aip,
 amsmath,amssymb,
 reprint,%
]{revtex4-1}

\usepackage{graphicx}
\usepackage{dcolumn}
\usepackage{bm}

\usepackage[utf8]{inputenc}
\usepackage[T1]{fontenc}
\usepackage{mathptmx}
\usepackage{etoolbox}
\usepackage{bigints}
\usepackage[dvipsnames]{xcolor}
\usepackage{comment} 

\makeatletter
\def\@email#1#2{%
 \endgroup
 \patchcmd{\titleblock@produce}
  {\frontmatter@RRAPformat}
  {\frontmatter@RRAPformat{\produce@RRAP{*#1\href{mailto:#2}{#2}}}\frontmatter@RRAPformat}
  {}{}
}%
\makeatother
\begin{document}

\preprint{AIP/123-QED}

\title[Solubility of carbon dioxide in water]{Solubility of carbon dioxide in water: some useful results for hydrate nucleation}

\author{Jes\'us Algaba}
\affiliation{Laboratorio de Simulaci\'on Molecular y Qu\'imica Computacional, CIQSO-Centro de Investigaci\'on en Qu\'imica Sostenible and Departamento de Ciencias Integradas, Universidad de Huelva, 21006 Huelva Spain}

\author{Iv\'an M. Zer\'on}
\affiliation{Laboratorio de Simulaci\'on Molecular y Qu\'imica Computacional, CIQSO-Centro de Investigaci\'on en Qu\'imica Sostenible and Departamento de Ciencias Integradas, Universidad de Huelva, 21006 Huelva Spain}

\author{Jos\'e Manuel M\'{\i}guez}
\affiliation{Laboratorio de Simulaci\'on Molecular y Qu\'imica Computacional, CIQSO-Centro de Investigaci\'on en Qu\'imica Sostenible and Departamento de Ciencias Integradas, Universidad de Huelva, 21006 Huelva Spain}

\author{Joanna Grabowska}
\affiliation{Department of Physical Chemistry, Faculty of Chemistry and BioTechMed Center, Gdansk University of Technology, ul. Narutowicza 11/12, 80-233 Gdansk, Poland}
\affiliation{Dpto. Qu\'{\i}mica F\'{\i}sica I, Fac. Ciencias Qu\'{\i}micas, Universidad Complutense de Madrid, 28040 Madrid, Spain}

\author{Samuel Blazquez}
\affiliation{Dpto. Qu\'{\i}mica F\'{\i}sica I, Fac. Ciencias Qu\'{\i}micas, Universidad Complutense de Madrid, 28040 Madrid, Spain}

\author{Eduardo Sanz}
\affiliation{Dpto. Qu\'{\i}mica F\'{\i}sica I, Fac. Ciencias Qu\'{\i}micas, Universidad Complutense de Madrid, 28040 Madrid, Spain}

\author{Carlos Vega}
\affiliation{Dpto. Qu\'{\i}mica F\'{\i}sica I, Fac. Ciencias Qu\'{\i}micas, Universidad Complutense de Madrid, 28040 Madrid, Spain}

\author{Felipe J. Blas}
\affiliation{Laboratorio de Simulaci\'on Molecular y Qu\'imica Computacional, CIQSO-Centro de Investigaci\'on en Qu\'imica Sostenible and Departamento de Ciencias Integradas, Universidad de Huelva, 21006 Huelva Spain}
\email{felipe@uhu.es}

\begin{abstract}
In this paper, the solubility of carbon dioxide (CO$_{2}$) in water along the isobar of 400 bar is determined by computer simulations using the well-known TIP4P/Ice force field for water and TraPPE model for CO$_{2}$. In particular, the solubility of CO$_{2}$ in water when in contact with the CO$_{2}$ liquid phase, and the solubility of CO$_{2}$ in water when in contact with the hydrate have been determined. The solubility of CO$_{2}$ in a liquid-liquid system decreases as temperature increases. The solubility of CO$_{2}$ in a hydrate-liquid system increases with temperature. The two curves intersect at a certain temperature that determines the dissociation temperature of the hydrate at $400\,\text{bar}$ ($T_{3}$). We compare the predictions with the $T_{3}$ obtained using the direct coexistence technique in a previous work. The results of both methods agree and we suggest $290(2)\,\text{K}$ as the value of $T_{3}$ for this system using the same cutoff distance for dispersive interactions. We also propose a novel and alternative route to evaluate the change in chemical potential for the formation of hydrate along the isobar. The new approach is based on the use of the solubility curve of CO$_{2}$ when the aqueous solution is in contact with the hydrate phase. It considers rigorously the non-ideality of the aqueous solution of CO$_{2}$, providing reliable values for driving force for nucleation of hydrates in good agreement with other thermodynamic routes used. It is shown that the driving force for hydrate nucleation at $400\,\text{bar}$ is larger for the methane hydrate than for the carbon dioxide hydrate when compared at the same supercooling. We have also analyzed and discussed the effect of the cutoff distance of the dispersive interactions and the occupancy of CO$_{2}$ on the driving force for nucleation of the hydrate.
\end{abstract}

\maketitle

%

\section{Introduction}

At ambient conditions of temperature and pressure ($298\,\text{K}$ and $1\,\text{bar}$), the thermodynamically stable phase of water is the liquid phase. If the temperature is decreased at a constant pressure of $1\,\text{bar}$, the liquid is no longer the most stable phase and a first-order phase transition takes place at $273.15\,\text{K}$. Consequently, and according to the Thermodynamics laws, water must freeze. The new thermodynamically stable phase is the well-known ordinary ice, also known as Ih or hexagonal ice. This solid phase is formed by a crystalline structure characterized by the oxygen atoms forming hexagonal symmetry with nearly tetrahedral bonding angles. The same happens if the pressure is above ambient conditions up to $2100\,\text{bar}$, approximately. Above this pressure, water can freeze into other ices, including ice III, V, VI, among others, as the pressure is increased.~\cite{Eisenberg1969a,Petrenko1999a,Sanz2004a} These are only some of the solid crystalline phases of the well-known polymorphic phases of water. However, this only happens if the original liquid phase is formed from pure water. When liquid water is mixed with another substance the story can be different.

There exist aqueous solutions of small compounds that exhibit different behavior when cooled down at constant pressure. Particularly, aqueous solutions of methane (CH$_{4}$), carbon dioxide (CO$_{2}$), nitrogen (N$_{2}$), hydrogen (H$_{2}$) or larger organic molecules, among many other different compounds, do not transform into a crystalline ice phase when the temperature is lowered. In fact, all these aqueous solutions freeze into new crystalline solid compounds named clathrate hydrates or simply hydrates.~\cite{Sloan2008a} Hydrates are non-stoichiometric crystalline inclusion compounds consisting of a network of hydrogen-bonding water molecules forming cages in which solutes (for instance, CH$_{4}$, CO$_{2}$, N$_{2}$ or H$_{2}$) are enclathrated at appropriate thermodynamic conditions of temperature and pressure.

Fundamental and applied research on hydrates and clathrates has been motivated by several reasons. First at all, hydrates are potential alternative sources of energy since huge amounts of CH$_{4}$ have been identified in hydrate deposits, either in the sea floor or in the permafrost frozen substrates, but their exploitation is not technically accessible yet due to a poor physicochemical characterization and various engineering issues.~\cite{Kvenvolden1988a,Koh2012a} Another remarkably relevant aspect of hydrates from both the scientific point of view and practical interest is the possibility to capture~\cite{Yang2014a,Ricaurte2014a} and store CO$_{2}$.~\cite{Kvamme2007a} This places gas hydrates at the center of environmental concerns regarding atmospheric greenhouse gases. Sequestration and capture of CO$_{2}$ in hydrates constitute a technological breakthrough which is seen as a promising alternative to other conventional methodologies for CO$_{2}$ capture, such as reactive absorption using amines and selective adsorption using adsorbent porous materials including sieves and zeolites.~\cite{Alessandro2009a,Choi2009a}

It is clear from the previous discussion that an accurate knowledge of  the thermodynamics and kinetics of the formation and growth of hydrates is necessary from the fundamental and practical points of view.  The thermodynamics of hydrates has been relatively well-established experimentally for years.~\cite{Sloan2008a} In addition, it is also possible to describe theoretically the phase equilibria of hydrates using the van der Waals and Platteeuw (vdW\&P) formalism.~\cite{Platteeuw1957a,Platteeuw1959a} This approach, combined with an equation of State (EOS) allows us to satisfactorily determine the phase equilibrium of both pure hydrates and mixtures.~\cite{Sloan2008a} Additionally, from the point of view of molecular simulation, there has been an enormous development in techniques and methodologies for the study of the formation and dissociation of a huge variety of hydrates.~\cite{Koh2002a,Tsimpanogiannis2018a,Michalis2015a,Costandy2015a,Michalis2016a,Waage2017a,Conde2010a,Conde2013a} Particularly, several research groups have determined the phase equilibrium of CO$_{2}$\cite{Miguez2015a,Perez-Rodriguez2017a} and CH$_{4}$ hydrates under oceanic crust conditions\cite{Fernandez-Fernandez2019a,Thoutam2021a} using the direct coexistence technique. The precise knowledge of phase equilibria of hydrates, and particularly their phase boundaries, is essential to provide a detailed description of the kinetic and nucleation processes of these systems.

Unfortunately, a complete description from a molecular perspective of the mechanisms of growth and hydrate formation is far from being satisfactory. In the last few years, some of the authors of this work have been working on the development and use of the Seeding Technique,~\cite{Espinosa2016c} in combination with the Classical Nucleation Theory (CNT),~\cite{Debenedetti1996a} to deal with several systems including the hard-sphere and Lennard-Jones models and more complex systems such as water and salty water.~\cite{Soria2018a} More recently, we have extended the study to deal with methane hydrates.~\cite{Grabowska2022a,Grabowska2022b} It is important to recall here that Molinero and collaborators used the Seeding Technique to estimate nucleation rates of hydrates~\cite{Knott2012a} modeled through the well-known mW water model.~\cite{Molinero2009a} Other authors have also contributed significantly to the understanding of the dynamics of nucleation and dissociation of hydrates from computer simulation.~\cite{Jacobson2010a,Jacobson2010b,Jacobson2011a,Sarupria2011a,Sarupria2012a,Yuhara2015a,Liang2013a,Barnes2014a,Walsh2011a,Warrier2016a,Zhang2016b,Lauricella2017a,Arjun2019a,Karmakar2019a,Arjun2020a,Arjun2021a,Guo2021a,Kvamme2020a} This work constitutes the extension of our most recent study~\cite{Grabowska2022a} on methane hydrates to deal with CO$_{2}$ hydrates. Before undertaking nucleation studies of CO$_{2}$ hydrates it is necessary to account for several issues, including the solubility of CO$_{2}$ in the aqueous solution when it is in contact with the CO$_{2}$-rich liquid phase and with the hydrate, an accurate prediction of the dissociation temperature, and the driving force for the nucleation.

\begin{figure}
\hspace*{-0.2cm}
\includegraphics[width=1.1\columnwidth]{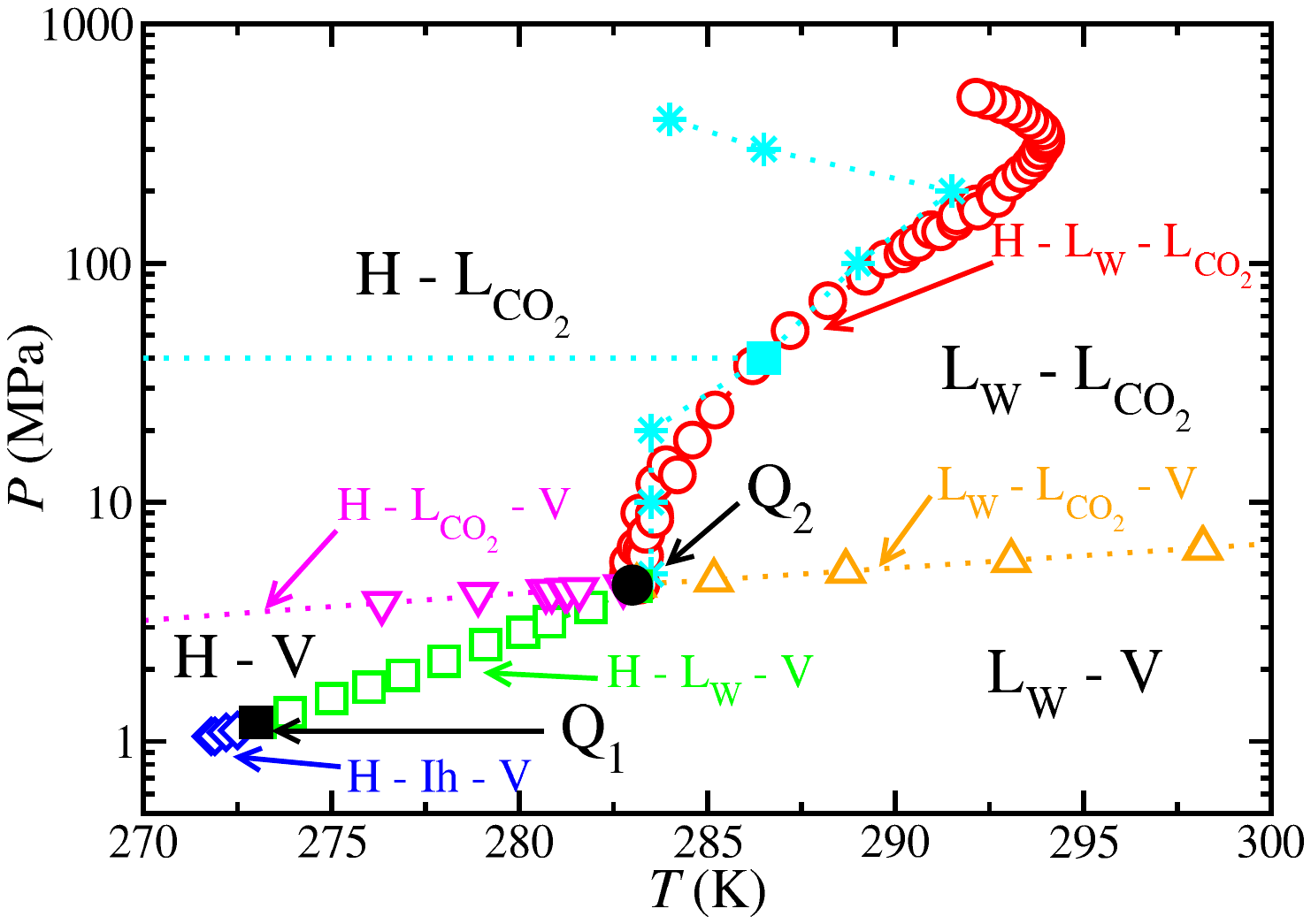}
\caption{$PT$ projection of the phase diagram of the CO$_{2}$ + water mixture. The open symbols represent the experimental three-phase lines for the H-L$_{\text{w}}$-L$_{\text{CO}_{2}}$ (red circles),~\cite{Sloan2008a} H-L$_{\text{w}}$-V (green squares),~\cite{Sloan2008a} H-Ih-V (blue diamonds),~\cite{Sloan2008a} L$_{\text{w}}$-L$_{\text{CO}_{2}}$-V (orange up triangles),~\cite{Sloan2008a} and L$_{\text{w}}$-L$_{\text{CO}_{2}}$-V (magenta down triangles)~\cite{Sloan2008a} equilibria. The black square and the black circle are the experimental quadruple points Q$_{1}$ and Q$_{2}$,~\cite{Sloan2008a} respectively. Cyan stars correspond to the simulation results obtained by M\'{\i}guez~\emph{et al.}~\cite{Miguez2015a} for the H-L$_{\text{w}}$-L$_{\text{CO}_{2}}$ three-phase line. Cyan filled square represents the state at $T_{3}$ and $400\,\text{bar}$ as obtained in this work. The horizontal dotted cyan line represents the $400\,\text{bar}$ isobar at which the driving force for nucleation is evaluated in this work. The rest of lines are guides to the eye.
}
\label{figure1}
\end{figure}

The phase behavior of the CO$_{2}$ + water binary mixture is dominated by a large region of liquid-liquid (L$_{\text{w}}$-L$_{\text{CO}_{2}}$) immiscibility.$^{51,52}$ Since the critical point of pure CO$_{2}$ and a liquid-liquid-vapor (L$_{\text{w}}$-L$_{\text{CO}_{2}}$-V) three-phase line are at conditions similar to those at which the CO$_{2}$ hydrates are found, between $270-295\,\text{K}$ and $10-5000\,\text{bar}$, approximately, another three-phase coexistence line involving a hydrate phase (i.e., a triple point that occurs at a certain temperature $T_{3}$ for each pressure) exhibits two branches. This is contrary to what happens with methane hydrates, that only exhibit one branch.~\cite{Sloan2008a} Fig.~\ref{figure1} shows the pressure-temperature ($PT$) projection of the phase diagram of the CO$_{2}$ + water binary mixture. At pressures below $44.99\,\text{bar}$, the dissociation line is a H-L$_{\text{w}}$-V three-phase line at which the hydrate, the aqueous solution of CO$_{2}$, and the vapor phases coexist. Above that pressure, the hydrate and the solution coexist with a CO$_{2}$ liquid phase and a three-phase H-L$_{\text{w}}$-L$_{\text{CO}_{2}}$ line where the hydrate, the aqueous solution of CO$_{2}$, and the liquid phase of CO$_{2}$ coexist starts. Both branches meet at a Q$_{2}$ quadruple point located at $283\,\text{K}$ and $44.99\,\text{bar}$ (black filled circle) at which the hydrate, the aqueous solution, the CO$_{2}$ liquid, and the vapor phases coexist,~\cite{Sloan2008a} as can be seen in Fig.~\ref{figure1}. Note that at Q$_{2}$ the L$_{\text{w}}$-L$_{\text{CO}_{2}}$-V three-phase line also meets with another H-L$_{\text{CO}_{2}}$-V three-phase in which the hydrate, the CO$_{2}$ liquid, and the vapor phases coexist at lower temperatures. In addition to this, there exists another quadruple point $Q_{1}$, located at $273\,\text{K}$ and $12.56\,\text{bar}$ (black filled square), at which the hydrate, the Ih ice, the solution, and the vapor phases coexist. This quadruple point connects the H-L$_{\text{w}}$-V three-phase line with a new three-phase H-Ih-V line involving the hydrate, the Ih ice, and the vapor phases that runs towards lower temperatures and pressures.

In this work, we concentrate on $400\,\text{bar}$ of pressure (see the $400\,\text{bar}$ isobar in Fig.~1 along which all the simulations are performed). At these conditions, the key solubility curves in the context of nucleation of CO$_{2}$ hydrates are the solubility of CO$_{2}$ in water when the aqueous solution is in contact with the CO$_{2}$ liquid phase and with the hydrate phase. In the first case, the solubility of CO$_{2}$ increases as the temperature is decreased. In the second case, as it occurs for the methane hydrate, there is little or no information from computer simulations or experiments. Here we determine the solubility of CO$_{2}$ in water from the hydrate along the isobar of $400\,\text{bar}$. This will allow us to estimate the dissociation line of the hydrate at this pressure, as we have done in our previous work for the case of the methane hydrate.~\cite{Grabowska2022a}

The dissociation line of the CO$_{2}$ hydrate has been already determined by us several years ago.~\cite{Miguez2015a} It is important to mention that also other authors have obtained similar results using computer simulations~\cite{Constandy2015a} and free energy calculations.~\cite{Waage2017a} Our previous results are slightly different than those found by Costandy and coworkers and Waage an collaborators since unlike dispersive interactions between water and CO$_{2}$ are different. However, we follow our previous work~\cite{Grabowska2022a} and determine the dissociation line of the hydrate using the solubility curve of CO$_{2}$ in the aqueous solution when it is in contact with the CO$_{2}$ liquid phase and the hydrate. We have found that the new estimations agree with the initial prediction of M\'{\i}guez \emph{et al.}~\cite{Miguez2015a} within the corresponding uncertainties.

The formation of the CO$_{2}$ hydrate can be viewed as a chemical reaction in which water and CO$_{2}$ molecules ``react'' in the aqueous solution phase to form hydrate molecules.~\cite{Kashchiev2000a,Kashchiev2002a,Kashchiev2002b,Kashchiev2003a} The change in chemical potential of this reaction is the driving force for nucleation, $\Delta\mu_{N}$. It is difficult to get good estimates of $\Delta\mu_{N}$ from experiments since it requires accurate values for a number of thermodynamic properties, including enthalpies and volumes of reactions, among other magnitudes.~\cite{Kashchiev2002a} Here we use the three independent routes introduced in our previous paper~\cite{Grabowska2022a} to deal with the nucleation driving force for the nucleation of CO$_{2}$ hydrates. Particularly, we calculate the driving force for nucleation with respect to the state on the H-L$_{\text{w}}$-L$_{\text{CO}_{2}}$ three-phase line at $400\,\text{bar}$. Note that this point is well above the two quadruple points Q$_{1}$ and Q$_{2}$ shown in Fig.~\ref{figure1}. In addition to this, we also propose a novel and alternative thermodynamic route based on the use of the solubility curve of CO$_{2}$ with the hydrate. This new route, that considers rigorously the non-ideality of the aqueous solution of CO$_{2}$ and provides reliable results of the driving force for nucleation, can be also used to determine $\Delta\mu_{N}$ of other hydrates.

The organization of this paper is as follows: In Sec. II, we describe the methodology used in this work. The results obtained, as well as their discussion, are described in Sec. III. Finally, conclusions are presented in Sec. IV.

\section{Methodology}

We use the GROMACS simulation package~\cite{VanDerSpoel2005a} to perform MD simulations. Computer simulations have been performed using three different versions of the $NPT$ or isothermal-isobaric ensemble. For pure systems which exhibit fluid phases (pure water and pure CO$_{2}$) and aqueous solutions of CO$_{2}$ which exhibit bulk phases, we use the standard isotropic $NPT$ ensemble, i.e., the three sides of the simulation box are changed proportionally to keep the pressure constant. For the hydrate phase, we use the anisotropic $NPT$ ensemble in which each side of the simulation box is allowed to fluctuate independently to keep the pressure constant. This ensures that the equilibrated solid phase has no stress and that the thermodynamic properties are correctly estimated. The same ensemble is used to simulate the two-phase equilibrium between the hydrate and the aqueous solution of CO$_{2}$ (SL coexistence). Finally, the two-phase equilibrium between the solution and the CO$_{2}$ liquid phase is obtained using the $NP_{z}\mathcal{A}T$ ensemble in which only the side of the simulation box perpendicular to the LL planar interface is allowed to change, with the interface area kept constant, to keep the pressure constant. For simulations involving LL and SL interfaces, we have used sufficiently large values of interfacial areas $\mathcal{A}$. The thermodynamics and interfacial properties obtained from simulations of LL interfaces do not show a dependence on the surface area for systems with $\mathcal{A}>10\times 10\,\sigma^{2}$.~\cite{Chen1995a,Gonzalez-Melchor2005a,Janecek2009a} Here $\sigma$ is the largest Lennard-Jones diameter of the intermolecular potentials. In all simulations, the $\mathcal{A}$ values used are higher than this value for LL and SL interfaces. See Sections III.A and III.C for the particular values used in this work.

In all simulations, we use the Verlet leapfrog\cite{Cuendet2007a} algorithm with a time steps of $2\,\text{fs}$. We use a Nos\'e-Hoover thermostat,~\cite{Nose1984a} with a coupling time of $2\,\text{ps}$ to keep the temperature constant. In addition to this, we also use the Parrinello-Rahman barostat~\cite{Parrinello1981a} with a time constant equal to $2\,\text{ps}$ to keep the pressure constant. We use two different cutoff distances for the dispersive and coulombic interactions, $r_{c}=1.0$ and $1.9\,\text{nm}$.
We use periodic boundary conditions in all three dimensions. The water-water, CO$_{2}$-CO$_{2}$, and water-CO$_{2}$ long-range interactions due to coulombic forces are determined using the three-dimensional Ewald technique.~\cite{Essmann1995a} Particularly, the real part of the coulombic potential is truncated at the same cutoff as the dispersive interactions. The Fourier term of the Ewald sums is evaluated using the particle mesh Ewald (PME) method. The width of the mesh is $0.1\,\text{nm}$, with a relative tolerance of $10^{-5}$. In some calculations, we also use the standard long-range corrections for the LJ part of the potential to energy and pressure with $r_{c}=1.0\,\text{nm}$. Water molecules are modeled using the TIP4P/Ice model~\cite{Abascal2005b} and the CO$_{2}$ molecules are described using the TraPPE model.~\cite{Potoff2001a} The H$_{2}$O–CO$_{2}$ unlike dispersive energy value is given by the modified Berthelot combining rule, $\epsilon_{12}=\xi(\epsilon_{11}\,\epsilon_{22})^{1/2}$, with $\xi=1.13$. This is the same used by M\'{\i}guez~\emph{et al},~\cite{Miguez2015a} which allows us to predict accurately the three-phase hydrate–water–carbon dioxide coexistence or dissociation line of the CO$_{2}$ hydrate, particularly the coexistence temperature at the pressure considered in this work, $400\,\text{bar}$ (see Fig.~10 and Table II of the work of M\'{\i}guez and co-workers for further details). Very recently, we have demonstrated that the same molecular parameters are able to predict accurately the CO$_{2}$ hydrate-water interfacial free energy.~\cite{Algaba2022b,Zeron2022a}

Finally, uncertainties are estimated using standard deviation of the mean values or sub-block average method. Particularly, bulk densities in the LL and SL coexistence studies are obtained by averaging the corresponding density profiles over the appropriate regions sufficiently away from the interfacial regions. The statistical uncertainties of these values are estimated from the standard deviation of the mean values. Solubilities of CO$_{2}$ in all the liquid phases are calculated as molar fractions from the densities of both components and the corresponding errors are obtained from propagation of uncertainty formulae. Uncertainties associated to LL interfacial tension values, molar enthalpies, and partial molar enthalpies are estimated using the standard sub-block average method. Particularly, the production periods are divided into $10$ (independent) blocks. The statistical errors are estimated from the standard deviation of the average.

\section{Results}

\subsection{Solubility of carbon dioxide in water from the CO$_{2}$ liquid phase}

We first concentrate on the solubility of CO$_{2}$ in the aqueous solution when the system exhibits LL immiscibility. In this case, there exists a coexistence between the water-rich and CO$_{2}$ liquid phases. We have used the direct coexistence technique to determine the solubility of CO$_{2}$ in the aqueous solution from the CO$_{2}$ liquid phase at several temperatures along the $400\,\text{bar}$ isobar. Particularly, we have performed MD $NP_{z}\mathcal{A}T$ simulations to ensure that temperature and pressure are constant. According to this, the planar interfacial area $\mathcal{A}=L_{x}\times L_{y}$ is kept constant and only $L_{z}$ is varied along each simulation. Here, $L_{x}$, $L_{y}$, and $L_{z}$ are the dimensions of the simulation box along the $x$-, $y$-, and $z$-axis, respectively. In this work, the $z$-axis is chosen to be perpendicular to the planar interface. The initial simulation box is prepared in the following way. We build a slab of $2800$ water molecules in contact, via a planar interface, with a second slab of $1223$ molecules of CO$_{2}$. The dimensions of $L_{x}$ and $L_{y}$ of all the simulation boxes used in this part of the work are kept constant with $L_{x}=L_{y}=3.8\,\text{nm}$ ($\mathcal{A}\simeq 12\times 12\,\sigma^{2}$). Since the pressure is constant, $L_{z}$ varies along each simulation for all the temperatures considered. In this work, $L_{z}$ varies from $11.06$ to $12.29\,\text{nm}$. Simulations to calculate solubilities are run during $100\,\text{ns}$. The first $20\,\text{ns}$ are used to equilibrate the system and the last $80\,\text{ns}$ are used as the production period to obtain the properties of interest. We have also determined the LL interfacial tension and details of the simulations are explained later in this section.

\begin{figure}
\hspace*{-0.2cm}
\includegraphics[width=1.1\columnwidth]{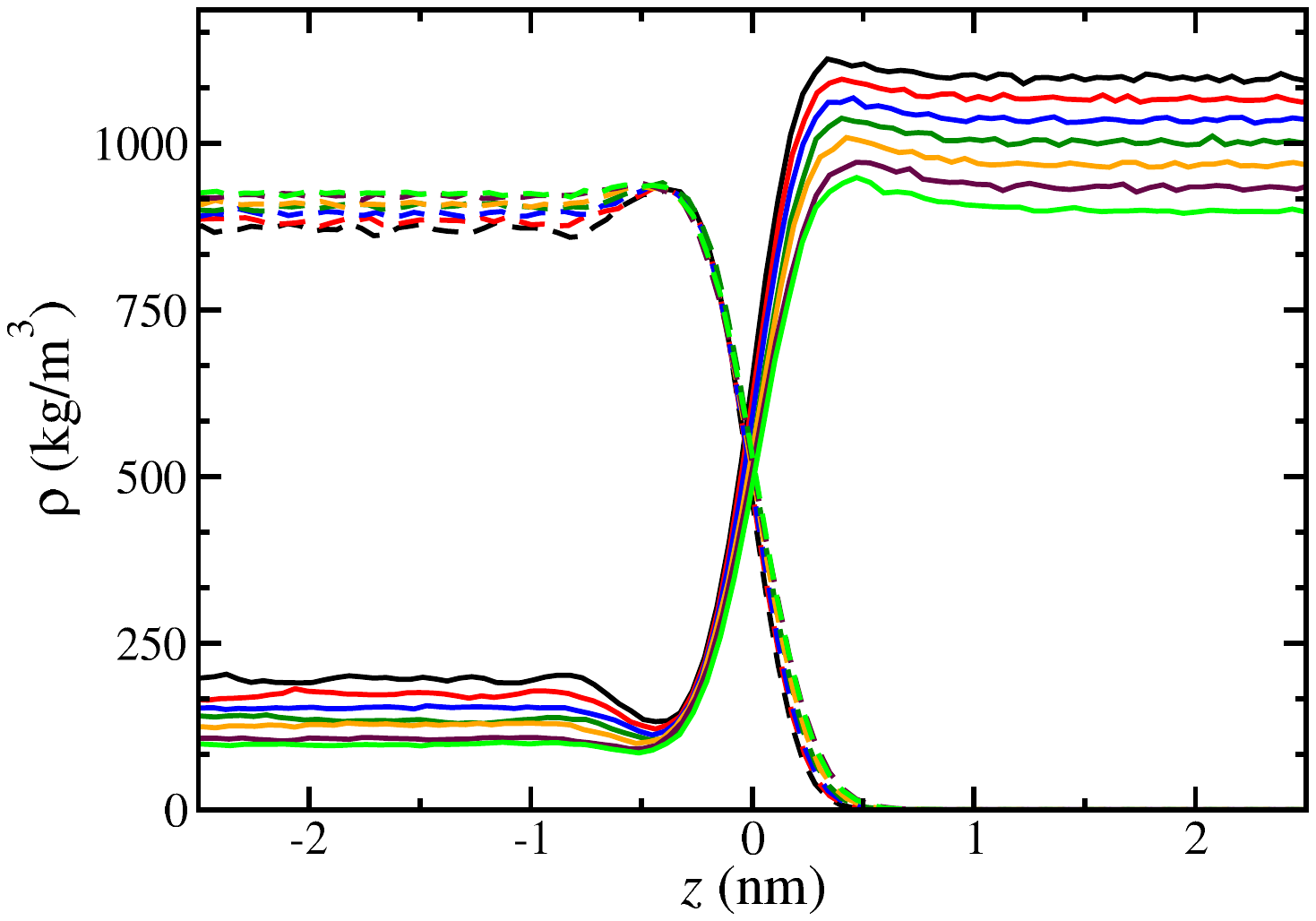}
\caption{Simulated equilibrium density profiles, $\rho(z)$, across the LL interface of CO$_{2}$ (continuous curves) and water (dashed curves) as obtained from MD $NP_{z}\mathcal{A}T$ simulations at $400\,\text{bar}$ and $250$ (black), $260$ (red), $270$ (blue), $280$ (dark green), $290$ (orange), $300$ (maroon), and $310\,\text{K}$ (green).}
\label{figure2}
\end{figure}

Figure~\ref{figure2} shows the density profiles of water and CO$_{2}$ as obtained by MD $NP_{z}\mathcal{A}T$ simulations at $400\,\text{bar}$ and temperatures from $250$ to $310\,\text{K}$. For a better visualization for the reader, we only plot half of the profiles corresponding to one of the interfaces exhibited by the system. The right side of the figure corresponds to the CO$_{2}$ liquid phase and the left side to the aqueous liquid phase. We divide the inhomogeneous simulation box into 200 parallel slabs along the $z$--direction, perpendicular to the planar LL interface, to study the density profiles. Following the standard approach, density profiles are obtained assigning the position of each interacting site to the corresponding slab and constructing the molecular density from mass balance considerations.

As can be seen, density profiles of water (dashed curves) exhibit preferential adsorption at the interface at all temperatures. Particularly, water molecules are accumulated at the aqueous phase side of the interface. The relative maximum, which is identified with the accumulation of the water molecules, increases as the temperature of the system is decreased. The bulk density of water in the aqueous solution of CO$_{2}$ (left side of the figure) slightly decreases as the temperature is lower, especially in the range of $250-290\,\text{K}$. As can be seen, density profiles of water are nearly equal to zero at the bulk CO$_{2}$ liquid phase, indicating that solubility of water in that phase is completely negligible. M\'{\i}guez \emph{et al.}~\cite{Miguez2014a} have previously studied the LL interface of aqueous solutions of CO$_{2}$ at similar temperatures ($287$ and $298\,\text{K}$) but at lower pressures ($P\le 55\,\text{bar}$). These authors have found similar behavior for the density profiles of water but with an important exception: they exhibit the traditional shape of the hyperbolic tangent function in which water density decreases monotonically from the bulk density of water in the aqueous phase to zero in the CO$_{2}$ liquid phase.

The behavior and structure of the density profiles of CO$_{2}$ (continuous curves) along the interface are similar to those exhibited by other mixtures but with an important exception: CO$_{2}$ molecules exhibit activity on both sides of the liquid--liquid interface of the system. Particularly, there is an accumulation of CO$_{2}$ molecules at the CO$_{2}$ liquid phase side of the interface. This accumulation increases as the temperature of the system is decreased, as it happens with water molecules on the other side of the interface. The bulk density of CO$_{2}$ in both phases increases as the temperature is increased. This variation is more important in the CO$_{2}$ liquid phase (right side of Fig.~\ref{figure2}). Contrary to what happens with water density in the CO$_{2}$ phase, the density of CO$_{2}$ in the aqueous solution is not negligible. This indicates that although the solubility of CO$_{2}$ in water is small (molar fraction of CO$_2$ between $0.04$ and $0.09$ in the range $310$--$250\,\text{K}$, respectively), its value is not so low as in the case of the solubility of water in CO$_{2}$. 

It is interesting to mention that density profiles of CO$_{2}$ also exhibit depressions in the aqueous solution side of the interface indicating desorption of CO$_{2}$ molecules in this region. The desorption of CO$_{2}$ molecules at the interface is correlated with the preferential adsorption of water molecules since the relative maxima and minima occur at the same position ($z\approx -0.45\,\text{nm}$). Note that preferential adsorption and desorption of CO$_{2}$ molecules at the LL interface of this kind of aqueous solutions has not been previously seen in the literature. Particularly, M\'{\i}guez \emph{et al.}~\cite{Miguez2014a} only observe density profiles that exhibit preferential adsorption at the interface.

\begin{figure}
\hspace*{-0.2cm}
\includegraphics[width=1.1\columnwidth]{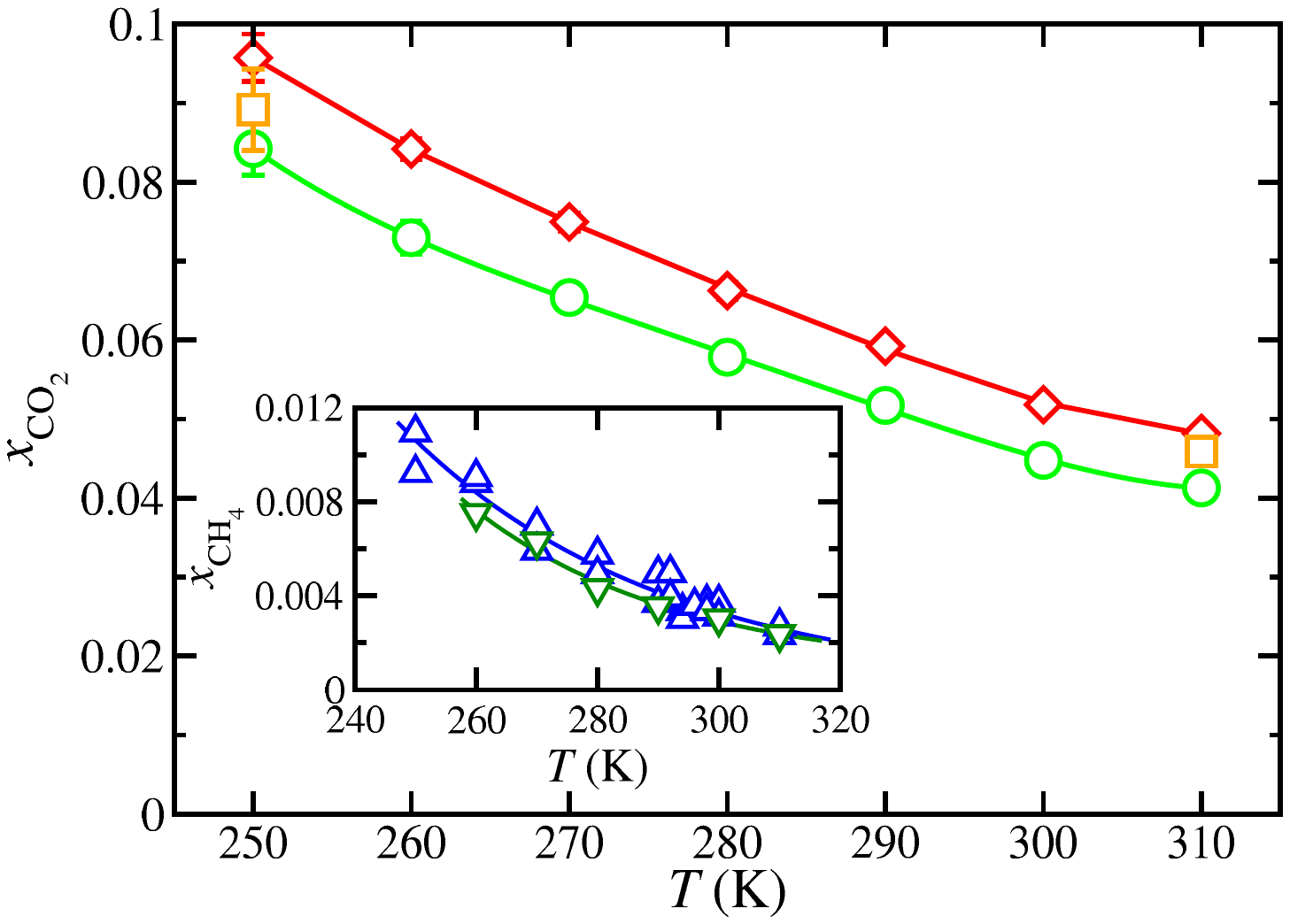}
\caption{Solubility of CO$_{2}$ in the aqueous phase, as a function of temperature, at $400\,\text{bar}$, when the solution is in contact with the CO$_{2}$ liquid phase via a planar interface. The symbols correspond to solubility values obtained from MD $NP_{z}\mathcal{A}T$ simulations using a cutoff of $1.0$ (green circles) and of $1.9\,\text{nm}$ (red diamonds). Orange squares correspond to simulation results using a cutoff of $1.0\,\text{nm}$ and long-range corrections. Inset: solubility of methane in water, as a function of temperature, at $400\,\text{bar}$ when the solution is in contact with the methane gas phase via a planar interface. Solubility values of methane in water are taken from our previous work.~\cite{Grabowska2022a} Simulations are performed at the same conditions using a cutoff of $0.9\,\text{nm}$ (blue trianges up) and  $1.7\,\text{nm}$ (dark green triangles down). In all cases, the curves are included as a guide to the eyes.}
\label{figure3}
\end{figure}

The solubilities of CO$_{2}$ in the aqueous phase have been determined from the information of the density profiles presented in Fig.~\ref{figure2} at the corresponding temperatures.
Fig.~\ref{figure3} shows the solubilities of CO$_{2}$ along the isobar at the temperatures considered. We have also included in the figure (inset) the same results obtained previously by us corresponding to the methane + water system.~\cite{Grabowska2022a} As can be seen, the solubility decreases as temperature increases. This result is in agreement with our previous work in which we considered the solubility of methane in water along the same isobar ($400\,\text{bar}$) and in contact with the gas phase (see the inset).~\cite{Grabowska2022a} In this work, the study is firstly done using a relatively short cutoff distance for dispersive interactions, $r_{c}=1.0\,\text{nm}$, which corresponds to a reduced cutoff value of $r^{*}=r_{c}/\sigma\approxeq 3.16$, with $\sigma=0.31668\,\text{nm}$. Here $\sigma$ is the length scale of the Lennard-Jones intermolecular interactions associated with the water model (TIP4P/ice).~\cite{Abascal2005b} In order to evaluate the effect of the cutoff distance, we have also determined the solubility of CO$_{2}$ using a larger cutoff value for the dispersive interactions ($1.9\,\text{nm}$ instead of $1.0\,\text{nm}$). As can be seen, the effect of increasing the cutoff is important in the whole range of temperatures considered. In particular, solubility increases between $17\,\%$, at high temperatures ($310\,\text{K}$), and $13\,\%$ at low temperatures ($250\,\text{K}$). This is an expected result according to previous studies of the effect of the cutoff distance on fluid-fluid coexistence.~\cite{Trokhymchuk1999a,Miguez2013a,Martinez-Ruiz2014a}

We have checked that there is no a priori temperature limit to perform the simulations as the temperature is decreased. From this point of view, the solubility of CO$_{2}$ in the aqueous solution can be computed without any difficulty since we do not observe nucleation of the hydrate at low temperatures. This is in agreement with previous results obtained by Grabowska \emph{et al.}~\cite{Grabowska2022a} However, as the temperature is decreased the dynamics of the system slows down and the equilibration of the LL interface becomes more difficult and longer simulation runs are required to achieve equilibrated density profiles.

We have also determined the solubility using the standard long-range corrections to energy and pressure
to the Lennard-Jones part of the potential (dispersive interactions). According to our results,
although long-range corrections are able to improve the solubility results, differences between these results and those obtained with a cutoff of  $1.9\,\text{nm}$  are still noticeable. In particular, the solubilities predicted using this approach are underestimated between $4$ and $6\,\%$ along the isobar
at the temperatures simulated. It is interesting to compare the behavior of the CO$_{2}$ solubility, as a function of the temperature along the $400\,\text{bar}$, with that corresponding to methane obtained by us previously.~\cite{Grabowska2022a} As can be seen in the inset of Fig.~\ref{figure3}, the effect of the long-range correction on the dispersive interactions is slightly larger in the case of CO$_{2}$ than in methane. This is an expected result since the CO$_{2}$ molecules are modeled using three Lennard-Jones interaction sites and methane only with one, and also because the solubility of CO$_{2}$ is about ten times higher than the CH$_4$ solubility at the same thermodynamic conditions. It is also remarkable that the use of longer cutoff distances has a contrary effect on the solubility of CO$_{2}$ than in that of methane, i.e., the solubility of CO$_{2}$ increases with an increase of the cutoff whereas it decreases in the case of the solubility of methane. This is probably due to the presence of the quadrupolar moment of the CO$_{2}$ molecule and to the water--CO$_{2}$ interactions.

In the previous paragraphs, we have presented and discussed the results corresponding to the solubility of CO$_{2}$ in the aqueous solution when it is in contact with the CO$_{2}$ liquid phase. Since both phases are in contact through a planar LL interface and are in equilibrium at the same $P$ and $T$, the chemical potential of water and CO$_{2}$ in both phases must satisfied that,

\begin{equation}
\mu_{\text{CO}_{2}}^{I}(P,T,x_{\text{CO}_{2}}^{I})= \mu_{\text{CO}_{2}}^{II}(P,T,x_{\text{CO}_{2}}^{II})
\label{cp-co2}
\end{equation}

\noindent
and

\begin{equation}
\mu_{\text{H}_{2}\text{O}}^{I}(P,T,x_{\text{CO}_{2}}^{I})= \mu_{\text{H}_{2}\text{O}}^{II}(P,T,x_{\text{CO}_{2}}^{II})
\label{cp-water}
\end{equation}

\noindent
Here the superscripts $I$ and $II$ label the aqueous and the
CO$_{2}$ liquid phases, respectively. Note that we have expressed the chemical potential of water in each phase in terms of the corresponding CO$_{2}$ molar fractions. It is important to note that this is consistent from the thermodynamic point of view and it is always possible since we are dealing with a binary system that exhibits two-phase equilibrium. Following the Gibbs phase rule, such a system has two degrees of freedom, which according to Eqs.~\eqref{cp-co2} and \eqref{cp-water} are $P$ and $T$. Consequently, the thermodynamic behavior of the system is fully described solving the previous equations since the composition of water in both phases, $x_{\text{H}_{2}\text{O}}^{I}$ and $x_{\text{H}_2\text{O}}^{II}$, can be readily obtained as $x_{\text{H}_{2}\text{O}}^{I}=1-x_{\text{CO}_{2}}^{I}$ and 
$x_{\text{H}_{2}\text{O}}^{II}=1-x_{\text{CO}_{2}}^{II}$.

According to our previous results shown in Fig.~\ref{figure2}, the density of water in the CO$_{2}$ liquid phase is $\rho_{\text{H}_{2}\text{O}}^{II}\approx 0$, and consequently, $x_{\text{H}_2\text{O}}^{II}=\rho_{\text{H}_2\text{O}}^{II}/(\rho_{\text{H}_2\text{O}}^{II}+\rho_{\text{CO}_{2}}^{II})\approx 0$
and $x_{\text{CO}_{2}}^{II}=1-X_{\text{H}_2\text{O}}^{II}\approx 1$.

Following the approximations of the previous paragraph combined with Eq.~\eqref{cp-co2}, the chemical potential of CO$_{2}$ in the aqueous solution can be obtained from the chemical potential of pure CO$_{2}$ at the same $P$ and $T$,

\begin{equation}
\mu_{\text{CO}_{2}}^{I}(P,T,x_{\text{CO}_{2}}^{I})\approx
\mu_{\text{CO}_{2}}^{II}(P,T,x_{\text{CO}_{2}}^{II}\approx 1)\approx
\mu_{\text{CO}_{2}}^{II}(P,T)
\end{equation}

\begin{figure}
\hspace*{-0.2cm}
\includegraphics[width=1.1\columnwidth]{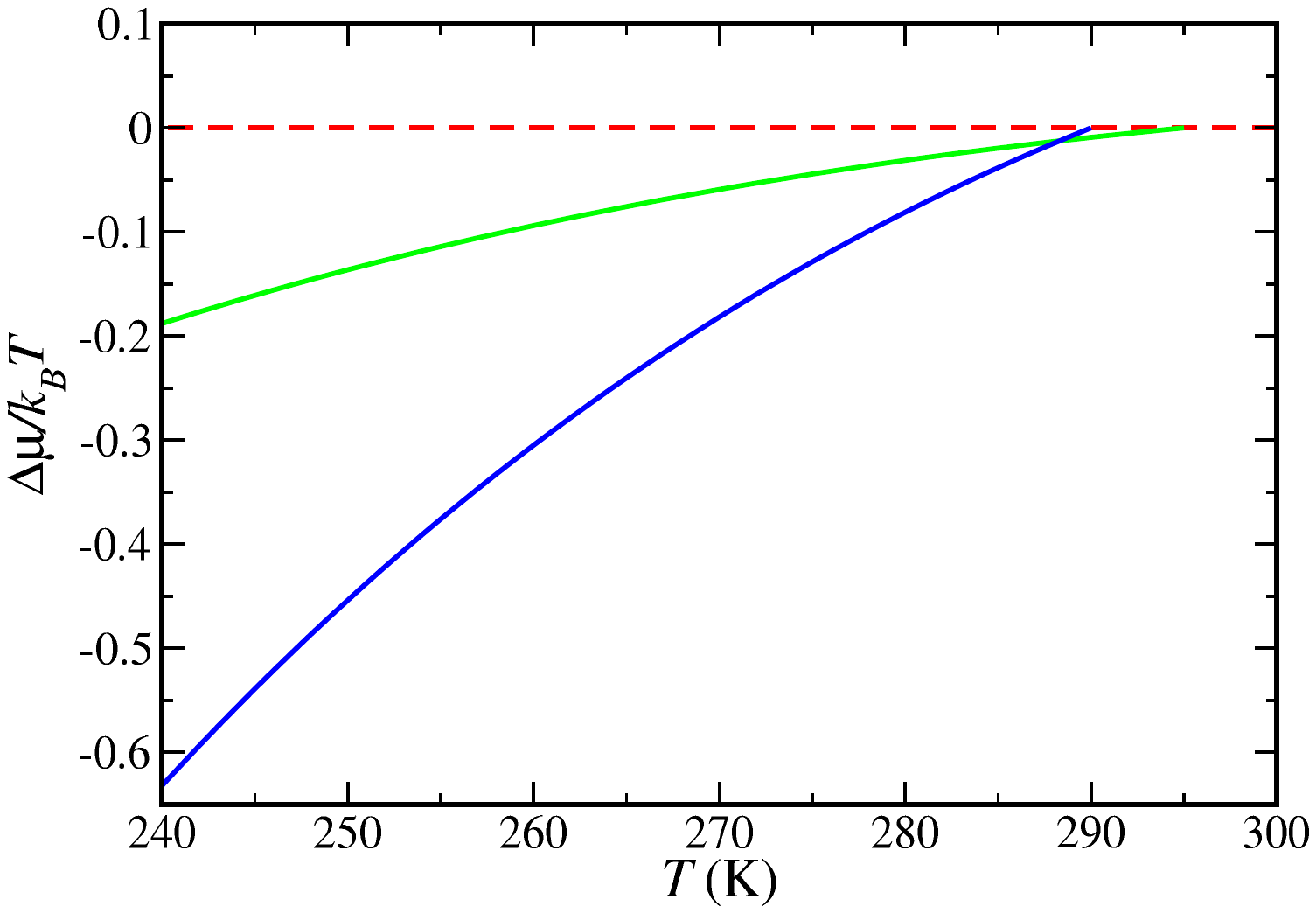}
\caption{Change of chemical potential, $\Delta\mu$, as a function of temperature, of the bulk CO$_{2}$ (blue curve) and the bulk methane (green curve) obtained along the isobar $400\,\text{bar}$. We have arbitrarily set the value of the chemical potential of CO$_{2}$ and methane to zero at $290$ and $295\,\text{K}$, respectively. Chemical potential values of methane are taken from our previous work.~\cite{Grabowska2022a}}
\label{figure4}
\end{figure}

The chemical potential of CO$_{2}$ along the isobar can be obtained from the thermodynamic relation,

\begin{equation}
\Biggl(\dfrac{\partial(\mu_{\text{CO}_{2}}/T)}{\partial T}\Biggr)_{P,N_{\text{H}_{2}\text{O}},N_{\text{CO}_{2}}}=-\dfrac{h_{\text{CO}_{2}}}{T^{2}}
\label{enthalpy}
\end{equation}

\noindent
where $h_{\text{CO}_{2}}=h_{\text{CO}_{2}}(P,T)$ is the partial molar enthalpy of CO$_{2}$, and the derivative is performed at constant pressure, $P$, and number of water and CO$_{2}$ molecules, $N_{\text{H}_{2}\text{O}}$ and $N_{\text{CO}_{2}}$, respectively. Since in our case the CO$_{2}$ liquid phase is essentially a pure CO$_{2}$ liquid, $h_{\text{CO}_{2}}$ is simply the molar enthalpy. Consequently, the chemical potential of CO$_{2}$, as a function of the temperature, along the $400\,\text{bar}$ isobar can be obtained by integrating the Eq.~\eqref{enthalpy} as,

\begin{equation}
\dfrac{\mu_{\text{CO}_{2}}(T)}{k_{B}T}=\dfrac{\mu_{\text{CO}_{2}}(T_{0})}{k_{B}T_{0}}-\bigintsss_{T_{0}}^{T} \dfrac{h_{\text{CO}_{2}}(T')}{k_{B}T'^{2}}\,dT'
\label{integral_co2}
\end{equation}

\noindent
where $k_{B}$ is the Boltzmann constant and $T_{0}$ is a certain reference temperature. Following our previous work~\cite{Grabowska2022a}, we set $\mu_{\text{CO}_{2}}(T_{0})=0$. According to Eq.~\eqref{integral_co2}, $\mu_{\text{CO}_{2}}(T)$ can be obtained by performing MD $NPT$ simulations of pure CO$_{2}$ along the $400\,\text{bar}$ isobar. In this case, since we are simulating a bulk phase, the standard $NPT$ is used in such a way that the three dimensions of the simulation box are allowed to fluctuate isotropically. We use a cubic simulation box with $1000$ CO$_{2}$ molecules. The dimensions of the simulation box, $L_{x}$, $L_{y}$, and $L_{z}$ vary depending on the temperature from $3.9$ to $4.2\,\text{nm}$. Simulations to calculate the molar enthalpy, at each temperature, are run during $100\,\text{ns}$, $20\,\text{ns}$ to equilibrate the system and $80\,\text{ns}$ as the production period to obtain $h_{\text{CO}_{2}}$. As in our previous work, we have not included the kinetic energy contribution (i.e. $5/2k_{B}T$ in the case of a rigid diatomic molecule such as CO$_{2}$). Note that this contribution is canceled out since we are evaluating chemical potential differences at constant $P$ and $T$. In this work we choose as the reference temperature $T_{0}=290\,\text{K}$. The reason for this selection will be clear later in the manuscript. Figure~\ref{figure4} shows the chemical potential of CO$_{2}$ as a function of the temperature (blue curve). We have also included in the same figure the chemical potential values of the bulk methane taken from our previous work~\cite{Grabowska2022a} (green curve) in order to compare both chemical potentials. Note that the reference temperature at which $\mu$ of the bulk methane is set to zero is $295\,\text{K}$.

\subsection{LL interfacial free energy}

From the same simulations we have also obtained the LL interfacial tension, $\gamma$, from the
diagonal components of the pressure tensor. The vapour pressure corresponds to the normal component,
$P\equiv P_{zz}$, of the pressure tensor. The interfacial tension is obtained using the well-known combination of the normal component and the tangential components, $P_{xx}$ and $P_{yy}$ through the mechanical route as, \cite{Hulshof1901a,Rowlinson1982b,deMiguel2006a,deMiguel2006b}

\begin{equation}
\gamma=\frac{L_{z}}{2}\left[\left<P_{zz}\right>-\frac{\left<P_{xx}\right>+\left<P_{yy}\right>}{2}\right]
\label{iftmd}
\end{equation}

\begin{figure}
\hspace*{-0.2cm}
\includegraphics[width=1.1\columnwidth]{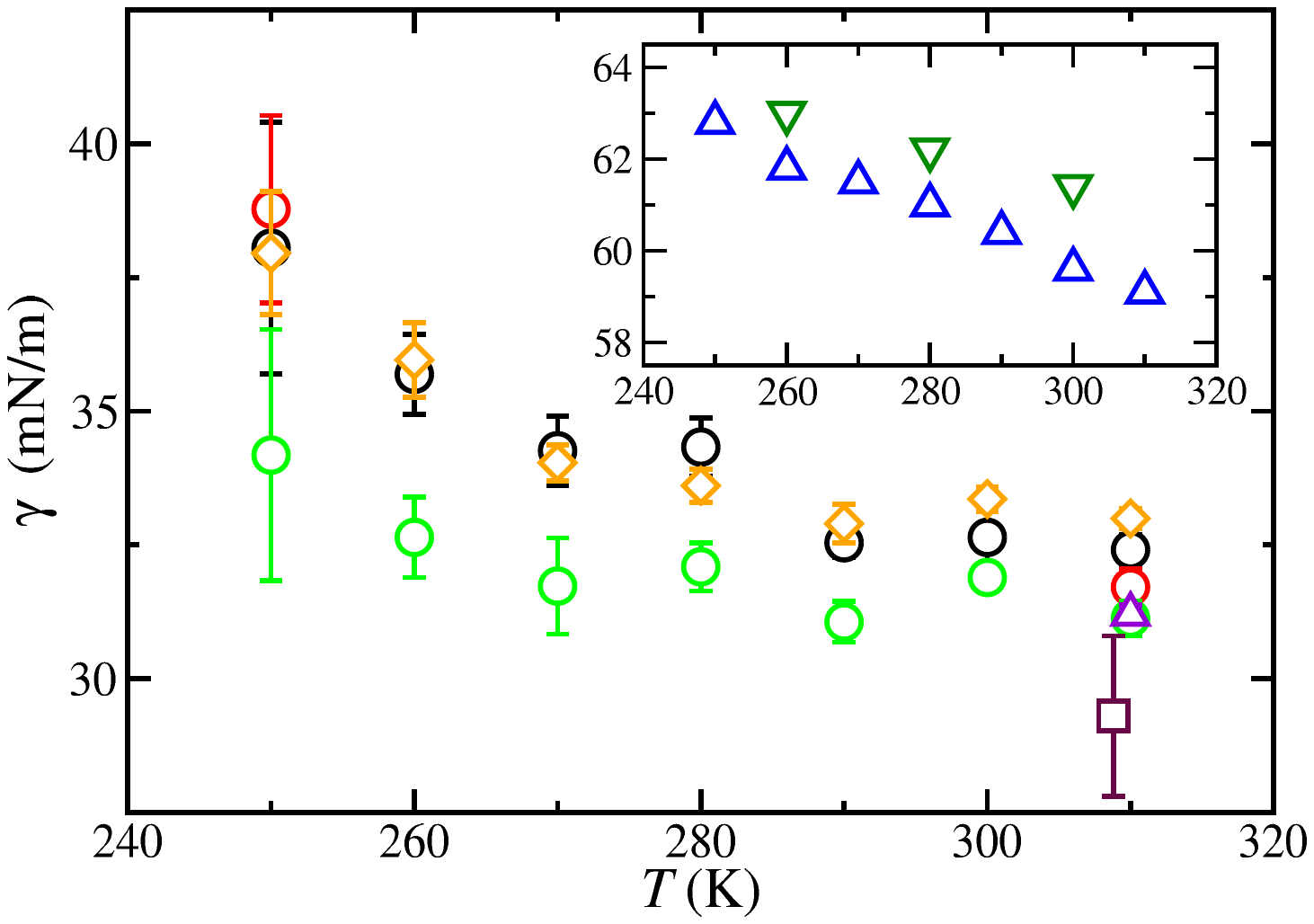}
\caption{LL interfacial tension values, $\gamma$, as a function of the temperature between the CO$_{2}$ liquid phase and the aqueous solution at $400\,\text{bar}$ obtained from MD $NP_{z}\mathcal{A}T$ simulations using a cutoff of $1.0\,\text{nm}$ (black circles for $50\,\text{ns}$ of equilibration and $50\,\text{ns}$ of production; orange diamonds for $150\,\text{ns}$ of equilibration and $150\,\text{ns}$ of production) and of $1.9\,\text{nm}$ (green circles for $50\,\text{ns}$ of equilibration and $50\,\text{ns}$ of production; violet triangles up for $150\,\text{ns}$ of equilibration and $150\,\text{ns}$ of production). Red circles correspond to simulation results using a cutoff of $1.0\,\text{nm}$ and long-range corrections ($50\,\text{ns}$ of equilibration and $50\,\text{ns}$ of production). The maroon square represents the experimental data taken from the literature.~\cite{Chiquet2007a} Inset: LL interfacial tension values of the methane + water mixture at $400\,\text{bar}$ taken from our previous work.~\cite{Grabowska2022a} Simulations are performed at the same pressure and similar temperature conditions using a cutoff of $0.9\,\text{nm}$ (blue trianges up) and  $1.7\,\text{nm}$ (dark green triangles down).
}
\label{figure5}
\end{figure}

\noindent
In Eq.~\eqref{iftmd}, the factor $1/2$ reflects that during the simulations there exist two LL interfaces in the system, being $L_{z}$ the size of the simulation box in the $z$ direction perpendicular to the planar interface. Fig.~\ref{figure5} shows the LL interfacial tension value as obtained from MD $NP_{z}\mathcal{A}T$ simulations. Results obtained in our previous work corresponding to the LL interfacial tension of the methane + water mixture are also shown in the inset of the figure.~\cite{Grabowska2022a} The interfacial tension decreases as the temperature is increased. We first calculate the interfacial tension using a cutoff value of $1.0\,\text{nm}$ (green circles). In this case, we have used $50\,\text{ns}$ for the equilibration period and $50\,\text{ns}$ more for the production period in which the averages are calculated. Our results indicate that the values exhibit large fluctuations, especially at low temperatures. To improve our results, we have extended the simulations. The first $150\,\text{ns}$ correspond to the equilibration period and the extra $150\,\text{ns}$ are used to obtain the corresponding average values (production period). As can be seen (green diamonds), although the mean values obtained in both cases are similar, the error bars decrease, especially at the lowest temperature.

It is well-known that the equilibrium interfacial tension value associated with an interface critically depends on the molecular details. In particular, its value is very sensitive to the cutoff due to the dispersive interactions used during the simulations.~\cite{Trokhymchuk1999a,Janecek2006a,Shen2007a,Blas2008a,MacDowell2009a,Miguez2013a} To account for the long-range interactions associated with the dispersive interactions, we have performed simulations using a cutoff distance of $1.9\,\text{nm}$, with $50\,\text{ns}$ for equilibration and another $50\,\text{ns}$ for production time (red circles). Note that in this case we are not using long-range corrections to energy and pressure. This value corresponds to a reduced cutoff distance $r_{c}^{*}=r_{c}/\sigma=6$, which is nearly the double value used in the first set of simulations. As can be seen, the main effect of increasing the cutoff distance is to decrease the interfacial tension values. Particularly, the effect is larger at low temperatures where the difference is about $3-4$ $\text{mJ/m}^{2}$. However, at high temperatures, differences are about $1$ $\text{mJ/m}^{2}$.

In order to account for the effect of the simulation length, we have extended the simulation at $310\,\text{K}$. As in the other set of simulations, we have equilibrated the system during the first $150$ and used the next $150\,\text{ns}$ to perform the corresponding averages (red triangles up). As can be seen, the new results are in practice identical to those obtained using only $50\,\text{ns}$ for production time.

To be consistent with the calculations of the solubility of CO$_{2}$ in the aqueous solution, we have also determined the interfacial tension using the traditional LRC to energy and pressure. As in the previous case, we have only simulated using these corrections at $250$ and $310\,\text{K}$ (orange circles). As can be seen, this approximation is not able to provide consistent results at low temperatures when compared with the data obtained using $r_{c}=1.9\,\text{nm}$. In particular, the interfacial tension is overestimated by more than $4.5$ $\text{mJ/m}^{2}$, which represents more than $13\%$ with respect to the value obtained using the larger cutoff distance. At the highest temperature, however, the overestimation of the interfacial tension is only about $0.5$ $\text{mJ/m}^{2}$ ($2\%$). Discrepancies at low temperatures between the results obtained using the largest cutoff distance ($1.9\,\text{nm}$) and those using the traditional energy and pressure long-range corrections are probably due to the differences in densities in both liquid phases at these conditions.

We have also compared the predictions obtained from MD simulations with experimental data taken from the literature (magenta squares).~\cite{Chiquet2007a} Unfortunately, to the best of our knowledge, there are no experimental data below $310\,\text{K}$. Our results are in good agreement with experimental measurements, although we slightly overestimate it by about 6\%.

Finally, it is also interesting to compare the LL interfacial tension values obtained in this work for the CO$_{2}$ + water system with those obtained in our previous work~\cite{Grabowska2022a} for the methane + water binary mixture. Computer simulation values of the former system are shown in the inset of Fig.~\ref{figure5}. As can be seen, LL interfacial tension values of the methane + water system are approximately twice than those corresponding to the mixture containing CO$_{2}$. But perhaps the most interesting feature is that, as it happens with the solubilities (see Fig.~\ref{figure3}), an increase of the cutoff distance due to the dispersive interactions has the opposite effect in the systems which containing methane and CO$_2$: an increase of the cutoff distance in the CO$_{2}$ system lowers the LL interfacial tension values, while in the methane mixture the interfacial tension values increase when the cutoff is larger (blue triangles up correspond to $r_{c}=0.9\,\text{nm}$ and dark green triangles down to $r_{c}=1.7\,\text{nm}$). Also note that the effect of the long-range dispersive contributions is more important in the CO$_{2}$ + water mixture than in the system containing methane. As we have discussed previously, this could be due to the presence of the electric quadrupole of CO$_{2}$.

\subsection{Solubility of carbon dioxide in  water from the hydrate phase}

We have also determined the solubility of CO$_{2}$ in water when the aqueous solution is in contact with the hydrate along the $400\,\text{bar}$ isobar at several temperatures. We first prepare a simulation box of CO$_{2}$ hydrate replicating a unit cell of hydrate four times along each spatial direction ($4\times 4 \times 4$), using $2944$ water and $512$ CO$_{2}$ molecules. This corresponds to a hydrate with the cages (8 cages per unit cell) fully occupied by CO$_{2}$ molecules. We equilibrate the simulation box for $40\,\text{ns}$ using an anisotropic barostat along the three axes. This allows  the dimensions of the simulation box to change independently. The pressure is the same along the three directions and equal to $400\,\text{bar}$ to allow the solid to relax and avoid any stress. In order to help the system to reach the equilibrium, we also prepare boxes of aqueous solutions with different concentrations of CO$_{2}$ depending on the temperature. This allows to reach the equilibrium as fast as possible in the last stage of the simulations when the hydrate and liquid phases are put in contact (see below). Particularly, the hydrate phase will grow or melt depending on the initial conditions, releasing/absorbing water and CO$_{2}$ molecules to/from the aqueous solution, until the solution phase reaches the equilibrium condition. Although the final state is independent of the initial CO$_{2}$ concentration in the aqueous phase, care must be taken in finite systems as those studied in this work. Initial conditions must be close enough to coexistence so that the system is able to reach equilibrium before exhaustion of any of the phases at coexistence. In this particular work, we have checked that density profiles of water and CO$_{2}$ in the aqueous phase reach the equilibrium value. This is practically done monitoring the averages profiles every $100\,\text{ns}$ until no significant variations in their bulk region are observed. Once densities of water and CO$_{2}$ are obtained, the molar fraction of CO$_{2}$ in the aqueous solution is calculated from the corresponding averaged density values.

We use simulation boxes of solutions containing $4000$ water molecules and varying the number of CO$_{2}$ molecules depending on temperature: $50$ ($250$, $260$, and $270\,\text{K}$), $120$ ($280$ and $290\,\text{K}$), and $240$ ($295\,\text{K}$) CO$_{2}$ molecules. We equilibrate each simulation box during $40\,\text{ns}$ using the isothermic-isobaric or $NP_{z}\mathcal{A}T$ ensemble. In this case, two of the dimensions of the simulation boxes, arbitrarily named $L_{x}$ and $L_{y}$, are kept constant ($L_{x}=L_{y}$ vary between $4.77$ and $4.82\,\text{nm}$ ($\mathcal{A}\simeq 15\times 15\,\sigma^{2}$) depending on the temperature) and equal to the values of two lengths of the simulation box of the hydrate. $L_{z}$ is however allowed to vary to achieve the equilibrium pressure of $400\,\text{bar}$. Particularly, $L_{z}$ varies from $10.28$ to $10.53\,\text{nm}$ depending on the temperature. Finally, the equilibrated hydrate and aqueous solution simulation boxes are assembled along the $z$ direction sharing a planar solid-liquid interface with interfacial area $\mathcal{A}=L_{x}\times L_{y}$. We then perform simulations in the $NPT$ ensemble using an anisotropic barostat with pressures identical in the three directions and equal to $400\,\text{bar}$. This allows the solid to relax and avoid any stress and obtain the correct value of the solubility at each temperature. Systems are equilibrated during $100\,\text{ns}$. After this, we run additional $300\,\text{ns}$ to obtain the equilibrium density profiles of the system from $250$ up to $295\,\text{K}$.

\begin{figure}
\hspace*{-0.2cm}
\includegraphics[width=1.1\columnwidth]{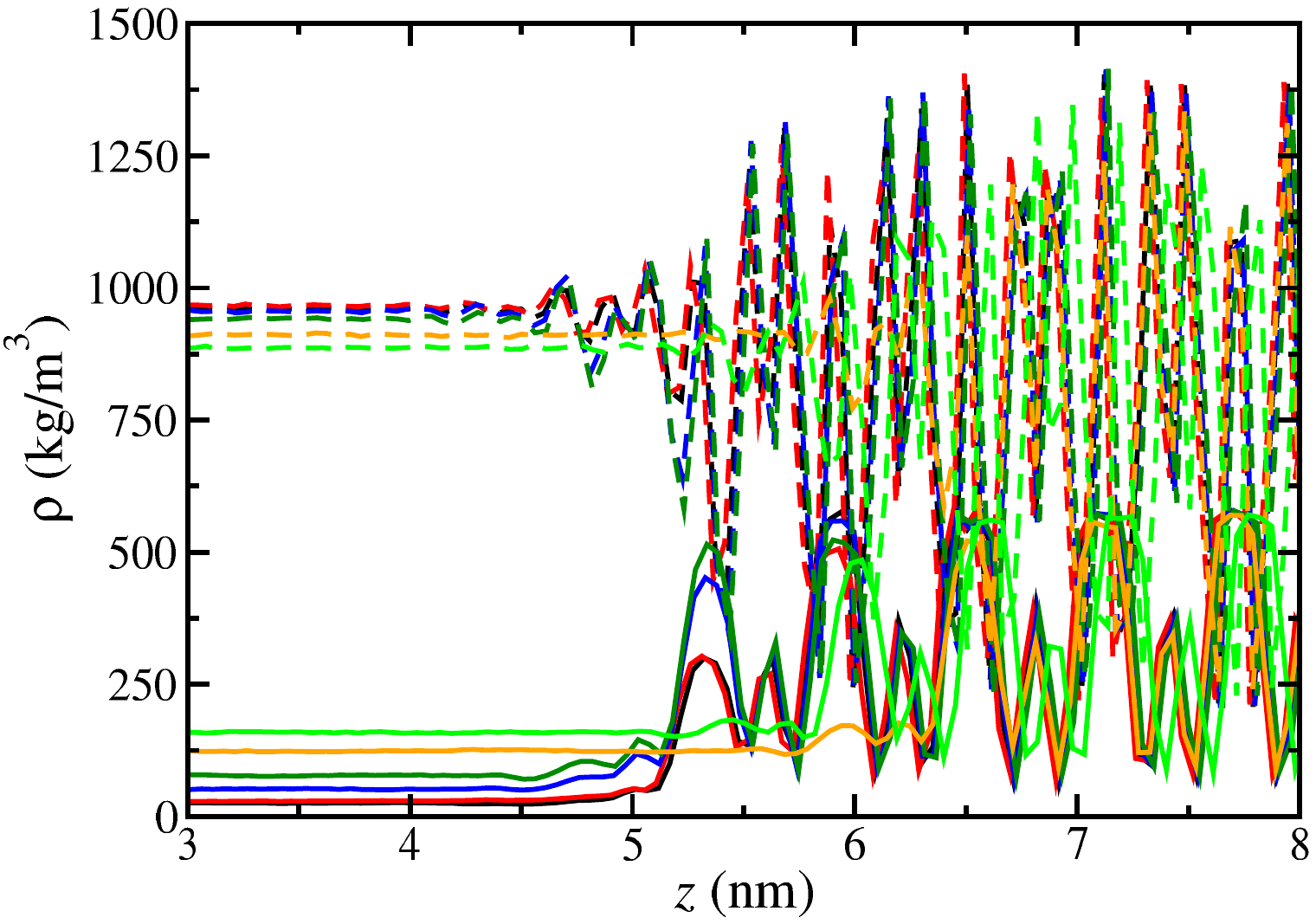}
\caption{Simulated equilibrium density profiles, $\rho(z)$, across the hydrate--CO$_{2}$liquid interface, of CO$_{2}$ (continuous curves) and water (dashed curves) as obtained from MD anisotropic $NPT$ simulations at $400\,\text{bar}$ and $250$ (black), $260$ (red), $270$ (blue), $280$ (dark green), $290$ (orange), and $295\,\text{K}$ (green).}
\label{figure6}
\end{figure}

Fig.~\ref{figure6} shows the density profiles of water and CO$_{2}$ molecules as obtained from anisotropic $NPT$ simulations at $400\,\text{bar}$ and temperatures ranging from $250$ to $295\,\text{K}$. The density profiles have been obtained as explained in section III.A. Note that at temperatures above $295\,\text{K}$ it is not possible to determine the solubility because the hydrate melts. In other words, there is a kinetic limit at high temperatures to determine the solubility of CO$_{2}$ from the hydrate.

As in the case of the LL coexistence described in section III.A, we only plot half of the profiles corresponding to one of the interfaces exhibited by the system. The right side of the figure corresponds to the hydrate phase and the left side to the aqueous solution phase. The density profiles in the hydrate phase exhibit the usual solid-like behavior for water and CO$_{2}$ molecules, with peaks at the corresponding crystallographic equilibrium position at which molecules are located in the hydrate. As can be seen, the density profiles at the lowest temperatures, from $250$ up to $280\,\text{K}$ show nearly the same structure, and only small differences are  observed at the hydrate-solution interface, as it is expected.

It is also interesting to analyze the behavior of profiles of water and CO$_{2}$ in the aqueous phase. The density profiles near the interface show some structural order due to the presence of the hydrate phase. Note that the positional order of the molecules is more pronounced at low temperatures, below $T\le 280\,\text{K}$. The bulk density of water (left side of the figure) slightly decreases as the temperature is increased, especially close to temperatures at which the hydrate melts. It is interesting to mention that bulk density profiles vary with temperature in the opposite way that when the aqueous solution is in contact with the CO$_{2}$ liquid phase (see Fig.~\ref{figure2}).

The bulk density of CO$_{2}$ (in the aqueous solution phase) increases as the temperature is increased. From the inspection of Fig.~\ref{figure6} it is clearly seen that the hydrate phase becomes less stable as the temperature approaches to $290-295\,\text{K}$. At lower temperatures, the hydrate-solution interface is located at $z\approx 5\,\text{nm}$, approximately, with the hydrate phase showing $6-7$ well-defined CO$_{2}$ layers. However, at $290$ and $295\,\text{K}$ only $5-6$ layers can be observed in the hydrate phase, with the interface located at $z\approx 6\,\text{nm}$. In fact, as we have previously mentioned, it is not possible to keep stable the hydrate at temperatures above $295\,\text{K}$, which eventually melts at higher temperatures.

\begin{figure}
\hspace*{-0.2cm}
\includegraphics[width=1.1\columnwidth]{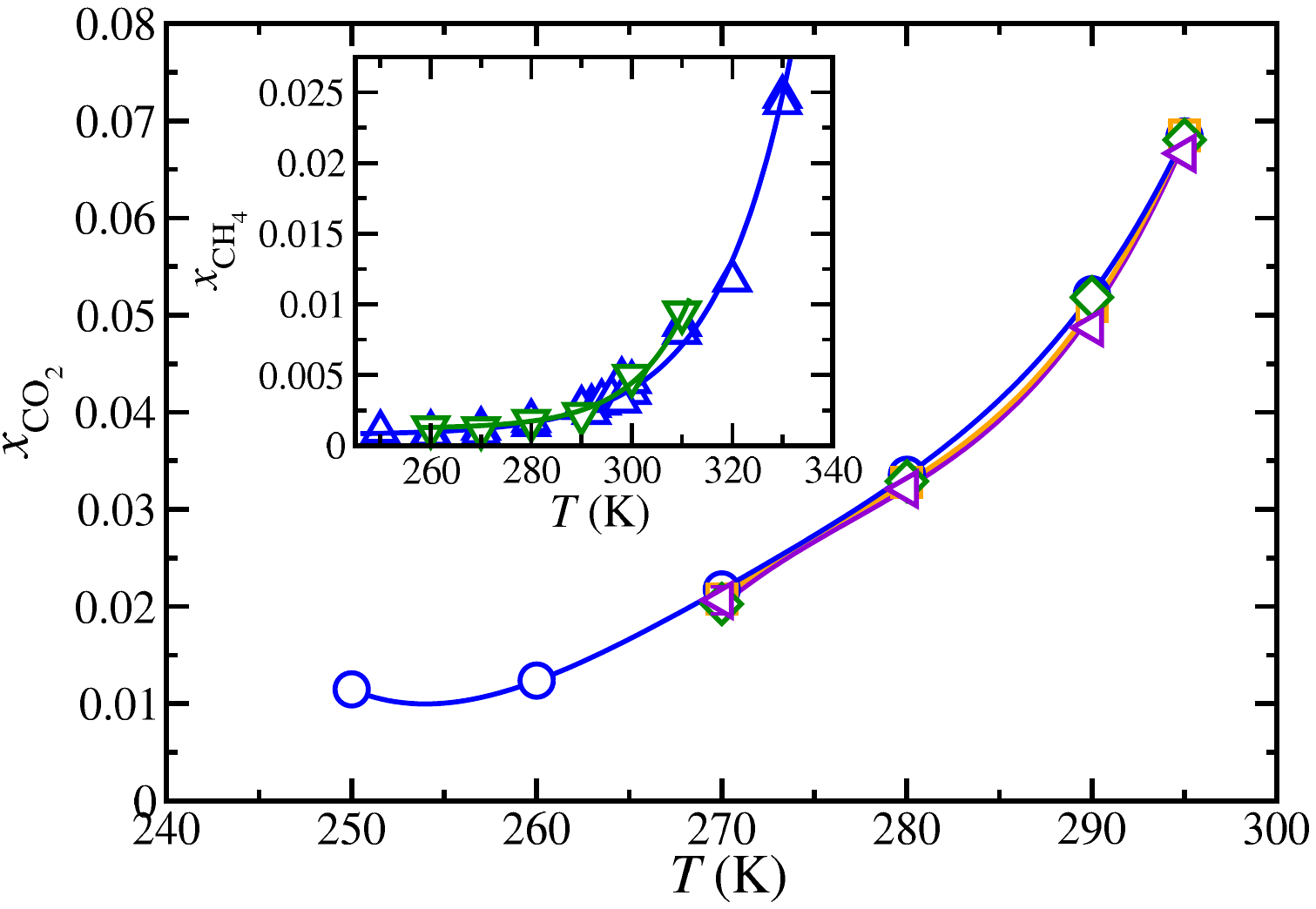}
\caption{Solubility of CO$_{2}$ in the aqueous phase, as a function of temperature, at $400\,\text{bar}$, when the solution is in contact with the hydrate phase via a planar interface. The symbols correspond to solubility values obtained from MD anisotropic $NPT$ simulations using a cutoff of $1.0$ (blue circles) and of $1.9\,\text{nm}$ (dark green diamonds). Orange squares correspond to simulation results using a cutoff of $1.0\,\text{nm}$ and long-range corrections. In all the previous cases, the hydrate is fully occupied by CO$_{2}$ molecules. Violet triangles left correspond to simulation results using a cutoff of $1.0\,\text{nm}$ and an occupancy of $50\%$ of the small or D cages ($87.5\%$ of overall occupancy). Inset: solubility of methane in water, as a function of temperature, at $400\,\text{bar}$ when the solution is in contact with the hydrate phase via a planar interface. Solubility values of methane in water are taken from our previous work.~\cite{Grabowska2022a} Simulations are performed at the same conditions using a cutoff of $0.9\,\text{nm}$ (blue triangles up) and $1.7\,\text{nm}$ (dark green triangles down). In all cases, the curves are included as a guide to the eyes.}
\label{figure7}
\end{figure}

We have calculated, from the information obtained from the density profiles, the solubility of CO$_{2}$ in the aqueous solution when it is in contact with the hydrate. As can be seen in Fig.~\ref{figure7}, the solubility of CO$_{2}$ increases as the temperature is raised. We have also included in Fig.~\ref{figure7} (inset) the same results obtained previously by us corresponding to the methane + water system.~\cite{Grabowska2022a} It can be seen that our results are in agreement with the results of the solubility of methane. Contrary to what happens in the case of the methane hydrate, it is only possible to calculate the solubility up to a temperature of $295\,\text{K}$.
This is just a few degrees (around five) above the values of the three point temperature $T_{3}$ of the CO$_{2}$ hydrate at this pressure (see section III.D). In the case of our previous work, the hydrate was kept in metastable equilibrium at about $35\,\text{K}$ over the dissociation temperature of the methane hydrate ($295\,\text{K}$).~\cite{Grabowska2022a}

We have also considered the impact of using different cutoff distances on solubilities. As in the case of the results presented in Section III.A, we have used three different cutoff distances for the dispersive interactions. Particularly, we use the same two values, $1.0$ and $1.9\,\text{nm}$. We have also performed simulations using the standard long-range corrections to energy and pressure with a cutoff value of $1.0\,\text{nm}$. Contrary to what happens with the solubility of CO$_{2}$ in water when the solution is in contact with the other liquid phase (CO$_{2}$), the solubility does not depend on the cutoff distance, as can be seen in Fig.~\ref{figure7}. Our results indicate that the long-range corrections due to the dispersive interactions have little or negligible effect on solubilities in aqueous solutions in contact with the hydrate. However, according to Figs.~\ref{figure3} and \ref{figure5}, long-range interactions play a key role in the thermodynamic and interfacial properties in systems involving fluid phases.~\cite{Trokhymchuk1999a,Janecek2006a,Shen2007a,Blas2008a,MacDowell2009a,Miguez2013a}

We have considered the effect of the CO$_{2}$ occupancy in the hydrate on the solubility in the aqueous solution. We have prepared the initial simulation boxes in a similar way to the case of full occupancy but with CO$_{2}$ occupying half of the small or D cages. Particularly, we use $2944$ water ($46\times 4\times 4\time 4$) and 448 CO$_{2}$ molecules. This means that the occupancy of the large or T cages is $100\%$ ($384$ CO$_{2}$ molecules) and the occupancy of the small or D cages is $50\%$ ($64$ CO$_{2}$ molecules). This represents a $87.5\%$ of occupancy of D and T cages. According to the experimental data,~\cite{Henning2000a,Udachin2001a,Ikeda1999a,Ripmeester1998a} the equilibrium occupancy of large or T cages of the CO$_{2}$ hydrate is nearly $100\%$. However, although there is a large discrepancy in measurements and predictions of the small cage occupancy, it is generally accepted that occupancy of small or D cages is approximately $30-60\%$ depending on thermodynamic conditions. Notice that with this occupancy (i.e $87.5\%$) the ratio of water to CO$_{2}$ molecules in the hydrate is not $5.75$ (as when the occupancy is $100\%$) but its value is now $46/7 \approx 6.57$. We follow the same procedure explained in the previous paragraph, with a cutoff distance of $1.0\,\text{nm}$. As a result, we obtain similar density profiles to those shown in Fig.~\ref{figure6}. The solubility of CO$_{2}$ in water when the aqueous solution is in contact with the hydrate with $87.5\%$ of occupancy is also shown in Fig.~\ref{figure7}. As can be seen, the solubility of CO$_{2}$ when it is in contact with the hydrate with an occupancy of $87.5\%$ is the same as that when it is in contact with the hydrate fully occupied within the error bars.

Finally, it is important to remark that, contrary to what we have found in our previous work for the solubility of methane in water,~\cite{Grabowska2022a} we do not find a melting of the hydrate in a two-step process, i.e., a bubble of pure methane appears in the liquid phase as a first step and then the methane of the aqueous solution moves to the bubble and the methane from the hydrate moves to the aqueous solution as a second step. We find here that the hydrogen bonds of the layer of the hydrate in contact with the aqueous solutions break and the hydrate starts to melt. Particularly, when the temperature is increased the concentration of CO$_{2}$ in the aqueous solution increases in order to stabilize the hydrate phase. In the case of the methane hydrate, the amount of methane molecules releases to the aqueous phase to achieve the new equilibrium state is small (the solubility of methane in water is very small) and the metastable hydrate phase can exists above the $T_{3}$. However, in the case of the CO$_{2}$ hydrate, the CO$_{2}$ saturates the aqueous phase (the solubility of CO$_{2}$ in water increases greatly with the temperature). The hydrate becomes unstable, the hydrogen bonds of the hydrate layer next to the aqueous solution breaks, and the hydrate finally melts.

\subsection{Three-phases coexistence from solubility calculations}

\begin{figure}
\hspace*{-0.2cm}
\includegraphics[width=1.1\columnwidth]{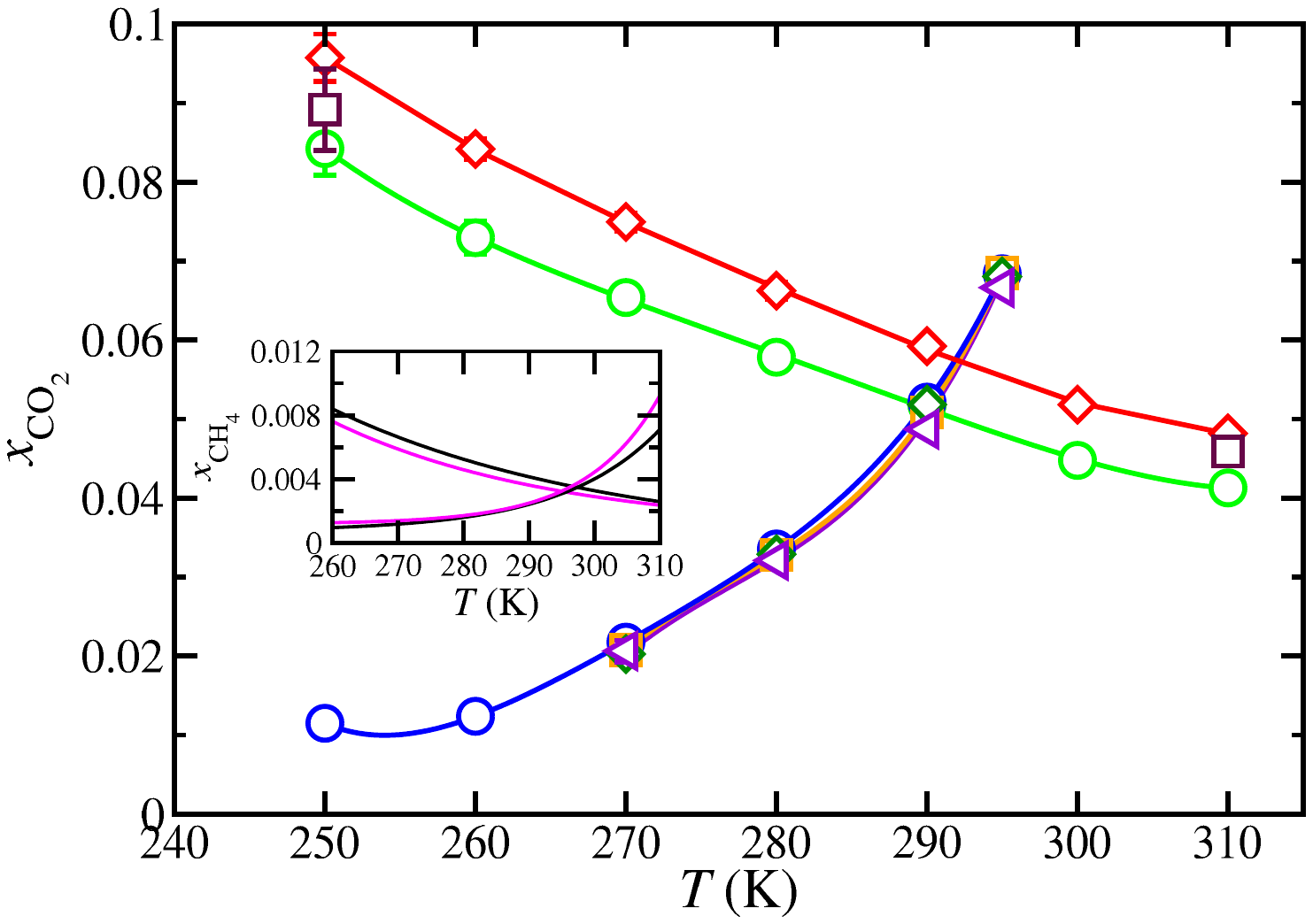}
\caption{Solubilities of CO$_{2}$ from the CO$_{2}$ liquid and the hydrate along the isobar $P=400\,\text{bar}$. The crossing of the two curves determines the dissociation temperature of the hydrate, $T_{3}$, at $400\,\text{bar}$. The symbols and colors are the same as those in Figs.~\ref{figure3} and \ref{figure7}. In all the previous cases, the hydrate is fully occupied by CO$_{2}$ molecules. Violet triangles left correspond to simulation results using a cutoff of $1.0\,\text{nm}$ and an occupancy of $50\%$ of the small or D cages ($87.5\%$ of overall occupancy).
Inset: solubilities of methane from the gas and the hydrate phase, as a function of temperature, at $400\,\text{bar}$. Solubility values of methane in water are taken from our previous work.~\cite{Grabowska2022a} Simulations are performed at the same conditions using a cutoff of $0.9\,\text{nm}$ (black curves) and $1.7\,\text{nm}$ (violet curves). In all cases, the curves are included as a guide to the eyes.}
\label{figure8}
\end{figure}

We have obtained the solubility of CO$_{2}$ in the aqueous solution, as a function of the temperature at a fixed pressure of $400\,\text{bar}$, when it is in contact with the CO$_{2}$ liquid phase (Section III.A) and with the hydrate (Section III.C). In both cases, the system exhibits two-phase coexistence. It is interesting to represent both solubilities in the same plot as we did in our previous study,~\cite{Grabowska2022a} and as it is shown in Fig.~\ref{figure8}. Since one of the solubility curves is a decreasing function of the temperature and the other an increasing function of the temperature, there exists a certain temperature, that we will call $T_{3}$ for reasons that will be clear soon, at which both solubilities are equal at $400\,\text{bar}$.

The points of the solubility curve of CO$_{2}$ in the aqueous solution from the CO$_{2}$ liquid phase correspond to thermodynamic states at which the pressure and the chemical potentials of water and CO$_{2}$ in the aqueous phase are equal to those in the CO$_{2}$ liquid phase. In addition to this, the points of the solubility of CO$_{2}$ in water from the hydrate phase correspond to states at which the pressure is also the same and at which the chemical potentials of both components in the aqueous phase are equal to those in the hydrate phase. Consequently, at $T_{3}$, the temperature, pressure, and chemical potentials of water and CO$_{2}$ in the aqueous solution, CO$_{2}$ liquid, and hydrate phases are the same. This means that the point at which the two solubility curves cross represents a three-phase coexistence state of the system at $400\,\text{bar}$. This is also known as the dissociation temperature of the CO$_{2}$ hydrate at the corresponding pressure ($400\,\text{bar}$).

The value obtained in this work for the $T_{3}$ is $290(2)\,\text{K}$ when the occupancy of the hydrate is $100\%$ (all the T and D cages are occupied by CO$_{2}$ molecules). We assume here an uncertainty of $2\,\text{K}$ for the dissociation temperature of the hydrate, following the same estimation of the $T_{3}$ error of the methane hydrate determined in our previous work.~\cite{Grabowska2022a} We have also determined the dissociation temperature of the hydrate when the occupancy of the small or D cages is $50\%$ ($87.5\%$ overall occupancy) using $r_{c}=1.0\,\text{nm}$. In this case, $T_{3}=290.5(2)\,\text{K}$, which is the same value obtained for the fully occupied hydrate within the error. Both dissociation temperature results seem to be occupancy-independent with the employed methodology and are in good agreement (within the corresponding uncertainties) with the value obtained by M\'{\i}guez \emph{et al.} using the direct coexistence technique,\cite{Miguez2015a} $287(2)\,\text{K}$. It is important to remark here that we are using the same models for water (TIP4P/ice)\cite{Abascal2005b} and CO$_{2}$ (TraPPE),\cite{Potoff2001a} the same unlike dispersive interactions between both components, and cutoff distance for dispersive interactions ($r_c{}=1.0\,\text{nm}$) than in the work of  M\'{\i}guez \emph{et al.}~\cite{Miguez2015a} At this point it is important to remark that the system sizes of this work are different than that used in the work of M\'{\i}guez \emph{et al.},~\cite{Miguez2015a} and this could have a subtle effect in the $T_{3}$ because of finite-size effects, as it has been found for the melting point of ice Ih.~\cite{Conde2017a} The experimental value of $T_{3}$ at $400\,\text{bar}$ is $286\,\text{K}$ so that the force field used in  this work provides a quite reasonable prediction. 

Other authors have determined the $T_{3}$  for this system from computer simulation. Costandy \emph{et al.}\cite{Constandy2015a} have calculated the dissociation temperature of the CO$_{2}$ hydrate at $400\,\text{bar}$ using the direct coexistence technique. They obtained a value of $283.5(1.7)\,\text{K}$. Although they also used the same water and CO$_{2}$ models, a number of differences lead to a slightly different value of $T_{3}$: different unlike dispersive interactions between water and CO$_{2}$ and cutoff distance for dispersive interaction ($1.1\,\text{nm}$). Waage and collaborators~\cite{Waage2017a} have also determined the dissociation line of the hydrate using free energy calculations. They also use the same models for water and CO$_{2}$ but different unlike dispersive interactions between them. In this case, the cutoff distance for dispersive interactions is $r_{c}=1.0\,\text{nm}$. These authors calculate the dissociation temperature of the hydrate at $200$ and $500\,\text{bar}$. The values obtained are $283.9(1.7)$ and $284.8(0.9)\,\text{K}$, respectively. Interpolating to $400\,\text{bar}$, the $T_{3}$ is $284.5\,\text{K}$, in good agreement with the results of Costandy \emph{et al.}\cite{Constandy2015a} Unfortunately, the result obtained here can not be compared with the predictions of Costandy \emph{et al.}\cite{Constandy2015a} and Waage and collaborators~\cite{Waage2017a} since dispersive interactions are not the same as those used here.

Finally, it is important to focus on the effect of the cutoff distance used to evaluate the long-range dispersive interactions. The $T_{3}$ value of $290(2)\,\text{K}$ has been obtained using a cutoff distance of $1.0\,\text{nm}$. We have also analyzed the solubilities of CO$_{2}$ from the CO$_{2}$ liquid and the hydrate phases using a much larger cutoff distance (i.e., $1.9\,\text{nm}$). As we have previously shown, the solubility of CO$_{2}$ from the CO$_{2}$ liquid phase increases when the value of the cutoff is increased. On the other hand, the solubility of CO$_{2}$ in the hydrate phase is not affected by the use of larger cutoff values. Consequently, the combined effect of the increase of the cutoff distance of the dispersive interactions is an increase of the $T_{3}$ since it is the intersection of the two solubility curves shown in Fig.~\ref{figure8}. Particularly, the dissociation temperature of the hydrate is now found at $292(2)\,\text{K}$, $2\,\text{K}$ above the $T_{3}$ observed with a cutoff distance of $1.0\,\text{nm}$, approximately. 

It is interesting to compare the effect of the cutoff due to the dispersive interaction in both CO$_{2}$ and methane hydrates. As can be seen, in the case of the methane hydrate, $T_{3}$ is shifted towards lower temperatures, by $2\,\text{K}$, when the cutoff is increased. In the current case (CO$_{2}$ hydrate), we observed the opposite effect, i.e., $T_{3}$ increases when the cutoff distance is increased. This is the same effect as observed for the solubility curve of CO$_{2}$ in the aqueous solution in contact with the CO$_{2}$ liquid phase. This effect, contrary to that observed in the methane hydrate, could be due to the electrostatic interactions of the quadrupole of CO$_{2}$ with other CO$_{2}$ molecules and also with water molecules, that it is not present in the case of the methane. We think this issue deserves a more detailed study but this is out of the scope of the current work.

We have determined the dissociation line of the CO$_{2}$ hydrate at $400\,\text{bar}$ from the calculation of the solubility of CO$_{2}$ when the aqueous solution is in contact with the other two phases in equilibrium, the CO$_{2}$ liquid phase and the hydrate phase. Grabowska and collaborators~\cite{Grabowska2022a} have already demonstrated that this route allows to determine $T_{3}$ of hydrates. This work confirms that this methodology is a good alternative to the direct coexistence method. Particularly, it shows a slightly better efficiency compared with the other technique (lesser simulation times are required) and provides consistent values of $T_{3}$.

\subsection{Driving force for nucleation of hydrates}

The dissociation line of the CO$_{2}$ hydrate separates its phase diagram in two parts in which two different two-phase coexistence regions exist.~\cite{Sloan2008a} At a certain pressure, for instance, $400\,\text{bar}$, at temperatures above $T_{3}$ the system exhibits LL immiscibility between an aqueous solution and a CO$_{2}$ liquid phase. Note that the solubility of water in CO$_{2}$ is very small and the CO$_{2}$ liquid phase can be considered pure CO$_{2}$ in practice. However, at temperatures below $T_{3}$ the system exhibits SL phase equilibrium between the hydrate and a fluid phase (water or CO$_{2}$ depending on the global composition of the system). This is consistent with the nature of the dissociation or three-phase line at which the hydrate, aqueous solution, and CO$_{2}$ liquid phases coexist.

The fluid phase in equilibrium with the hydrate below $T_{3}$ depends on the global composition of the system. Here we assume that the hydrate is fully occupied by CO$_{2}$ molecules, i.e., 8 CO$_{2}$ molecules for every 46 water molecules according to the stoichiometry of hydrates type sI. Let be $N_{\text{H}_{2}\text{O}}$ and $N_{\text{CO}_{2}}$ the number of water and CO$_{2}$ molecules used in the fluid phases during the simulations, respectively. If the ratio $N_{\text{H}_{2}\text{O}}/N_{\text{CO}_{2}}> 5.75$, one should have hydrate--water phase separation (below $T_{3}$). However, if $N_{\text{H}_{2}\text{O}}/N_{\text{CO}_{2}}< 5.75$, one should have a hydrate -- CO$_{2}$ phase system for $T<T_{3}$.

As described by Kashchiev and Firoozabadi~\cite{Kashchiev2002a} and Grabowska and collaborators,~\cite{Grabowska2022a} the formation of a hydrate in the aqueous solution phase can be viewed as a chemical reaction that takes place at constant $P$ and $T$,


\begin{equation}
\text{CO}_{2} (\text{aq},x_{\text{CO}_{2}}) +
5.75\,\text{H}_{2}\text{O} (\text{aq},x_{\text{CO}_{2}}) 
\rightarrow [\text{CO}_{2}(\text{H}_{2}\text{O})_{5.75}]_{\text{H}}
\label{reaction}
\end{equation}

Since we work at constant pressure in this work ($P=400\,\text{bar}$), we drop the dependence of $P$ in the rest of equations. Assuming that all the cages of the hydrate are filled, a unit cell of CO$_{2}$ hydrate is formed by $46$ water molecules and $8$ CO$_{2}$ molecules, i.e., 1 CO$_{2}$ molecule per $46/8=5.75$ water molecules. According to this, Eq.~\eqref{reaction} considers the hydrate as a new compound formed from one molecule of CO$_{2}$ and 5.75 molecules of water. We can also associate to this compound one unique chemical potential for the hydrate at $T$, $\mu_{\text{H}}^{\text{H}}(T)$. Note that this chemical potential is simply the sum of the chemical potential of CO$_{2}$ in the solid plus $5.75$ times the chemical potential of water in the solid, i.e.,

\begin{equation}
\mu^{\text{H}}_{\text{H}}(T)=\mu_{\text{CO}_{2}}^{\text{H}}(T)+5.75\,\mu_{\text{H}_{2}\text{O}}^{\text{H}}(T)
\label{chempot_hydrate}
\end{equation}

\noindent
According to the previous discussion, the compound $[\text{CO}_{2}(\text{H}_{2}\text{O})_{5.75}]_{\text{H}}$ is simply the ``hydrate'' and we call one ``molecule'' of the hydrate in the solid the molecule $[\text{CO}_{2}(\text{H}_{2}\text{O})_{5.75}]$.

Following Kashchiev and Firoozabadi,~\cite{Kashchiev2002a} we denote the driving force for nucleation of the hydrate formed from the aqueous solution with a concentration $x_{\text{CO}_{2}}$ at $T$ as,

\begin{align}
\Delta\mu_{N}(T,x_{\text{CO}_{2}})&=\mu^{\text{H}}_{\text{H}}(T) \nonumber\\
& -\mu^{\text{aq}}_{\text{CO}_{2}}(T,x_{\text{CO}_{2}})-5.75\,\mu^{\text{aq}}_{\text{H}_{2}\text{O}}(T,x_{\text{CO}_{2}})
\label{driving_force}
\end{align}

\noindent
Note that $\Delta\mu_{N}$ in this paper is $\Delta\mu_{\text{nucleation}}$ in our previous paper.~\cite{Grabowska2022a} $\Delta\mu_{N}$ also depends on pressure but since we are working at constant pressure ($400\,\text{bar}$), we drop the pressure dependence from all the equations in this study. $\mu_{\text{H}}^{\text{H}}(T)$ has been previously defined in Eq.~\eqref{chempot_hydrate} as the chemical potential of the ``hydrate molecule'' in the hydrate phase, and $\mu^{\text{aq}}_{\text{CO}_{2}}(T,x_{\text{CO}_{2}})$ and  $\mu^{\text{aq}}_{\text{H}_{2}\text{O}}(T,x_{\text{CO}_{2}})$ are the chemical potentials of CO$_{2}$ and water in the
aqueous solution, respectively, at $T$ and molar fraction of CO$_{2}$, $x_{\text{CO}_{2}}$. Note that the composition of CO$_{2}$ in Eq.~\eqref{driving_force} is, a priori, independent of the pressure and temperature selected. In other words, one could have different driving forces for nucleation, at a given $P$ and $T$, changing the composition of the aqueous solution (for instance, in a supersaturated solution of CO$_{2}$). However, there exists a particular value of $x_{\text{CO}_{2}}$ which is of great interest from the experimental point of view. Experiments on the nucleation of hydrates are performed when the water phase is in contact with the CO$_{2}$ liquid phase through a planar interface. Since both phases are in equilibrium at $P$ and $T$, the solubility of CO$_{2}$ in water (molar fraction of CO$_{2}$ in the aqueous solution) is fully determined since $x^{\text{eq}}_{\text{CO}_{2}}\equiv x_{\text{CO}_{2}}^{\text{eq}}(T)$. Following the notation of Grabowska and coworkers,~\cite{Grabowska2022a} the driving force for nucleation at experimental conditions is given by,

\begin{align}
\Delta\mu^{\text{EC}}_{N}(T)&=\mu^{\text{H}}_{\text{H}}(T) -\mu^{\text{aq}}_{\text{CO}_{2}}(T,x_{\text{CO}_{2}}^{\text{eq}}(T)) \nonumber\\
& -5.75\,\mu^{\text{aq}}_{\text{H}_{2}\text{O}}(T,x_{\text{CO}_{2}}^{\text{eq}}(T))
\label{driving_force_EN}
\end{align}

\noindent
Note that $\Delta\mu^{\text{EC}}_{N}$ depends only on $T$ (and on $P$ but in this work we are working at the same $P=400\,\text{bar}$). We provide here valuable information for this magnitude when the molecules are described using the TIP4P/Ice and TraPPE models for water and CO$_{2}$, respectively.

In the next sections we concentrate on the driving force for nucleation at experimental conditions, $\Delta\mu^{\text{EC}}_{N}(T)$, obtained using four different routes. In the first one (route 1), we use the definition of the driving force of nucleation given by Eq.~\eqref{driving_force_EN}. In the second one (route 2), we use the solubility curves of CO$_{2}$ from the hydrate and the CO$_{2}$ liquid phase. In third one (route 3), we use the enthalpy of dissociation of the hydrate and assume that it does not change with temperature nor composition. In the fourth one (route 4), we propose a novel methodology based on the use of the solubility curve of CO$_{2}$ with the hydrate, valid not only for the CO$_{2}$ hydrate but also for other hydrates. This route can be used to determine $\Delta\mu_{N}$ at any arbitrary temperature and mixture composition and not only at experimental conditions. Finally, we discuss the results obtained using the different routes and compare the driving force for nucleation of the CO$_{2}$ hydrate with that of the methane hydrate previously obtained by us in a previous work.~\cite{Grabowska2022b}

\subsubsection{Route 1 for calculating $\Delta\mu^{\text{EC}}_{N}$.}

Route 1 was proposed and described in our previous work~\cite{Grabowska2022a} and we summarize here only the main approximations and the final expression of the driving force for nucleation. To evaluate $\Delta\mu^{\text{EC}}_{\text{N}}(T,x^{\text{eq}}_{\text{CO}_{2}})$ in Eq.~\eqref{driving_force_EN} we need to calculate the chemical potential of the ``hydrate molecule'' in the hydrate phase, and the chemical potentials of ${\text{CO}_{2}}$ and water in the aqueous phase  at a supercooled temperature $T$ below $T_{3}$. The change in the hydrate chemical potential when the temperature passes from $T_{3}$ to $T$ can be evaluated in a similar way as that for pure ${\text{CO}_{2}}$ from $T_{3}$ to $T$. In fact, this later change has been already calculated in Section III.A using Eq.~\eqref{integral_co2} and evaluated from the corresponding thermodynamic integration using computer simulations in the $NPT$ ensemble.

The chemical potential of water in the aqueous phase at $T$ can be estimated using the procedure of Grabowska and collaborators~\cite{Grabowska2022a} applied to the case of the CO$_{2}$ hydrate according to Eqs.~(20)-(26) of their paper. This is done in two steps. In the first step, the change in the chemical potential of the solution when its temperature passes from $T_{3}$ to $T$ is approximated by that of pure water calculated from thermodynamic integration (see Eqs.~(24) and (26) of the work of Grabowska and collaborators). The second step involves the change in the chemical potential of water in the solution when the composition of CO$_{2}$ changes from $x^{\text{eq}}_{\text{CO}_{2}}(T_{3})$ to $x^{\text{eq}}_{\text{CO}_{2}}(T)$ (see Eq.~(25) of our previous work).  The rigorous calculation of this contribution requires the knowledge of the activity coefficient of water in an aqueous solution with a given composition of water, $\gamma^{aq}_{\text{H}_{2}\text{O}}(T,x^{\text{eq}}_{\text{H}_{2}\text{O}})$, at $T$ and $T_{3}$. Grabowska and coworkers assume that this magnitude, in the case of an aqueous solution of methane, $\gamma^{aq}_{\text{H}_{2}\text{O}}\approx 1$ since the solution is very diluted. We follow here the same assumption.

Using these approximations it is possible to compute the driving force at experimental conditions for nucleation given by Eq.~\eqref{driving_force_EN}. The final expression is given by,

\begin{widetext}
\begin{equation}
\dfrac{\Delta\mu^{\text{EC}}_{\text{N}}(T,x^{\text{eq}}_{\text{CO}_{2}})}
{k_{B}T}=
-\bigintss_{T_{3}}^{T} 
\dfrac{h_{\text{H}}^{\text{H}}(T')- \Big\{h_{\text{CO}_{2}}^{\text{pure}}(T')+5.75
h_{\text{H}_{2}\text{O}}^{\text{pure}}(T')\Big\}}{k_{B}T'^{2}}dT'
- \Big[k_{B}T\ln\{x^{\text{eq}}_{\text{H}_{2}\text{O}}(T)\}-k_{B}T_{3}\ln\{x^{\text{eq}}_{\text{H}_{2}\text{O}}(T_{3})\}\Big]
\label{driving_force_route1}
\end{equation}
\end{widetext}

\noindent
Here $h_{\text{H}}^{\text{H}}=H/N_{\text{CO}_{2}}$ is the enthalpy $H$ of the hydrate per $\text{CO}_{2}$ molecule and $N_{\text{CO}_{2}}$ is the number of CO$_{2}$ molecules in the hydrate. Note that Eq.~\eqref{driving_force_route1} is consistent with the view of Kashchiev and Firoozabadi~\cite{Kashchiev2002a} of the hydrate as a new compound formed from one molecule of CO$_{2}$ and $5.75$
 molecules of water when the hydrate is fully occupied. Also note that it is necessary to use that the driving force for nucleation at $T_{3}$ is equal to zero, i.e.,

\begin{align}
\Delta\mu^{\text{EC}}_{\text{N}}(T_{3},x^{\text{eq}}_{\text{CO}_{2}})&=\mu^{\text{H}}_{\text{H}}(T_{3}) -\mu^{\text{aq}}_{\text{CO}_{2}}(T_{3},x^{\text{eq}}_{\text{CO}_{2}}(T_{3})) \nonumber\\
& -5.75\,\mu^{\text{aq}}_{\text{H}_{2}\text{O}}(T_{3},x^{\text{eq}}_{\text{CO}_{2}}(T_{3}))=0
\label{driving_force_t3}
\end{align}

\noindent
This is equivalent to arbitrary set to zero the chemical potentials of CO$_{2}$ and water in the hydrate at $T_{3}$. Eq.~\eqref{driving_force_route1} is similar to Eq.~\eqref{h_dissoc} (route 3 or dissociation route) but taking into account two effects: (1) the temperature dependence of molar enthalpies of hydrate, CO$_{2}$, and water; and (2) the change of composition of the solution when passing from $T_{3}$ to $T$ (see route 3 below for further details).

As we have previously mentioned, each change in the chemical potentials needed to compute the driving force for nucleation is obtained by evaluating molar enthalpies of pure CO$_{2}$, water, and hydrate phases involved in the integrals given by Eq.~\eqref{driving_force_route1}. Note that the chemical potential of CO$_{2}$ in the aqueous solution has been already obtained in Section III.A. Particularly since in this route one is interested in computing $\Delta\mu_{\text{N}}$ at experimental conditions, the chemical potential of CO$_{2}$ in the aqueous solution is equal to that of pure CO$_{2}$ at the same $P$ and $T$. Consequently, it has been obtained from the integration of the molar enthalpy at several temperatures according to Eq.~\eqref{integral_co2}.

In the case of water, the chemical potential can be obtained performing MD $NPT$ simulations of pure water along the $400\,\text{bar}$ isobar. As in the case of pure CO$_{2}$, we also use the standard $NPT$ is used in such a way that the three dimensions of the simulation box are allowed to fluctuate isotropically. We used a cubic simulation box with $1000$ H$_{2}$O molecules. The dimensions of the simulation box, $L_{x}$, $L_{y}$, and $L_{z}$ vary depending on the temperature from $3.14$ to $3.08\,\text{nm}$. Simulations to calculate the molar enthalpy, at each temperature, are run during $100\,\text{ns}$, $20\,\text{ns}$ to equilibrate the system, and $80\,\text{ns}$ as the production period to obtain $h^{\text{pure}}_{\text{H}_{2}\text{O}}$.

In the case of the pure hydrate, we have obtained the chemical potential in a similar way, performing simulations in the $NPT$ ensemble using an isotropic barostat at $400\,\text{bar}$. At the beginning of each simulation, we use a cubic box formed by 27 replicas of the unit cell in a $3\times 3\times 3$ geometry. The dimensions of the simulation box vary between $3.85$ and $3.62\,\text{nm}$ depending on the temperature. As in the rest of simulations, we calculate the enthalpy at different temperatures, from $260$ to $295\,\text{K}$. Simulations are run during $100\,\text{ns}$, $20\,\text{ns}$ to equilibrate the system and $80\,\text{ns}$ to calculate the molar enthalpy of the hydrate.

\subsubsection{Route 2 for calculating $\Delta\mu^{\text{EC}}_{N}$.}

Route 2 was also proposed and described in our previous work~\cite{Grabowska2022a}. This route is inspired by the work of Molinero and coworkers\cite{Knott2012a} and we summarize here only the main approximations and the final expression of the driving force for nucleation. According to this, it is possible to find a different, but an equivalent, thermodynamic route to calculate the driving force for the nucleation of methane hydrates. We check in this work whether this approach can also be used to deal with CO$_{2}$ hydrates. Let us consider Eq.~\eqref{driving_force_EN} at experimental conditions, i.e., at the equilibrium composition of CO$_{2}$ in the aqueous solution when it is in contact via a planar interface with a CO$_{2}$ liquid phase (L), $x^{\text{eq}}_{\text{CO}_{2}}(T|\text{L})$. Note that the vertical line represents flat interface equilibrium with
the liquid CO$_{2}$. To clarify the derivation of the final expression, we write explicitly the solubility of water in the solution, at experimental conditions, as $x^{\text{eq}}_{\text{H}_{2}\text{O}}(T|\text{L})$. Obviously, as we have mentioned previously in Section III.A, the solubility of water in the solution can be obtained readily as  $x^{\text{eq}}_{\text{H}_{2}\text{O}}(T|\text{L})=1-x^{\text{eq}}_{\text{CO}_{2}}(T|\text{L})$. 

We now assume that the chemical potential of the ideal solution's components can be expressed, in general, in terms of the chemical potentials of the pure components in the standard state and their molar fractions. In other words, since the molar fraction of CO$_{2}$ in the solution is small, we are assuming that water is the dominant component (solvent) and the CO$_{2}$ is the minor component (solute) in the mixture. Under these circumstances, the activity coefficients of water and CO$_{2}$ are close to one.~\cite{Levine2009a} According to this and following our previous work,\cite{Grabowska2022a} Eq.~\eqref{driving_force_EN} can be written as,

\begin{align}
\Delta\mu^{\text{EC}}_{N}(T)&=
-k_{B}T\text{ln}\Biggl[\dfrac{x^{\text{eq}}_{\text{CO}_{2}}(T|\text{L})}
{x^{\text{eq}}_{\text{CO}_{2}}(T|\text{H})}\Biggl] \nonumber \\
&-5.75k_{B}T\text{ln}\Biggl[\dfrac{x^{\text{eq}}_{\text{H}_{2}\text{O}}(T|\text{L})}
{x^{\text{eq}}_{\text{H}_{2}\text{O}}(T|\text{H})}\Biggl]
\label{driving_force4}
\end{align}

\noindent
$x^{\text{eq}}_{\text{CO}_{2}}(T|\text{H})$ and $x^{\text{eq}}_{\text{H}_{2}\text{O}}(T|\text{H})$ represent the molar fraction of CO$_{2}$ and H$_{2}$O in the solution when it is in equilibirum via a planar interface (vertical line) with the hydrate phase (H), respectively. Note that $\Delta\mu^{\text{EC}}_{N}(T)$ and all the molar fractions also depend on pressure but in this work we work at fixed $P=400\,\text{bar}$. This is the equation obtained previously by us considering the driving force for the nucleation of the methane hydrate.\cite{Grabowska2022a} As we will see later in this section, Eq.~\eqref{driving_force4} does not provide reliable values for the driving force of nucleation of the CO$_{2}$ hydrate, contrary to what happens with the methane hydrate. The solubilities of methane in the solution when is in contact with the methane phase and with the hydrate are one order of magnitude lower  than those of CO$_{2}$ in the case of the CO$_{2}$ hydrate. Consequently, this route can be useful only in cases in which the solubility of the guest is extremely low.

\subsubsection{Route 3 (dissociation) for calculating $\Delta\mu^{\text{EC}}_{N}$.}

It is possible to estimate the driving force for nucleation of a hydrate using a simple and approximate route based on the knowledge of the enthalpy of dissociation of the hydrate.~\cite{Kashchiev2002a} 
The dissociation enthalpy of the hydrate, $h_{\text{H}}^{\text{diss}}$, is defined as the enthalpy change of the process,~\cite{Grabowska2022a}

\begin{equation}
[\text{CO}_{2}(\text{H}_{2}\text{O})_{5.75}]_{\text{H}}
\rightarrow\text{CO}_{2} (\text{liq}) +
5.75\,\text{H}_{2}\text{O} (\text{liq}) 
\label{dissociation}
\end{equation}

\noindent
Dissociation enthalpies are usually calculated assuming that the hydrate dissociates into pure water and pure CO$_{2}$. Note that this corresponds to the definition of enthalpy of dissociation and that in reality CO$_{2}$ will be dissolved in water and an even smaller amount of water will be dissolved in the CO$_{2}$ liquid phase. We have determined the dissociation enthalpy of the hydrate simply by performing simulations of the pure phases (hydrate, water, and CO$_{2}$) at several temperatures at $400\,\text{bar}$.

According to our previous work,~\cite{Grabowska2022a} we evaluate the driving force for nucleation assuming the following approximations: (1) the enthalpy of dissociation of the hydrate, $h_{\text{H}}^{\text{diss}}$ does not change with the temperature; (2) its value can be taken from its value at $T_{3}$; and (3) enthalpy of dissociation does not vary with composition of the aqueous solution containing CO$_{2}$ when the temperature is changed. According to this, $\Delta\mu_{N}^{\text{EC}}$ is given by,

\begin{equation}
\Delta\mu_{N}^{\text{EC}}=k_{B}T
\bigintsss_{T_{3}}^{T}\dfrac{h^{\text{diss}}_{\text{H}}}{k_{B}T'^{2}}\,dT'\approxeq -h^{\text{diss}}_{\text{H}} (T_{3}) \bigg(1-\dfrac{T}{T_{3}}\bigg)
\label{h_dissoc}
\end{equation}

\noindent
Note that Eq.~\eqref{driving_force_route1} reduces to Eq.~\eqref{h_dissoc} under the approximations used in this route. 

\subsubsection{Route 4 for calculating $\Delta\mu^{\text{EC}}_{N}$.}

The driving force for nucleation of the CO$_{2}$ hydrate, at any arbitrary temperature, $T_{N}$, and molar fraction of CO$_{2}$ in the aqueous solution, $x_{\text{CO}_{2}}^{\text{(N)}}$, at fixed pressure is defined as,

\begin{align}
\Delta\mu_{\text{N}}(T_{\text{N}},x^{\text{N}}_{\text{CO}_{2}})&=\mu^{\text{H}}_{\text{H}}(T_{\text{N}}) -\mu^{\text{aq}}_{\text{CO}_{2}}(T_{\text{N}},x^{\text{N}}_{\text{CO}_{2}}) \nonumber\\
& -5.75\,\mu^{\text{aq}}_{\text{H}_{2}\text{O}}(T_{\text{N}},x^{\text{N}}_{\text{CO}_{2}})
\label{driving_force_route4}
\end{align}

\noindent
Note that $\Delta\mu_{\text{N}}$ also depends on pressure. However, since we work at constant pressure ($P=400\,\text{bar}$), we drop the pressure dependence from equations from this point. It is also important to recall that since we are assuming that all cages of the hydrate are filled, the chemical potential of a ``hydrate molecule'' in the hydrate phase depends only on temperature. Finally, the chemical potentials of CO$_{2}$ and water also depend on the molar fraction of water in the aqueous solution, $x_{\text{H}_{2}\text{O}}^{\text{N}}$. Since we are dealing with a binary mixture, $x_{\text{H}_{2}\text{O}}^{\text{N}}=1-x^{\text{N}}_{\text{CO}_{2}}$. For simplicity, we choose $x^{\text{N}}_{\text{CO}_{2}}$ as independent variable of the chemical potentials of CO$_{2}$ and water in the solution. 

\begin{figure}
\hspace*{0.0cm}
\hspace*{-0.25cm}
\includegraphics[width=1.25\columnwidth]{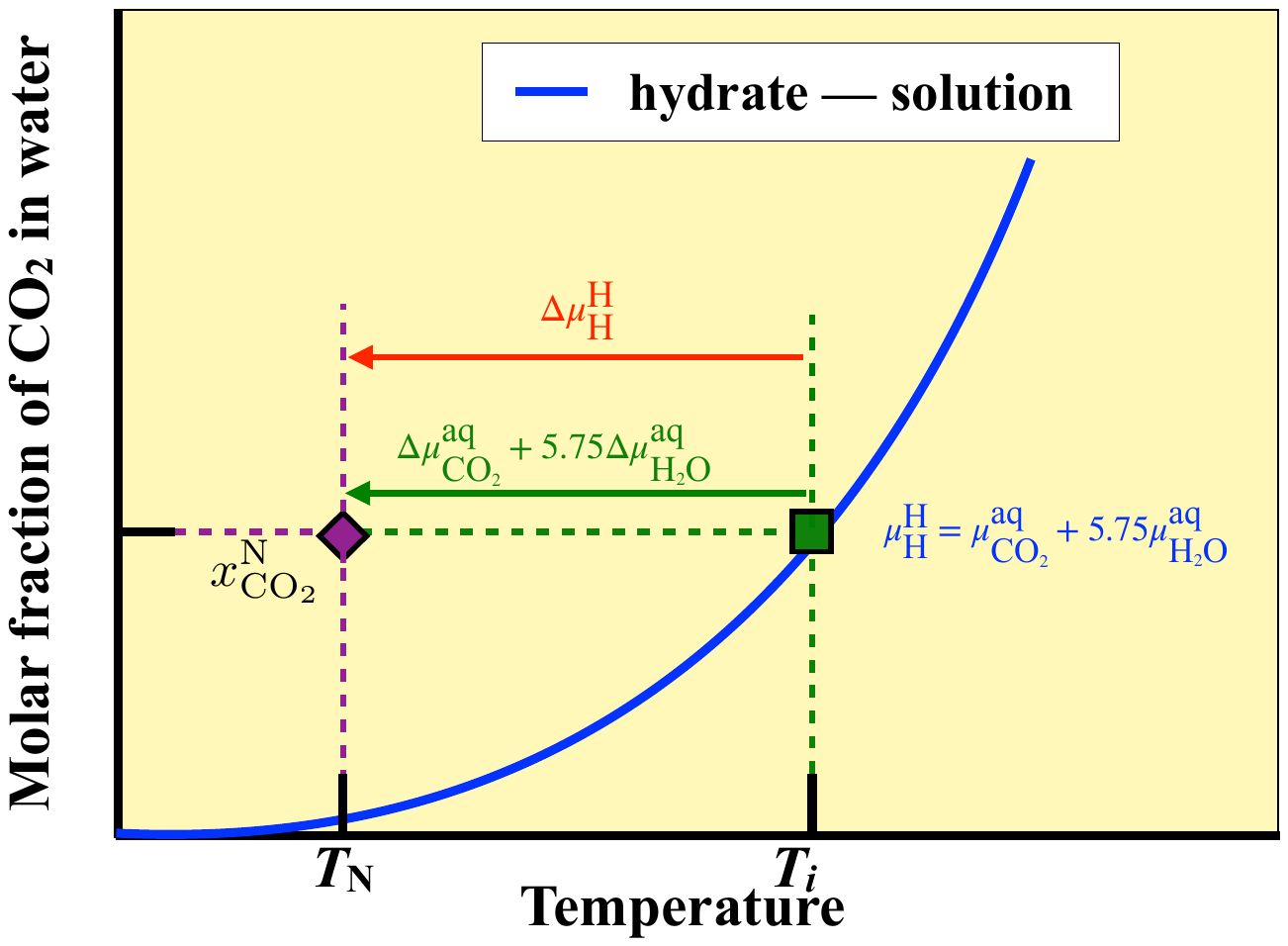}
\vspace*{-1.0cm}
\caption{Schematic depiction of route 4 for obtaining $\Delta\mu_{\text{N}}$. The blue solid curve represents the solubility curve of CO$_{2}$ with the hydrate (hydrate--solution equilibrium). The green square is a state at $T_{i}$ at which the aqueous solution with molar fraction $x_{\text{CO}_{2}}^{\text{N}}$ is in equilibrium with the hydrate phase and the magenta diamond the state at $T_{\text{N}}$ and $x_{\text{CO}_{2}}^{\text{N}}$ at which $\Delta\mu_{\text{N}}$ is calculated. The arrows in colors represent the paths followed by thermodynamic integration, from $T_{i}$ to $T_{\text{N}}$, of the molar enthalpy of the "hydrate molecule" (red arrow) and the partial molar enthalpies of CO$_{2}$ and water (green arrow) using Eq.~\eqref{final_driving_force1}.}
\label{figure9}
\end{figure}

The driving force for nucleation of the CO$_{2}$ hydrate, $\Delta\mu_{\text{N}}$, depends on $T_{\text{N}}$ and $x_{\text{CO}_{2}}^{\text{N}}$ and both are independent variables. This means that route 4 is valid for calculating the driving force for nucleation at any $T_{\text{N}}$ and $x_{\text{CO}_{2}}^{\text{N}}$. As we will see later, the method can be particularized to evaluate $\Delta\mu_{\text{N}}$ at experimental conditions. In this case, $\Delta\mu_{\text{N}}=\Delta\mu_{\text{N}}^{\text{EC}}(T_{\text{N}})=\Delta\mu_{\text{N}}^{\text{EC}}(T_{\text{N}},x_{\text{CO}_{2}}^{\text{eq}}(T_{\text{N}}|\text{L}))$, as we have previously mentioned.

To evaluate $\Delta\mu_{\text{N}}$ we need to calculate the chemical potential of the ``hydrate molecule'', $\mu^{\text{H}}_{\text{H}}(T_{\text{N}})$, at a supercooled temperature $T_{\text{N}}$, and the chemical potentials of CO$_{2}$ and water molecules of an aqueous solution of CO$_{2}$ with molar fraction $x_{\text{CO}_{2}}^{\text{N}}$ at the same temperature. This route is based on the use of the solubility curve of the hydrate with temperature, at constant pressure, previously described in Section III.C. A schematic depiction of the curve and the thermodynamic route for obtaining $\Delta\mu_{\text{N}}$ at arbitrary $T_{\text{N}}$ and $x_{\text{CO}_{2}}^{\text{N}}$ is presented in Fig.~\ref{figure9}. Let us consider a reference state in our calculations at temperature $T_{\text{ref}}$ on the solubility curve in contact with the hydrate. As it will be clear later, the particular value of $T_{\text{ref}}$ is not important since we are dealing with differences of chemical potentials and the final value of $\Delta\mu_{\text{N}}$ does not depend on the election of the reference state. Due to this, the reference state does not appear in Fig.~\eqref{figure9}.

The first contribution to $\Delta\mu_{\text{N}}$ in Eq.~\eqref{driving_force_route4} is the chemical potential of the ``hydrate molecule'' in the hydrate phase at $T_{\text{N}}$. The chemical potential $\mu_{\text{H}}^{\text{H}}$ can be obtained using the Gibbs-Helmholtz thermodynamic relation for pure systems,

\begin{equation}
\Biggl(\dfrac{\partial(\mu_{\text{H}}^{\text{H}}/T)}{\partial T}\Biggr)_{P, N_{\text{H}}}=-\dfrac{h_{\text{H}}^{\text{H}}}{T^{2}}
\label{enthalpy_hydrate}
\end{equation}

\noindent
where $h_{\text{H}}^{\text{H}}=h_{\text{H}}^{\text{H}}(T)$ is the molar enthalpy of the ``hydrate molecule'' and the derivative is performed at constant pressure, $P$, and number of ``hydrate molecules'', $N_{\text{H}}$. Note that here $h_{\text{H}}^{\text{H}}$ represents the enthalpy of the hydrate per molecule of CO$_{2}$ according to the definition in Section III.E.1. The chemical potential of the ``hydrate molecule'' in the hydrate phase at a supercooling temperature $T_{\text{N}}$ can be obtained by integrating the Eq.~\eqref{enthalpy_hydrate} from $T_{\text{ref}}$ to $T_{\text{N}}$ as,

\begin{equation}
\dfrac{\mu_{\text{H}}^{\text{H}}(T_{\text{N}})}{k_{B}T_{\text{N}}}=
\dfrac{\mu_{\text{H}}^{\text{H}}(T_{\text{ref}})}{k_{B}T_{\text{\text{ref}}}}
-\bigintsss_{T_{\text{ref}}}^{T_{\text{N}}} \dfrac{h_{\text{H}}^{\text{H}}(T)}{k_{B}T^{2}}\,dT
\label{integral_hydrate_route4}
\end{equation}

\noindent
where $k_{B}$ is the Boltzmann constant.

The last two contributions to the driving force for nucleation in Eq.~\eqref{driving_force_route4}, $\mu^{\text{aq}}_{\text{CO}_{2}}$ and $\mu^{\text{aq}}_{\text{H}_{2}\text{O}}$, need to be evaluated at temperature $T_{N}$
and molar fraction $x_{\text{CO}_{2}}^{\text{N}}$. Individual chemical potentials of CO$_{2}$ and water in the solution at a given temperature are not easy to evaluate, as we have seen in routes 1 and 2. However, it is possible to use the solubility curve of CO$_{2}$ with the hydrate to overcome this problem. 
 
Let be $T_{i}$ the temperature at which the aqueous solution with molar fraction $x_{\text{CO}_{2}}^{\text{N}}$ is in equilibrium with the hydrate phase as indicated in Fig.~\ref{figure9}. Since both phases are in equilibrium at these conditions, the chemical potentials of CO$_{2}$ and water in the hydrate phase and in the aqueous solution are equal,

\begin{equation}
\mu_{\text{CO}_{2}}^{\text{H}}(T_{i})=
\mu^{\text{aq}}_{\text{CO}_{2}}(T_{i},x_{\text{CO}_{2}}^{\text{N}})
\label{equal_cp_1_route4}
\end{equation}

\begin{equation}
\mu_{\text{H}_{2}\text{O}}^{\text{H}}(T_{i})=\mu^{\text{aq}}_{\text{H}_{2}\text{O}}(T_{i},x_{\text{CO}_{2}}^{\text{N}})
\label{equal_cp_2_route4}
\end{equation}

\noindent
Note that $x_{\text{CO}_{2}}^{\text{N}}=x^{\text{eq}}_{\text{CO}_{2}}(T_{i}|\text{H})$ according to the nomenclature used in Section III.E.2 (route 2) and in our previous paper.~\cite{Grabowska2022a} The vertical line here represents that the aqueous solution is in equilibrium with
the solid hydrate via a flat interface.

 Combining Eqs.~\eqref{equal_cp_1_route4} and \eqref{equal_cp_2_route4} with Eq.~\eqref{chempot_hydrate}, that gives the chemical potential of the ``hydrate molecule'' in terms of the chemical potentials of CO$_{2}$ and water in the hydrate phase, we obtain,

\begin{equation}
\mu_{\text{H}}^{\text{H}}(T_{i})=\mu_{\text{CO}_{2}}^{\text{aq}}(T_{i},x_{\text{CO}_{2}}^{\text{N}})
+5.75\,\mu_{\text{H}_{2}\text{O}}^{\text{aq}}(T_{i},x_{\text{CO}_{2}}^{\text{N}})
\label{cp_hydrate_route4}
\end{equation}

Eq.~\eqref{cp_hydrate_route4} is the heart of route 4. According to it, the combination $\mu_{\text{CO}_{2}}^{\text{aq}}+5.75\,\mu_{\text{H}_{2}\text{O}}^{\text{aq}}$ is known along the solubility curve of the hydrate at any temperature $T_{i}$: it is equal to the chemical potential of the "hydrate molecule" at the temperature considered. This apparently simple result allows to calculate accurately the driving force for nucleation at any temperature and composition of the solution using a one-step thermodynamic integration. As it will be clear at the end of this section, this method can be used to determine the driving force for nucleation of other hydrates. 

In the first step, we calculate the difference of $\mu_{\text{CO}_{2}}^{\text{aq}}
+5.75\,\mu_{\text{H}_{2}\text{O}}^{\text{aq}}$ between the reference state (ref) at $T_{\text{ref}}$ and a second state ($i$) at $T_{i}$, both on the solubility curve of CO$_{2}$ with the hydrate as indicated in Fig~\ref{figure9}. According to Eq.~\eqref{cp_hydrate_route4}, this is completely equivalent to evaluate the difference of $\mu_{\text{H}}^{\text{H}}$ between $T_{\text{ref}}$ and $T_{i}$ along the solublity curve. This change can be evaluated using again the Eq.~\eqref{enthalpy_hydrate} (Gibbs-Helmholtz relation) and integrating between the two temperatures,

\begin{equation}
\dfrac{\mu_{\text{H}}^{\text{H}}(T_{i})}{k_{B}T_{i}}=
\dfrac{\mu_{\text{H}}^{\text{H}}(T_{\text{ref}})}{k_{B}T_{\text{\text{ref}}}}
-\bigintsss_{T_{\text{ref}}}^{T_{i}} \dfrac{h_{\text{H}}^{\text{H}}(T)}{k_{B}T^{2}}\,dT
\label{integral_hydrate2_route4}
\end{equation}

In the second step, that involves the difference between the chemical potentials of CO$_{2}$ and water in solution at temperatures $T_{i}$ and $T$ at constant molar fraction $x_{\text{CO}_{2}}^{\text{N}}$, $\Delta\mu_{\text{CO}_{2}}^{\text{aq}}$ and $\Delta\mu_{\text{H}_{2}\text{O}}^{\text{aq}}$, can be obtained from the Gibbs-Helmholtz equation for CO$_{2}$ and water,

\begin{equation}
\Biggl(\dfrac{\partial(\mu_{\alpha}^{\text{aq}}/T)}{\partial T}\Biggr)_{P,x_{\alpha}}=-\dfrac{\overline{h}_{\alpha}^{\text{aq}}}{T^{2}}
\label{delta_mu_i}
\end{equation}


\noindent
Here $\alpha=\{$CO$_{2}$, H$_{2}$O$\}$ and represents one of the components of the mixture. Note that the partial derivative is calculated at constant composition. In this case, the composition corresponds to that of the aqueous solution in equilibrium with the hydrate phase at $T_{i}$. $\overline{h}_{\alpha}^{\text{aq}}$ is the partial molar enthalpy of component $\alpha$ in the aqueous solution. The partial molar enthalpy is defined as,

\begin{equation}
\overline{h}_{\alpha}^{\text{aq}}=N_{A}\Biggl(\dfrac{\partial H}{\partial N_{\alpha}}\Biggr)_{P,T,N_{\beta\ne\alpha}}=
\lim_{\Delta N_{\alpha}\rightarrow 0}N_{A}\Biggl(\dfrac{\Delta H}{\Delta N_{\alpha}}\Biggr)_{P,T,N_{\beta\ne\alpha}}
\label{partial_enthalpy_i}
\end{equation}

\noindent
where $N_{A}$ is the Avogadro's number and $H$ is the aqueous solution's enthalpy. The limit can be numerically evaluated computing the enthalpy for two systems that have the same number of water molecules and different number of CO$_{2}$ to evaluate the partial molar enthalpy of CO$_{2}$. The partial molar enthalpy of water can be estimated in a similar way, i.e., the number of molecules of CO$_{2}$ in the system is kept constant while the number of water molecules changes. According to this, it is possible to evaluate the variation of the chemical potential of CO$_{2}$ and water from $T_{i}$ to $T_{\text{N}}$, $\Delta\mu_{\text{CO}_2}^{\text{aq}}$  and $\Delta\mu_{\text{H}_2\text{O}}^{\text{aq}}$, from the knowledge of the partial molar enthalpies of both components. In particular, the combination of the chemical potentials of CO$_{2}$ and water, as a function of $T_{\text{N}}$, can be obtained by integrating Eq.~\eqref{delta_mu_i} as,

\begin{widetext}
\begin{equation}
\dfrac{\mu_{\text{CO}_{2}}^{\text{aq}}(T_{\text{N}},x_{\text{CO}_{2}}^{\text{N}})+5.75
\mu_{\text{H}_{2}\text{O}}^{\text{aq}}(T_{\text{N}},x_{\text{CO}_{2}}^{\text{N}})}{k_{B}T_{\text{N}}}=
\dfrac{\mu_{\text{CO}_{2}}^{\text{aq}}(T_{i},x_{\text{CO}_{2}}^{\text{N}})+5.75
\mu_{\text{H}_{2}\text{O}}^{\text{aq}}(T_{i},x_{\text{CO}_{2}}^{\text{N}})}{k_{B}T_{i}}-\bigintss_{T_{i}}^{T_{\text{N}}} 
\dfrac{\overline{h}_{\text{CO}_{2}}^{\text{aq}}(T,x_{\text{CO}_{2}}^{\text{N}})+5.75
\overline{h}_{\text{H}_{2}\text{O}}^{\text{aq}}(T,x_{\text{CO}_{2}}^{\text{N}})}{k_{B}T^{2}}dT
\label{enthalpy_co2_water}
\end{equation}
\end{widetext}

Now, it is possible to find a closed expression for evaluating $\Delta\mu_{\text{N}}$
at arbitrary $T_{\text{N}}$ and $x_{\text{CO}_{2}}^{\text{N}}$ in terms of the enthalpies of the "hydrate", CO$_{2}$, and water molecules. Using Eqs.~\eqref{integral_hydrate_route4}, \eqref{cp_hydrate_route4}, \eqref{integral_hydrate2_route4}, and \eqref{enthalpy_co2_water}, the driving force for nucleation can be written as,

\begin{widetext}
\begin{equation}
\dfrac{\Delta\mu_{\text{N}}(T_{\text{N}},x^{\text{N}}_{\text{CO}_{2}})}
{k_{B}T_{\text{N}}}=
-\bigintss_{T_{i}}^{T_{\text{N}}} 
\dfrac{h_{H}^{H}(T)- \Big\{\overline{h}_{\text{CO}_{2}}^{\text{aq}}(T,x_{\text{CO}_{2}}^{\text{N}})+5.75
\overline{h}_{\text{H}_{2}\text{O}}^{\text{aq}}(T,x_{\text{CO}_{2}}^{\text{N}})\Big\}}{k_{B}T^{2}}dT
\label{final_driving_force1}
\end{equation}
\end{widetext}

\begin{figure}
\hspace*{0.0cm}
\includegraphics[width=1.2\columnwidth]{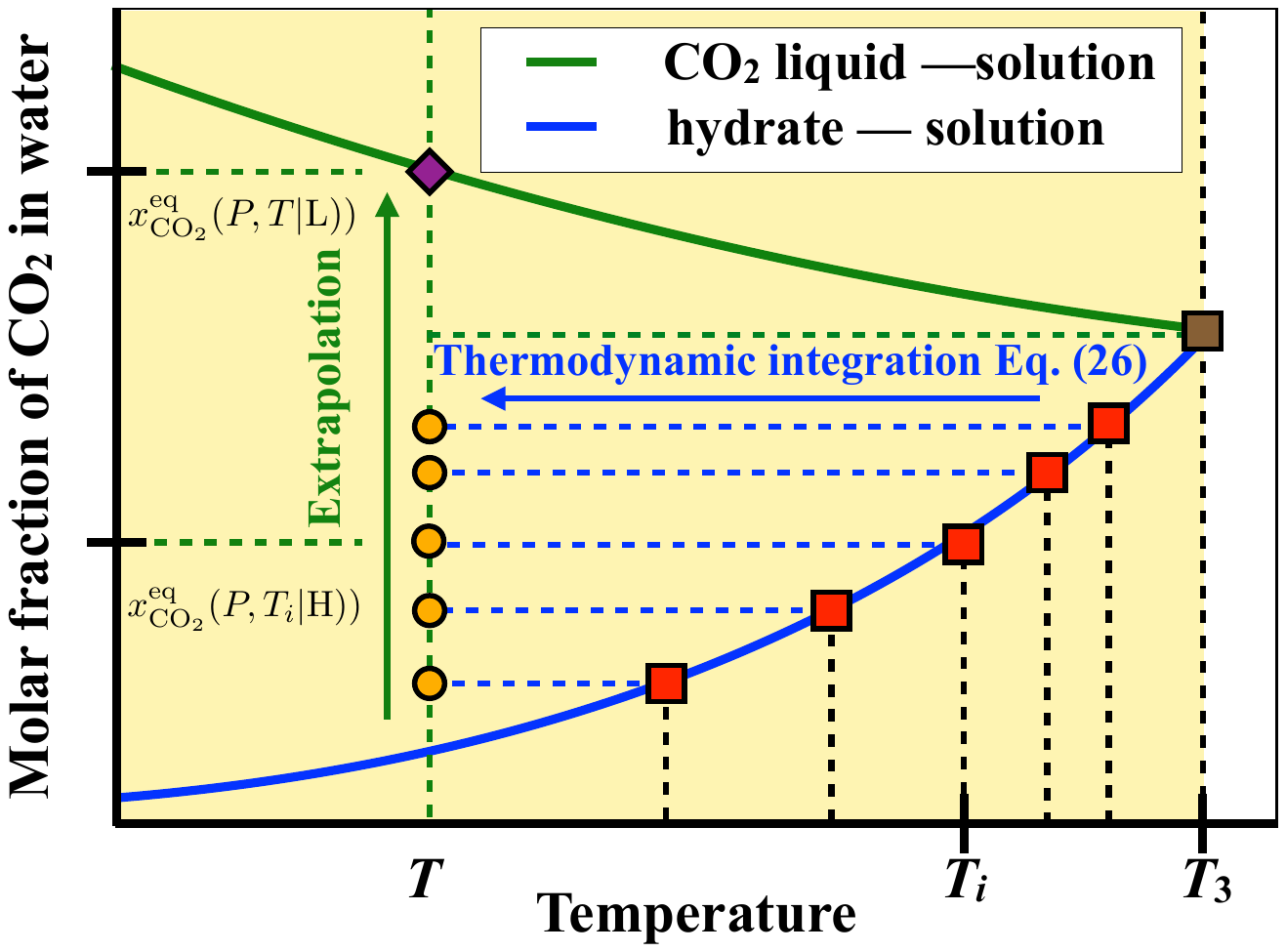}
\vspace*{-1.0cm}
\caption{Schematic depiction of route 4 for obtaining $\Delta\mu_{\text{N}}^{\text{EC}}$. Solid curves represent the solubility curves of CO$_{2}$ in water below $T_{3}$ for CO$_{2}$ liquid--solution (green curve) and hydrate--solution (blue curve). The red squares represent hydrate--solution coexistence states at $T_{i}$ and molar fractions $x_{\text{CO}_{2}}^{\text{eq}}(T_{i}|\text{H})$. The dark yellow circles are states at a lower $T$ and the same compositions. The magenta diamond is the CO$_{2}$ liquid--solution coexistence state obtained from extrapolation from states at $T$ and molar fraction  $x_{\text{CO}_{2}}^{\text{eq}}(T|\text{L})$. $\Delta\mu^{EC}_{N}(T,x_{\text{CO}_{2}}(T|\text{L}))$ is obtained from extrapolation according to Figs.~\ref{figure13} or \ref{figure14}.}
\label{figure10}
\end{figure}

\noindent
We recall here that $h_{\text{H}}^{\text{H}}$ is the enthalpy of the ``hydrate molecule'' per molecule of CO$_{2}$. Since we are assuming that the hydrate is fully occupied, the factor in the partial molar enthalpy of water must be $46/8=5.75$ to be consistent with the stoichiometry of the unit cell. It is important to remark two important aspects of this expression. As we have previously mentioned, $\Delta\mu_{\text{N}}$ does not depend on the reference state. Note that the two integrations of the molar enthalpy of the ``hydrate molecule'' between $T_{ref}$ and $T_{\text{N}}$, given by Eq.~\eqref{integral_hydrate_route4}, and between $T_{\text{ref}}$ to $T_{i}$, given by Eq.~\eqref{integral_hydrate2_route4}, are now expressed as a single integration of the molar enthalpy of the ``hydrate molecule`` between $T_{i}$ and $T_{\text{N}}$. In other words, since the driving force for nucleation does not depend on the reference state, the initial state of Eq.~\eqref{final_driving_force1} is simply $T_{i}$. This is also related with another important fact: the driving force for nucleation of the hydrate is zero not only at $T_{3}$ but also along the whole solubility curve of the hydrate.

The second interesting aspect of Eq.~\eqref{final_driving_force1} is that the integrand of the right term can be formally written as an enthalpy of dissociation of the hydrate that depends on $T_{\text{N}}$ and $x^{\text{N}}_{\text{CO}_{2}}$. Under this perspective, Eq.~\eqref{final_driving_force1} resembles Eq.~\eqref{h_dissoc} of route 3 since it has the same mathematical form.

Eq.~\eqref{final_driving_force1} is a rigorous and exact expression (within the statistical uncertainties of the simulation results) obtained only from thermodynamic arguments for calculating the driving force for nucleation of the CO$_{2}$ hydrate at any $T_{\text{N}}$ and $x_{\text{CO}_{2}}^{\text{N}}$. Obviously, this route is general and can be used to calculate driving forces for nucleation of other hydrates from the knowledge of the solubility curve of the corresponding guest with the hydrate.

Let us now apply this route to the particular case of the CO$_{2}$ hydrate and evaluate $\Delta\mu_{\text{N}}^{\text{EC}}(T)$ at experimental conditions, i.e., with $x_{\text{CO}_{2}}^{\text{N}}\equiv x_{\text{CO}_{2}}^{\text{eq}}(T|\text{L})$. Each of the chemical potential changes in Eqs.~\eqref{driving_force_EN} or \eqref{driving_force_route4} can be obtained evaluating the molar enthalpy of the ``hydrate molecule'' and the partial molar enthalpies of CO$_{2}$ and water in the aqueous solution in Eq.~\eqref{final_driving_force1}. In the case of the pure hydrate, the change in the chemical potential can be obtained performing simulations in the $NPT$ ensemble using an isotropic barostat at $400\,\text{bar}$. At the beginning of each simulation, we use a cubic box formed by 27 replicas of the unit cell in a $3\times 3\times 3$ geometry. The dimensions of the simulation box vary between $3.85$ and $3.62\,\text{nm}$ depending on the temperature. As in the rest of simulations, we calculate the enthalpy at different temperatures, from $260$ to $295\,\text{K}$. Simulations are run during $100\,\text{ns}$, $20\,\text{ns}$ to equilibrate the system and $80\,\text{ns}$ to calculate the molar enthalpy of the hydrate.

\begin{figure}
\hspace*{-0.5cm}
\includegraphics[width=1.1\columnwidth]{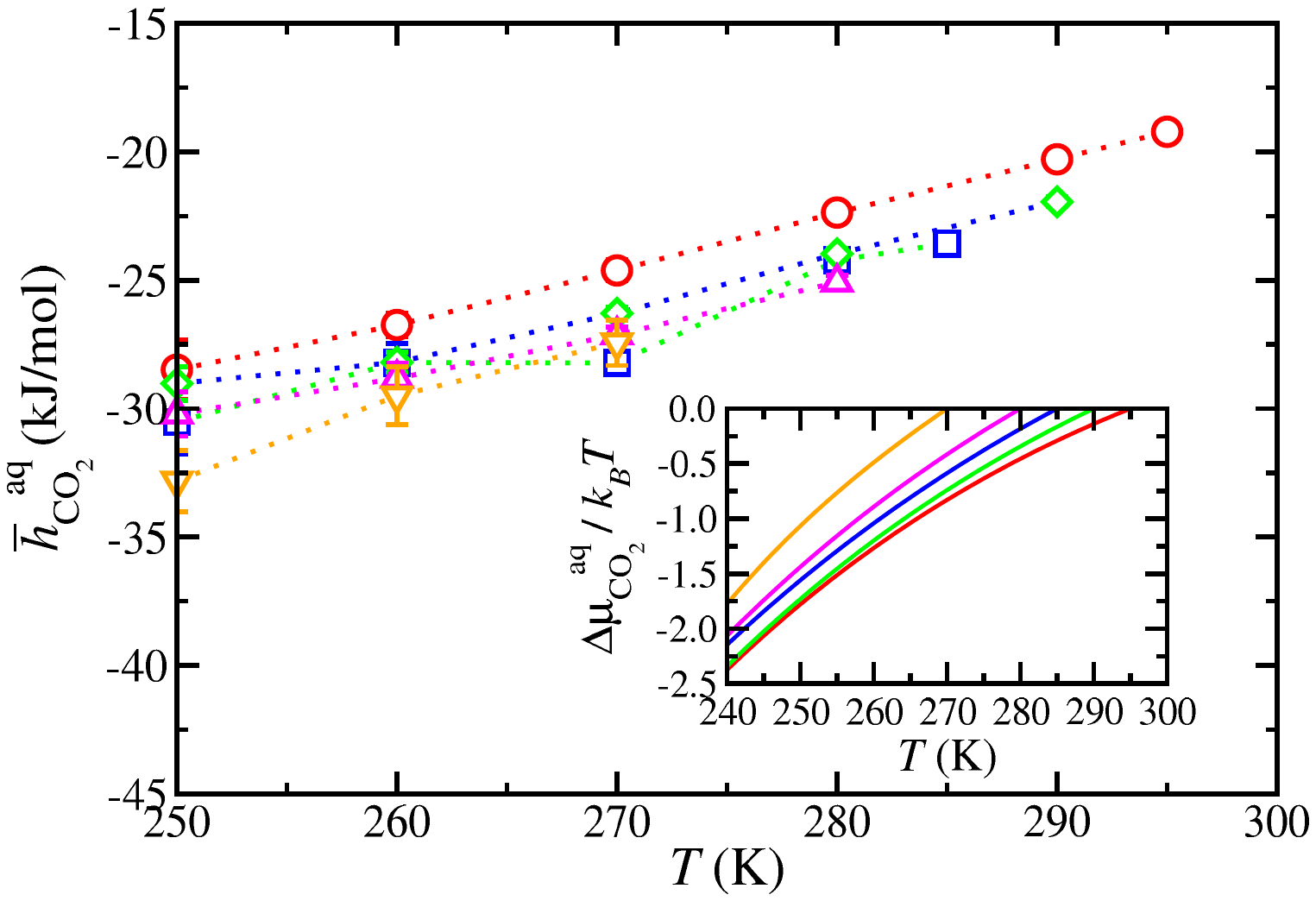}
\caption{Partial molar enthalpies of CO$_{2}$ in the aqueous solution, as a function of temperature, at different compositions: $x_{\text{CO}_2}=0.0680$ (red circles), $0.0521$ (green squares), $0.0413$ (blue diamonds), $0.0335$ (magenta triangles up), and $0.0215$ (orange triangles down). In the inset, difference of the chemical potential of CO$_{2}$, as a function of temperature, at the same compositions with respect to their values at the solubility curve of the hydrate at $295$ (red), $290$ (green), $285$ (blue), $280$ (magenta), and $270\,\text{K}$ (orange) in shown. The vertical lines in the symbols represent the error bars.}
\label{figure11}
\end{figure}

As can be seen in Fig.~\ref{figure8}, the solubility curve of CO$_{2}$ with the liquid phase is a convex and decreasing function of temperature. According to this, it is not possible to reach this curve from the solubility curve of CO$_{2}$ with the hydrate from a temperature $T_{i}$ below $T_{3}$ following the one-step thermodynamic path of route 4 (see also Fig.~\ref{figure9}). A feasible solution is to choose a $T_{i}$ value above $T_{3}$ where the hydrate--solution coexistence line is metastable. However, the $T_{3}$ of the CO$_{2}$ hydrate is located at $T_{3}=290\,\text{K}$ (with molar fraction $x_{\text{CO}_{2}}^{\text{eq}}(T_{3})=0.05$), and the solubility curve with the hydrate can be only obtained up to $295\,\text{K}$, as it is shown in Fig.~\ref{figure7}. This means that Eq.~\eqref{final_driving_force1} can be only applied to calculate $\Delta\mu^{\text{EC}}_{\text{N}}$ for supercoolings above $270-280\,\text{K}$, approximately (depending on the value of the cutoff). In the case of the methane hydrate it is possible to calculate the hydrate--solution equilibrium curve at temperatures significantly higher than $T_{3}$. This allows to evaluate $\Delta\mu^{\text{EC}}_{\text{N}}$ at lower tempertatures than in the case of the CO$_{2}$ hydrate (see the inset of Fig.~\ref{figure8} in this work and Fig.~4 of our previous work~\cite{Grabowska2022a}).

To overcome this problem, we propose to use Eq.~\eqref{final_driving_force1}, for several values of $T_{i}<T_{3}$, and to perform an extrapolation of the composition at the temperature at which we evaluate $\Delta\mu^{\text{EC}}_{\text{N}}$, as it is indicated schematically in Fig.~\ref{figure10}. According to this, $\mu_{\text{CO}_2}^{\text{aq}}$ and $\mu_{\text{H}_2\text{O}}^{\text{aq}}$, or the combination of both according to Eq.~\eqref{enthalpy_co2_water}, can be obtained by performing $NPT$ MD simulations of the solution along the $400\,\text{bar}$ isobar at constant composition. As in the case of the ``hydrate molecule'', since we are simulating bulk phases, the standard $NPT$ is used in such a way that the three dimensions of the simulation box are allowed to fluctuate isotropically. We evaluate the partial molar enthalpies of both components at five different concentrations: $x_{\text{CO}2}=0.0680$, $0.0521$, $0.0413$, $0.0335$, and $0.0215$. These values are the compositions of the aqueous solution along the solubility curve of CO$_{2}$ with the hydrate, $x_{\text{CO}_{2}}^{\text{eq}}(T_{i}|\text{H})$, at $T_{i}=295$, $290$, $285$, $280$, and $270$, respectively. We calculate numerically the derivative of Eq.~\eqref{partial_enthalpy_i} by computing the enthalpy for two different systems that have the same number of water molecules and different number of CO$_{2}$ molecules and the same number of CO$_{2}$ molecules and different number of water molecules to determine $\overline{h}^{\text{aq}}_{\text{CO}_2}$ and $\overline{h}^{\text{aq}}_{\text{H}_2\text{O}}$, respectively. Particularly, $\overline{h}^{\text{aq}}_{\text{H}_2\text{O}}$ is obtained from the difference of the enthalpies of the aqueous solution using $990$ and $1010$ water molecules ($\Delta N_{\text{H}_2\text{O}}=20$) for all the temperatures and compositions of the mixtures. In the case of $\overline{h}^{\text{aq}}_{\text{CO}_2}$, we have used different number of CO$_{2}$ molecules depending on the composition of the mixture: $20$ and $24$ ($\Delta N_{\text{CO}_2}=4$) for $x_{\text{CO}2}=0.0215$, $32$ and $38$ ($\Delta N_{\text{CO}_2}=6$) for $x_{\text{CO}2}=0.0335$, $40$ and $46$ ($\Delta N_{\text{CO}_2}=6$) for $x_{\text{CO}2}=0.0413$, $50$ and $60$ ($\Delta N_{\text{CO}_2}=10$) for $x_{\text{CO}2}=0.0521$, and $68$ and $78$ ($\Delta N_{\text{CO}_2}=10$) for $x_{\text{CO}2}=0.0680$. Simulations to calculate the enthalpy, at each temperature, are run during $300\,\text{ns}$, $50\,\text{ns}$ to equilibrate the system and $250\,\text{ns}$ as the production period. Dividing the enthalpy difference, $\Delta H$, by the difference of the number of CO$_{2}$ molecules, $\Delta N_{\text{CO}_2}$, and of the number of water molecules, $\Delta N_{\text{H}_2\text{O}}$, and multiplying by Avogadro's constant, we get estimations of $\overline{h}^{\text{aq}}_{\text{CO}_2}$ and $\overline{h}^{\text{aq}}_{\text{H}_2\text{O}}$ at several compositions of the mixture and temperatures.

\begin{figure}
\hspace*{-0.5cm}
\includegraphics[width=1.1\columnwidth]{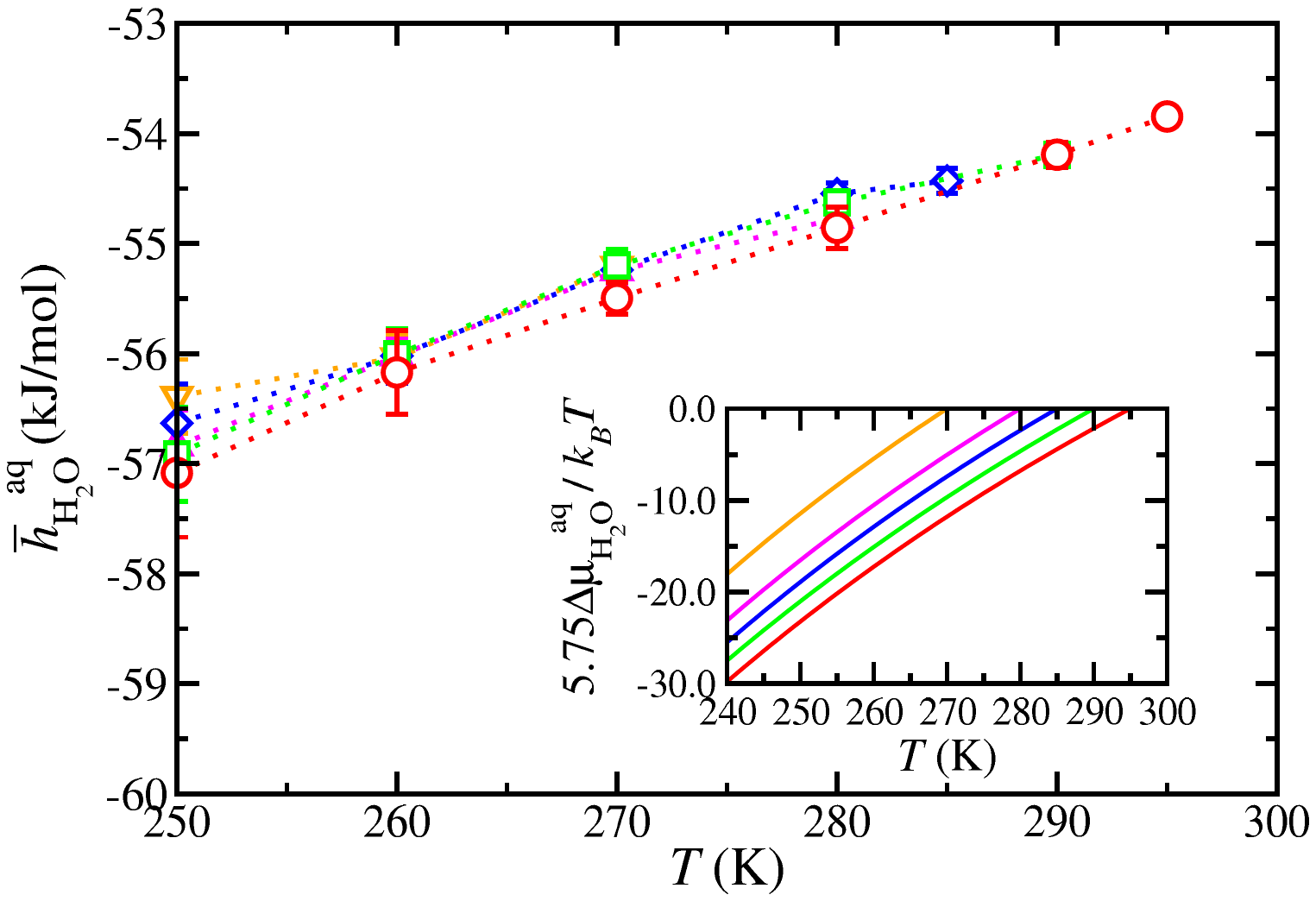}
\caption{Partial molar enthalpies of H$_{2}$O in the aqueous solution, as a function of temperature, at different compositions. In the inset, 5.75 times the difference of the chemical potential of H$_{2}$O, as a function of temperature, at the same compositions is shown. The symbols and colors are the same as those in Fig.~\ref{figure11}. The vertical lines in the symbols represent the error bars.}
\label{figure12}
\end{figure}

Fig.~\ref{figure11} shows the partial molar enthalpy of CO$_{2}$ at five constant compositions, as functions of the temperature. The composition of the mixture in each curve corresponds to the molar fraction of the solution at several temperatures, $T_{i}$, along the solubility curve of CO$_{2}$ with the hydrate, as indicated in the previous paragraph. The difference of the chemical potential of CO$_{2}$ along the integration path (see the Fig.~\ref{figure9}), $\Delta\mu_{\text{CO}_2}^{\text{aq}}$, is represented in the inset of Fig.~\ref{figure11}. Note that we have five different curves, one corresponding to each temperature value $T_{i}$. We follow the same approach and represent the partial molar enthalpy of water at the same compositions in Fig.~\ref{figure12}. In the inset is now depicted $5.75\Delta\mu_{\text{H}_2\text{O}}^{\text{aq}}$, as function of the temperature, which represents the change of the rest of the "hydrate molecule" dissociated in the solution according to the nomenclature used in this section. As can be seen, partial molar enthalpies of CO$_{2}$ and water show small variations with the composition and decrease as the temperature decreases. The variation of both chemical potentials also exhibit similar behavior, i.e., they decrease as the temperature is lowered when keeping the composition of the mixture constant. In addition, each $\Delta\mu_{\text{CO}_2}^{\text{aq}}\rightarrow 0$ and $\Delta\mu_{\text{H}_2\text{O}}^{\text{aq}}\rightarrow 0$ at the temperature associated with the corresponding composition on the solubility curve of CO$_{2}$ with the hydrate.

\begin{figure}
\hspace*{-0.2cm}
\includegraphics[width=1.1\columnwidth]{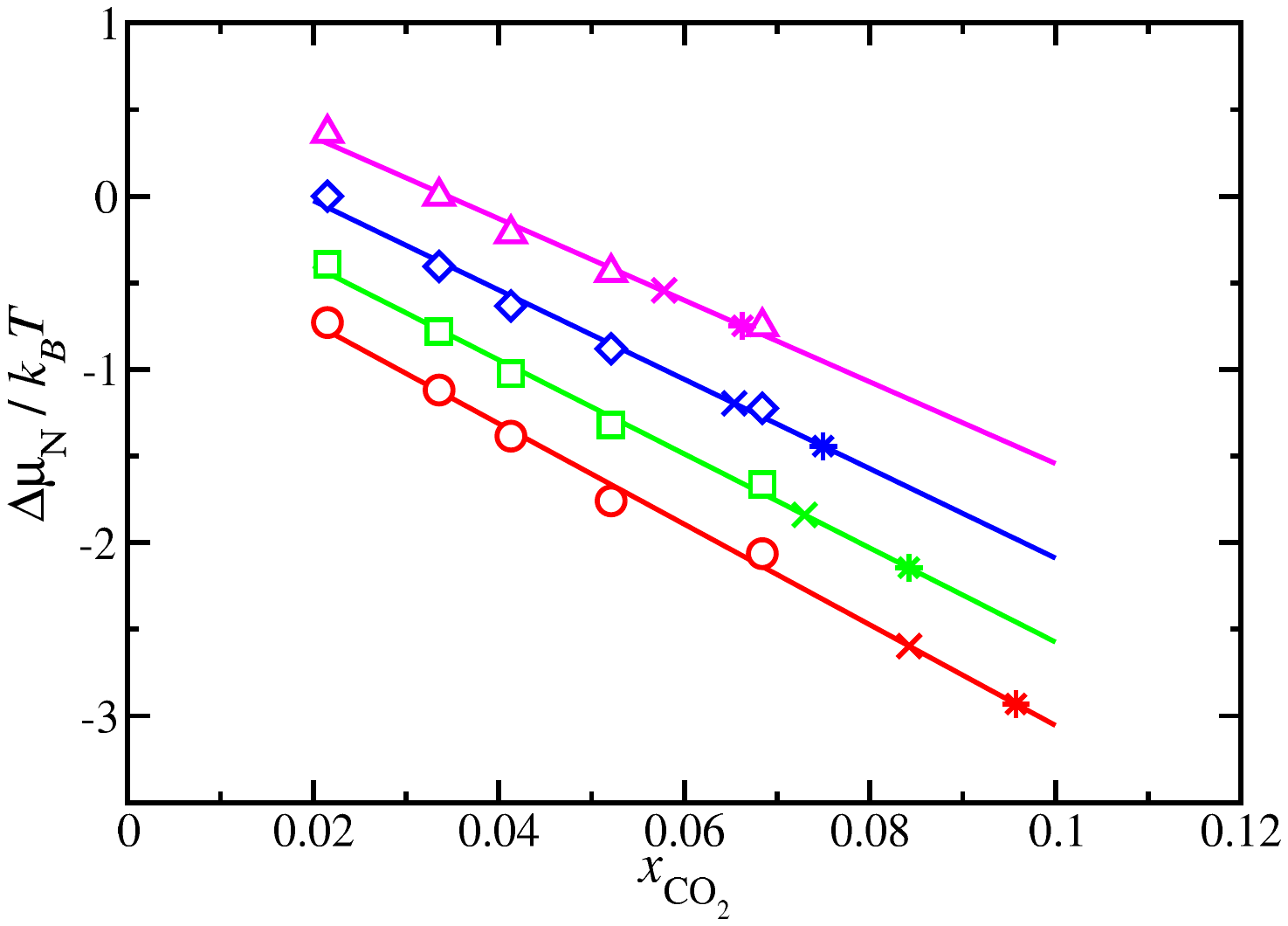}
\caption{$\Delta\mu_{\text{N}}$ values obtained from Eq.~\eqref{final_driving_force1} according to route 4, as functions of composition, at $250$ (red circles), $260$ (green squares), $270$ (blue diamonds), and $280\,\text{K}$ (magenta triangles up). The continuous lines are linear regressions of $\Delta\mu_{\text{N}}$ values with $x_{\text{CO}_{2}}$ and the values of $\Delta\mu^{\text{EC}}_{\text{N}}$ at the molecular fractions of the solution in equilibrium with the CO$_{2}$ liquid phase, $x^{\text{eq}}_{\text{CO}_{2}}(T|\text{L})$ using different cutoff distances $r_{c}$ are represented by crosses ($1.0\,\text{nm}$) and stars ($1.9\,\text{nm}$).
}
\label{figure13}
\end{figure}

Using the values of $\mu_{\text{CO}_{2}}^{\text{aq}}$ and $5.75\,\mu_{\text{H}_{2}\text{O}}^{\text{aq}}$, as functions of temperatures and at the molar fractions of CO$_{2}$ considered previously, it is possible to evaluate $\Delta\mu_{\text{N}}$ according to the scheme indicated in Fig.~\ref{figure10}. Figure~\ref{figure13} shows $\Delta\mu_{\text{N}}$, as a function of $x_{\text{CO}{2}}$, at four temperatures below $T_{3}$ ($250$, $260$, $270$, and $280\,\text{K}$). As can be seen, $\Delta\mu_{\text{N}}$ follows a linear dependence with the composition of the solution. We have performed linear regressions using the five values of $x_{\text{CO}_{2}}$, for each temperature, to obtain $\Delta\mu^{\text{EC}}_{\text{N}}$ at experimental conditions. We have also represented the compositions of the solution in equilibrium with the CO$_{2}$ liquid phase on the solubility curve of CO$_{2}$, $x^{\text{eq}}_{\text{CO}_{2}}(T|\text{L})$ at the four temperatures (crosses and stars). As can be seen, we have two different values of 
$x^{\text{eq}}_{\text{CO}_{2}}(T|\text{L})$ depending on the cutoff due to the dispersive interactions used, $r_{c}=1.0$ (crosses) and  $1.9\,\text{nm}$ (stars).

\begin{figure}
\hspace*{-0.5cm}
\includegraphics[width=1.1\columnwidth]{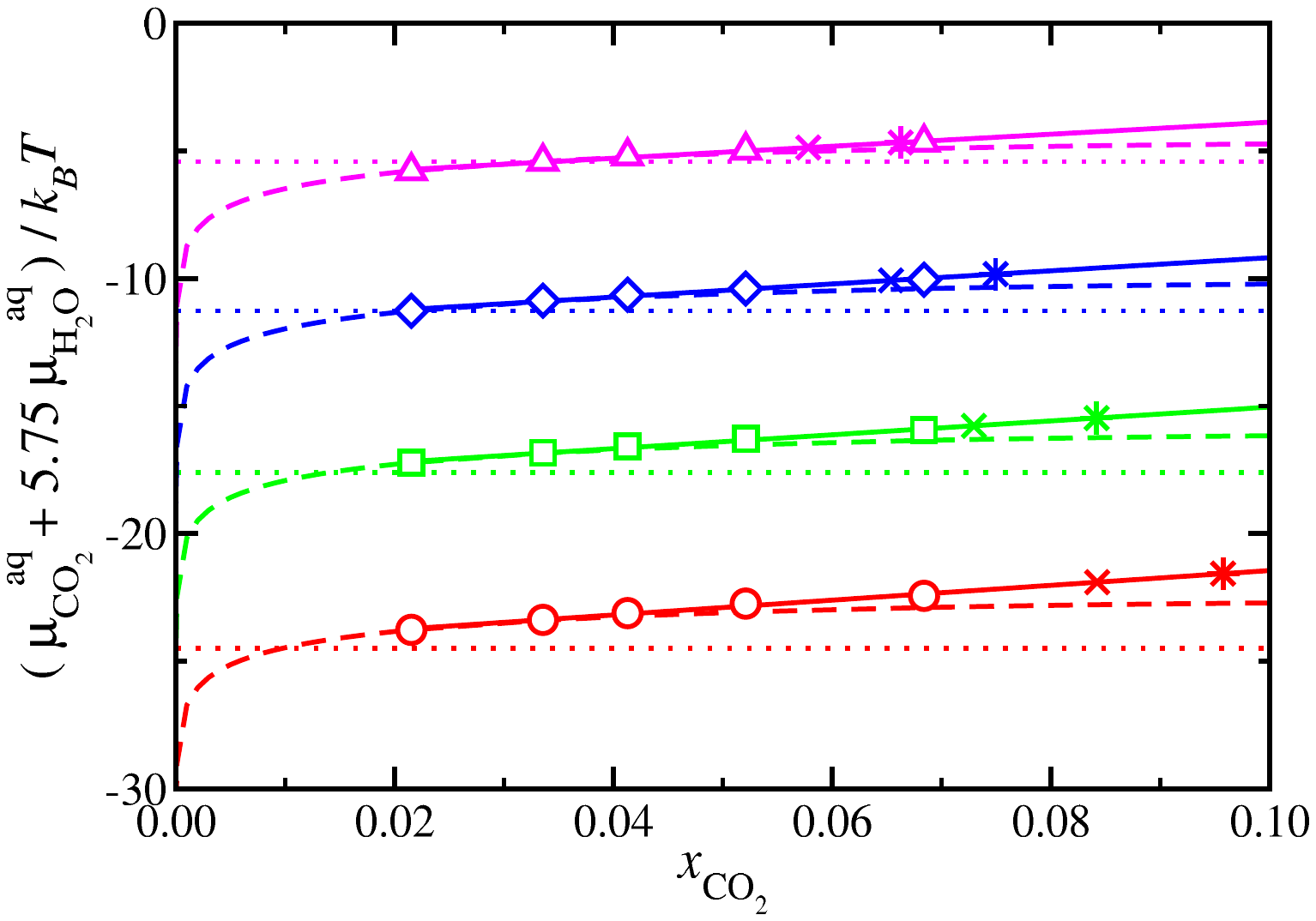}
\caption{$\mu_{\text{CO}_{2}}^{\text{aq}}+5.75\,\mu_{\text{H}_{2}\text{O}}^{\text{aq}}$ values obtained from the route 4, as functions of composition, at $250$ (red circles), $260$ (green squares), $270$ (blue diamonds), and $280\,\text{K}$ (magenta triangles up). The dotted horizontal lines represent the chemical potential values of the "hydrate molecules" in the hydrate phase, the continuous lines are linear regressions of the chemical potential values with $x_{\text{CO}_{2}}$, and the dashed curves are the chemical potential values at $x_{\text{CO}_{2}}=0.0215$ and assuming a variation with the composition given by Eq.~\eqref{driving_force4}. The composition of the solution in equilibrium with the CO$_{2}$ liquid phase using different cutoff distances $r_{c}$ are represented by crosses ($1.0\,\text{nm}$) and stars ($1.9\,\text{nm}$). We have set to zero the chemical potentials of CO$_{2}$ and water in the hydrate at $T=290\,\text{K}$.
}
\label{figure14}
\end{figure}

It is also interesting to show $\mu_{\text{CO}_{2}}^{\text{aq}}+5.75\,\mu_{\text{H}_{2}\text{O}}^{\text{aq}}$, as a function of composition, evaluated at several temperatures below the $T_{3}$ of the hydrate as shown in Fig.~\ref{figure14} (see also Fig.~\ref{figure10}). Note that we have set to zero the chemical potentials of CO$_{2}$ and water in the hydrate at $T_{3}=290\,\text{K}$. This representation contains the same information as Fig.~\ref{figure13} but allows to discuss important aspects related with the approximations of route 3. We have represented $\mu_{\text{CO}_{2}}^{\text{aq}}+5.75\,\mu_{\text{H}_{2}\text{O}}^{\text{aq}}$ at $T=250$, $260$, $270$, and $280\,\text{K}$ at the compositions $x^{\text{eq}}_{\text{CO}_{2}}$ previously selected (symbols). In addition to that, we have also represented the value of the change in the chemical potential of the hydrate, at the four temperatures, in equilibrium with the aqueous solution along the solubility curve of the hydrate. Note that the value of $\mu_{\text{CO}_{2}}^{\text{aq}}+5.75\,\mu_{\text{H}_{2}\text{O}}^{\text{aq}}$ at $270\,\text{K}$ is equal to that of the hydrate, $-11.2\,k_{B}T$, since it is in equilibrium with the solution with $x^{\text{eq}}_{\text{CO}_2}=0.0215$. The same is true at $280\,\text{K}$, i.e., $\mu_{\text{CO}_{2}}^{\text{aq}}+5.75\,\mu_{\text{H}_{2}\text{O}}^{\text{aq}}=\mu_{\text{H}}^{\text{H}}=-5.4\,k_{B}T$, state at which the solution with $x_{\text{CO}_2}=0.0335$ is in equilibrium with the hydrate.

The values of $\mu_{\text{CO}_{2}}^{\text{aq}}+5.75\,\mu_{\text{H}_{2}\text{O}}^{\text{aq}}$, at the corresponding temperatures, follow a linear dependence with the composition of the solution. We have performed linear regressions using the five values obtained in this work, for each temperature, and the corresponding lines are also shown in Fig.~\ref{figure14}. We have also represented the compositions of the solution in equilibrium with the CO$_{2}$ liquid phase on the solubility curve of CO$_{2}$, $x^{\text{eq}}_{\text{CO}_{2}}(T|\text{L})$ at the four temperatures (crosses and stars). Note that these values fit in an excellent way to the linear regression. As can be seen, we have two different values of 
$x^{\text{eq}}_{\text{CO}_{2}}(T|\text{L})$ depending on the cutoff due to the dispersive interactions used, $r_{c}=1.0$ (crosses) and  $1.9\,\text{nm}$ (stars). See Fig.~\ref{figure8} and Section III.D for further details. According to this, it is possible to know accurately the values of $\mu_{\text{CO}_{2}}^{\text{aq}}+5.75\,\mu_{\text{H}_{2}\text{O}}^{\text{aq}}$ by extrapolating the linear fits depending on the temperature and the composition (see Fig.~\ref{figure13}). These values, in combination with the values of $\mu_{H}^{H}(T)$ (already obtained in routes 1 and 2), can be used to predict with confidence the driving force for nucleation of the CO$_{2}$ hydrate using this new approach. Note that results obtained from Fig.~\ref{figure13} are the same than those obtained from Fig.~\ref{figure14} since they contain the same information.

Before finishing this section, Fig.~\ref{figure14} contains valuable information that deserves to be discussed in detail. Dashed curves represent the values of $\mu_{\text{CO}_{2}}^{\text{aq}}+5.75\,\mu_{\text{H}_{2}\text{O}}^{\text{aq}}$ obtained at the lowest concentration considered, $x_{\text{CO}2}=0.0215$, and assuming that the variation with composition follows the approximation used in route 2 proposed by us in our previous work{\cite{Grabowska2022a}} and also used by Molinero and co-workers~\cite{Knott2012a} and by us in this work. In other words, the difference of the chemical potentials of CO$_{2}$ and water when the composition is varied can be calculated assuming that the activity coefficients of both component are close to 1 or are similar in the solution. Unfortunately , this approximation is not valid for CO$_{2}$ hydrates. As can be seen in Fig.~\ref{figure14}, this approach (route 2) underestimates the value of $\mu_{\text{CO}_{2}}^{\text{aq}}+5.75\,\mu_{\text{H}_{2}\text{O}}^{\text{aq}}$ more than $0.8\,k_{B}T$ (using $r_{c}=1.0\,\text{nm}$ and more than $1\,k_{B}T$ using $r_{c}=1.9\,\text{nm}$) at $250\,\text{K}$ with respect to the value obtained from route 4. As we will see later in the next section, this result explains from a thermodynamic perspective why the route 2 can not be used with confidence to estimate the driving force for nucleation of the CO$_{2}$ hydrate.

\subsubsection{Evaluation of $\Delta\mu^{\text{EC}}_{N}$ using different routes.}

\begin{figure}
\hspace*{-0.2cm}
\includegraphics[width=1.1\columnwidth]{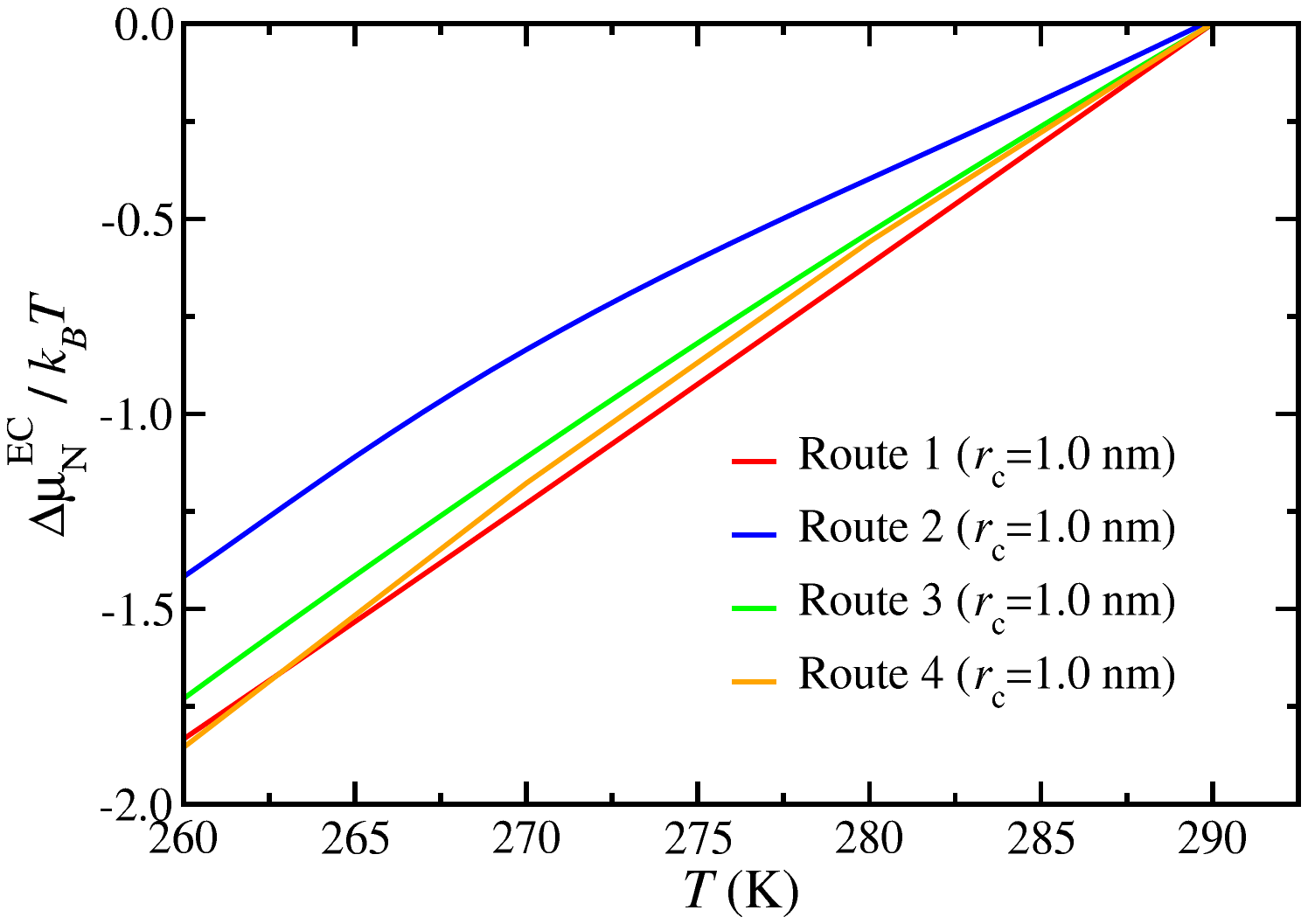}
\caption{Driving force for nucleation of the CO$_{2}$ hydrate at experimental conditions, as a function of the temperature along the $400\,\text{bar}$ isobar with $r_{c}=1.0\,\text{nm}$ and $T_{3}=290\,\text{K}$, using the route 1 (red curve), route 2 (blue curve), route 3 (green curve), and route 4 (orange curve).}
\label{figure15}
\end{figure}

We have obtained the driving force for nucleation of the CO$_{2}$ hydrate using the four routes presented in the previous sections. All the results have been obtained using a cutoff distance for the dispersive interactions $r_c{}=1.0\,\text{nm}$. As we have seen in the previous sections, this value of $r_{c}$ gives a dissociation temperature of the hydrate $T_{3}=290(2)\,\text{K}$. The results obtained using the different routes are presented in Fig.~\ref{figure15}. The route 1 given by Eq.~\eqref{driving_force_route1} predicts an almost linear behavior of $\Delta\mu_{\text{N}}^{\text{EC}}$ with the temperature. This results is in agreement with our previous results obtained for the driving force for nucleation of the methane hydrate.~\cite{Grabowska2022a}

We have also used the novel route proposed in this work (route 4), based on the use of the solubility curve of CO$_{2}$ with the hydrate, given by Eq.~\eqref{final_driving_force1}. As we have mentioned in the previous section, the route 4 should provide reliable values of $\Delta\mu_{\text{N}}^{\text{EC}}$ since it is based on rigorous thermodynamic integration calculations. The only approximation made are the extrapolations of $\Delta\mu_{\text{N}}$ to the $x^{\text{eq}}_{\text{CO}_{2}}$ on the solubility curve of CO$_{2}$ with the liquid at the corresponding temperatures. However, we think this is a good approach taken into account the low values of the concentration and the results presented in Figs.~\ref{figure13} and \ref{figure14}. As can be seen in Fig.~\ref{figure15}, small differences are seen between results obtained from routes 1 and 4. Route 1 slightly underestimates the driving force for nucleation in nearly the whole range of temperatures considered in this work, especially in the intermediate range of temperatures.

As in our previous work,~\cite{Grabowska2022a} we have also used the dissociation route (route 3) according to Eq.~\eqref{h_dissoc}, proposed by Kashchiev and Firoozabadi.~\cite{Kashchiev2002a} The agreement between the results from route 3 and 4 is good, especially at low supercoolings. The dissociation route overestimates the values of $\Delta\mu_{\text{N}}^{\text{EC}}$ obtained from the route 4 about $0.1\,k_{B}T$ at $260\,\text{K}$, approximately. This represents a value 6.5\% higher than the value obtained from the route 4, the maximum difference found between both approaches in the whole range of temperatures.

Finally, we have also obtained $\Delta\mu_{\text{N}}^{\text{EC}}$, as a function of the temperature, via the route 2 proposed by us in our previous work~\cite{Grabowska2022a} and inspired by the work of Molinero and collaborators.~\cite{Knott2012a} This route, given by Eq.~\eqref{driving_force4}, entails crude approximations. As can be seen, the route 2 is not able to provide reliable predictions of $\Delta\mu_{\text{N}}^{\text{EC}}$ in the whole range of supercoolings. In fact, it overestimates its value by $0.32\,k_{B}T$, a value 20\% higher than that obtained using the more rigorous route 4 in a wide range of temperatures. This result is in agreement with the findings observed in Figs.~\ref{figure13} and \ref{figure14} and it is a direct consequence of the main approximation made in route 2: that the activity coefficients of water and CO$_{2}$ are equal to one. Although this is a good approximation for the methane hydrate,~\cite{Grabowska2022a} it is not a realistic option for the CO$_{2}$ hydrate. The root of this behavior must be found in the large differences in solubility of methane in water compared with that of 
CO$_{2}$ in water (in contact with both, the gas/liquid phase and the hydrate phase). See the Figs.~\ref{figure3} and \ref{figure7} and the corresponding insets.

\begin{figure}
\hspace*{-0.2cm}
\includegraphics[width=1.1\columnwidth]{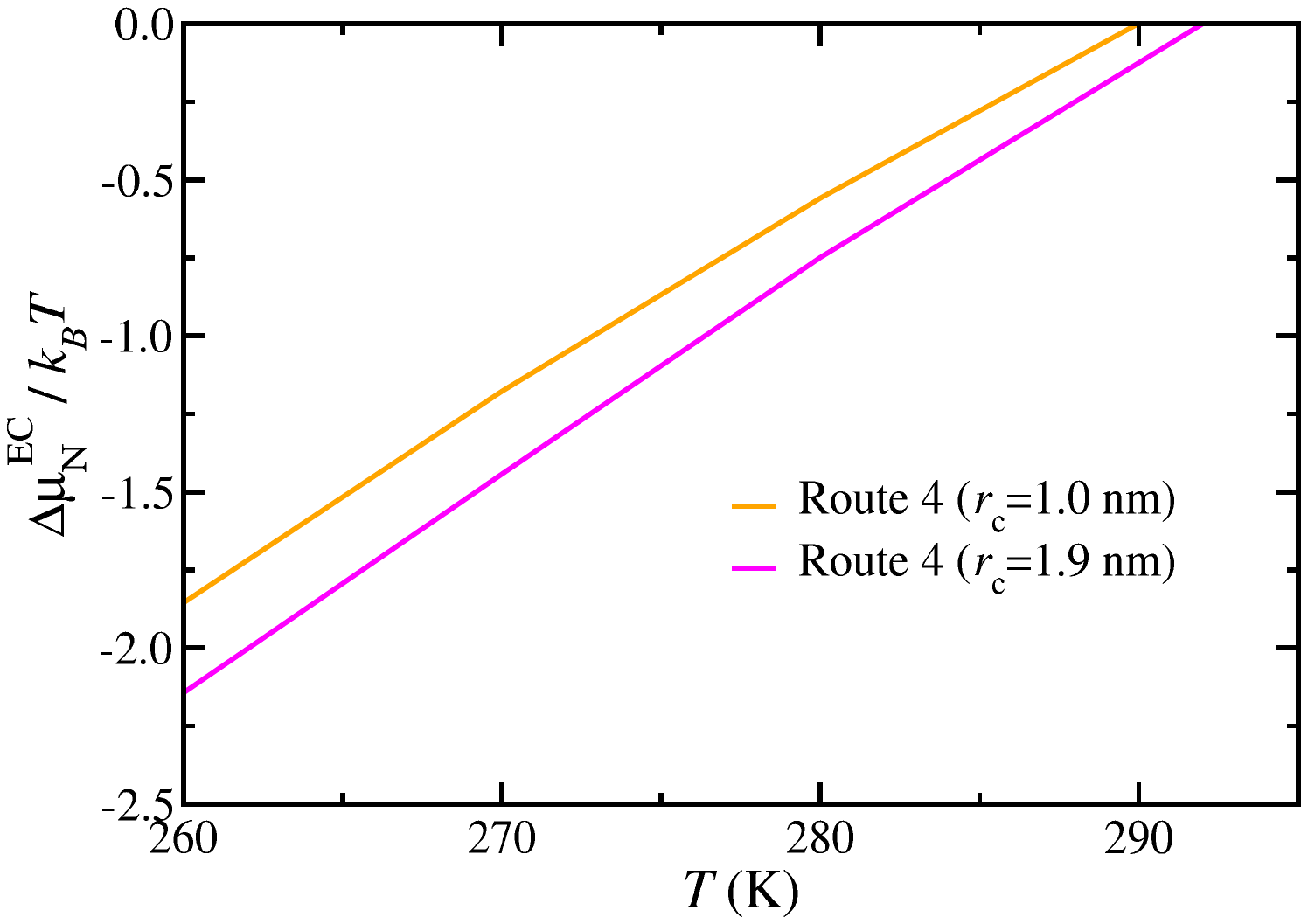}
\caption{Driving force for nucleation of the CO$_{2}$ hydrate at experimental conditions, as a function of the temperature along the $400\,\text{bar}$ isobar using the route 4 with cutoff distances $r_{c}=1.0$ (orange curve) and $1.9\,\text{nm}$ (magenta curve).
}
\label{figure16}
\end{figure}

\begin{figure}
\hspace*{-0.2cm}
\includegraphics[width=1.1\columnwidth]{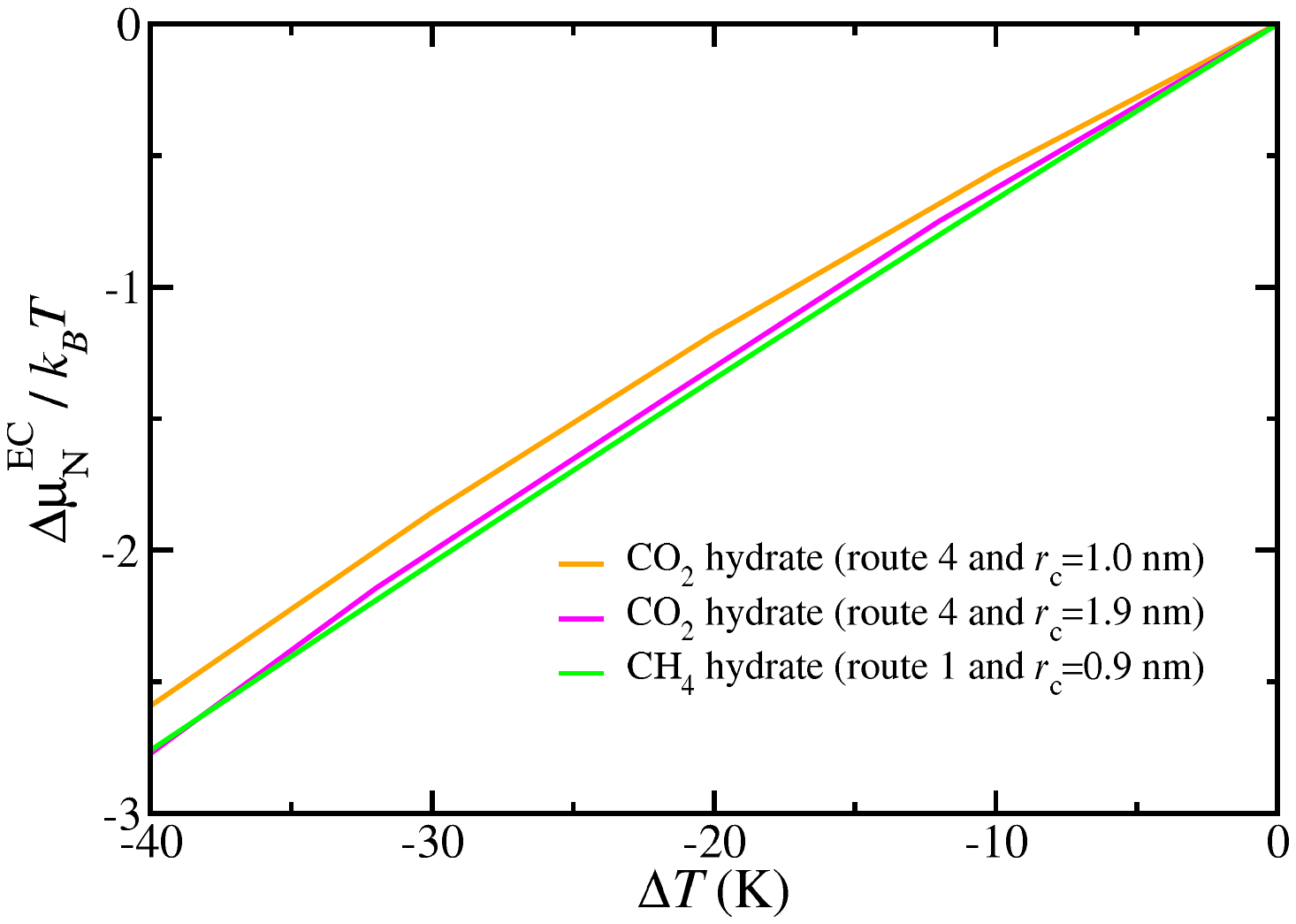}
\caption{Driving force for nucleation of the CO$_{2}$ hydrate at experimental conditions, as a function of the supercooling $\Delta T$ along the $400\,\text{bar}$ isobar, with $r_{c}=1.0\,\text{nm}$ (orange curve) and $1.9\,\text{nm}$ (magenta curve) obtained using the route 4 in both cases. The driving force for nucleation of the methane hydrate along the same isobar with $r_{c}=0.9\,\text{nm}$ is also shown (green curve).~\cite{Grabowska2022a} The dissociation temperatures in each case are $T_{3}=290$ (CO$_{2}$ hydrate with $r_{c}=1.0\,\text{nm}$), $292$ (CO$_{2}$ hydrate with $r_{c}=1.9\,\text{nm}$), and $295\,\text{K}$ (methane hydrate with $r_{c}=0.9\,\text{nm}$).
}
\label{figure17}
\end{figure}
  
In summary, the route 2 is not in general a good choice for calculating driving forces for nucleation of hydrates. This route can be used when the solubility of guest molecules in water is extremely low. The route 3 is an easy and fast way to estimate $\Delta\mu_{\text{N}}^{\text{EC}}$ values. However, we do not recommend this route in general except for cases in which solubility of the guest molecules in water is extremely low, as in the case of route 2. Finally, the route 4 proposed here in the most rigorous and nearly exact way to evaluate driving forces for nucleation of hydrates.

It is important to discuss the effect of the cutoff distance due to the dispersive interactions on $\Delta\mu_{\text{N}}^{\text{EC}}$. We have already analyzed the effect of $r_{c}$ on the solubility curve of CO$_{2}$ in contact with both, the CO$_{2}$ liquid (Fig.~\ref{figure3}) and the hydrate (Fig.~\ref{figure7}). Although the solubility curve of CO$_{2}$ with the hydrate is practically unaffected when $r_{c}$ is changed from $1.0$ to $1.9\,\text{nm}$, the situation is completely different for the solubility curve of CO$_{2}$ with the liquid. As a consequence of this, the $T_{3}$ of the CO$_{2}$ hydrate changes from $290(2)\,\text{K}$ when $r_{c}=1.0\,\text{nm}$ to $292(2)\,\text{K}$ when $r_{c}=1.9\,\text{nm}$. Obviously, this change must also affect to the values of $\Delta\mu_{\text{N}}^{\text{EC}}$. We have obtained the driving force for nucleation following the route 4 using a cutoff distance $r_{c}=1.9\,\text{nm}$ and results are compared with those using $r_{c}=1.0\,\text{nm}$. As can be seen in Fig.~\ref{figure16}, the main effect is to displace the curve towards higher temperatures. This is an expected result due to the difference in the $T_{3}$ values using different cutoff distances. However, it is clearly seen that differences between both curves increases as the temperature is decreased: the difference between both values in absolute value is $0.125\,k_{B}T$ at $290\,\text{K}$, approximately, but that difference increases up to $0.286\,k_{B}T$ at $260\,\text{K}$, approximately. This is more than double of the value of the difference predicted at $290\,\text{K}$, suggesting that the increase of the cutoff distance has a deep effect on the driving force for nucleation of the system.

To check the real impact of the cutoff distance of the dispersive interaction on the driving force for nucleation, we have plot $\Delta\mu_{\text{N}}^{\text{EC}}$, as a function of the supercooling $\Delta T$, instead of the absolute temperature $T$. This allows to compare both results at the same supercooling and to have a clearer picture of this effect. As can be seen in Fig.~\ref{figure17}, there is an important effect on $\Delta\mu_{\text{N}}^{\text{EC}}$ when $r_{c}$ is changed from $1.0$ to $1.9\,\text{nm}$. For instance, at $|\Delta T|\approx 25\,\text{K}$, $\Delta\mu_{\text{N}}^{\text{EC}}$ changes from $-1.5$ to $-1.65\,k_{B}T$ when the cutoff distance is increased. According to this, the driving force for nucleation of the hydrate is 10\% larger when $r_{c}=1.9\,\text{nm}$ than that obtained using $1.0\,\text{nm}$. This effect is not negligible. The origin of this displacement is due to the strong dependence of the solubility of CO$_{2}$ in water on $r_{c}$ (aqueous solution in contact with the CO$_{2}$ liquid phase). Due to the effect of the cutoff distance on $\Delta\mu_{\text{N}}^{\text{EC}}$, appropriate values of $r_{c}$ are required in order to obtain reliable values of this magnitude.

It is also very interesting to compare the driving force for the nucleation of the CO$_{2}$ and methane hydrates at the same pressure. We have determined in our previous work~\cite{Grabowska2022a} the driving force at experimental conditions along the same isobar. We also present these results in Fig.~\ref{figure17}, obtained using the route 1 and a cutoff distance of $r_{c}=0.9\,\text{nm}$. We compare these results with that obtained for the CO$_{2}$ hydrate using a cutoff distance of $1.0\,\text{nm}$. As can be seen, the driving force of the CO$_{2}$ hydrate is, in absolute value, lower than that of the methane hydrate along the isobar of $400\,\text{bar}$. For instance, at a supercooling of $|\Delta T|\approx 25\,\text{K}$, the driving force for the methane hydrate is $\Delta\mu^{\text{EC}}_{N}\approx -1.7\,k_{B}T$. At the same supercooling for the CO$_{2}$, $\Delta\mu^{\text{EC}}_{N}\approx -1.5\,k_{B}T$. This means that the driving force for the nucleation of the CO$_{2}$ hydrate, at $400\,\text{bar}$, is $13\%$ lower than that of the methane hydrate at the same supercooling ($\Delta T=25\,\text{K}$). According to this, the nucleation of the methane hydrate should be more favorable than that of the CO$_{2}$ hydrate. Obviously, this would be true if the other factors that affect the nucleation rate of the hydrates are equal, i.e., the water--hydrate interfacial energy.

We have also considered the effect of the occupancy of CO$_{2}$ in the hydrate on the driving force for nucleation. Particularly, we study hydrates with $7$ CO$_{2}$ molecules per unit cell, i.e., $50\%$ of occupancy in the small or D cages and $100\%$ of occupancy in the large or T cages, which is equivalent to $87.5\%$ of overall occupancy. According to the work of Kashchiev and Firoozabadi,~\cite{Kashchiev2002a} the formation of a hydrate in the aqueous solution phase can be described as the chemical reaction of Eq.~\eqref{reaction}. This reaction can be viewed as the formation of a ``hydrate molecule'' per each CO$_{2}$ molecule in the aqueous solution.

However, since we now calculate and compare driving forces for nucleation of hydrates with different occupancies, it is more convenient to write Eq.~\eqref{reaction} per cage of hydrate formed from the aqueous solution than per CO$_{2}$ molecule used to form the hydrate from the solution. In the case of a hydrate fully occupied, the reaction is the same in both descriptions since an unit cell of hydrate is formed from $8$ cages ($6$ T and $2$ D cages) and it is occupied by $8$ CO$_{2}$ molecules as well. Let's define the occupancy, $x_{\text{occ}}$ as the fraction of cages occupied by the CO$_{2}$, $x_{\text{occ}}=n_{\text{CO}_{2}}/n_{\text{cg}}$, where $n_{\text{CO}_{2}}$ and $n_{\text{cg}}$ are the number of CO$_{2}$ molecules and cages per unit cell. When the occupancy is $100\%$, $x_{\text{occ}}=8/8=1$ and $x_{\text{occ}}=7/8=0.875$ when 
it is $87.5\%$. According to this, the formation of one cage of hydrate, with occupancy $87.5\%$, from the aqueous solution phase can be viewed as a classical chemical reaction that takes place at constant $P$ and $T$,

\begin{equation}
x_{\text{occ}}\,\text{CO}_{2} (\text{aq}) +
5.75\,\text{H}_{2}\text{O} (\text{aq}) 
\longrightarrow [(\text{CO}_{2})_{x_{\text{occ}}}(\text{H}_{2}\text{O})_{5.75}]_{\text{H}}
\label{reaction2}
\end{equation}

\begin{figure}
\hspace*{-0.7cm}
\includegraphics[width=1.1\columnwidth]{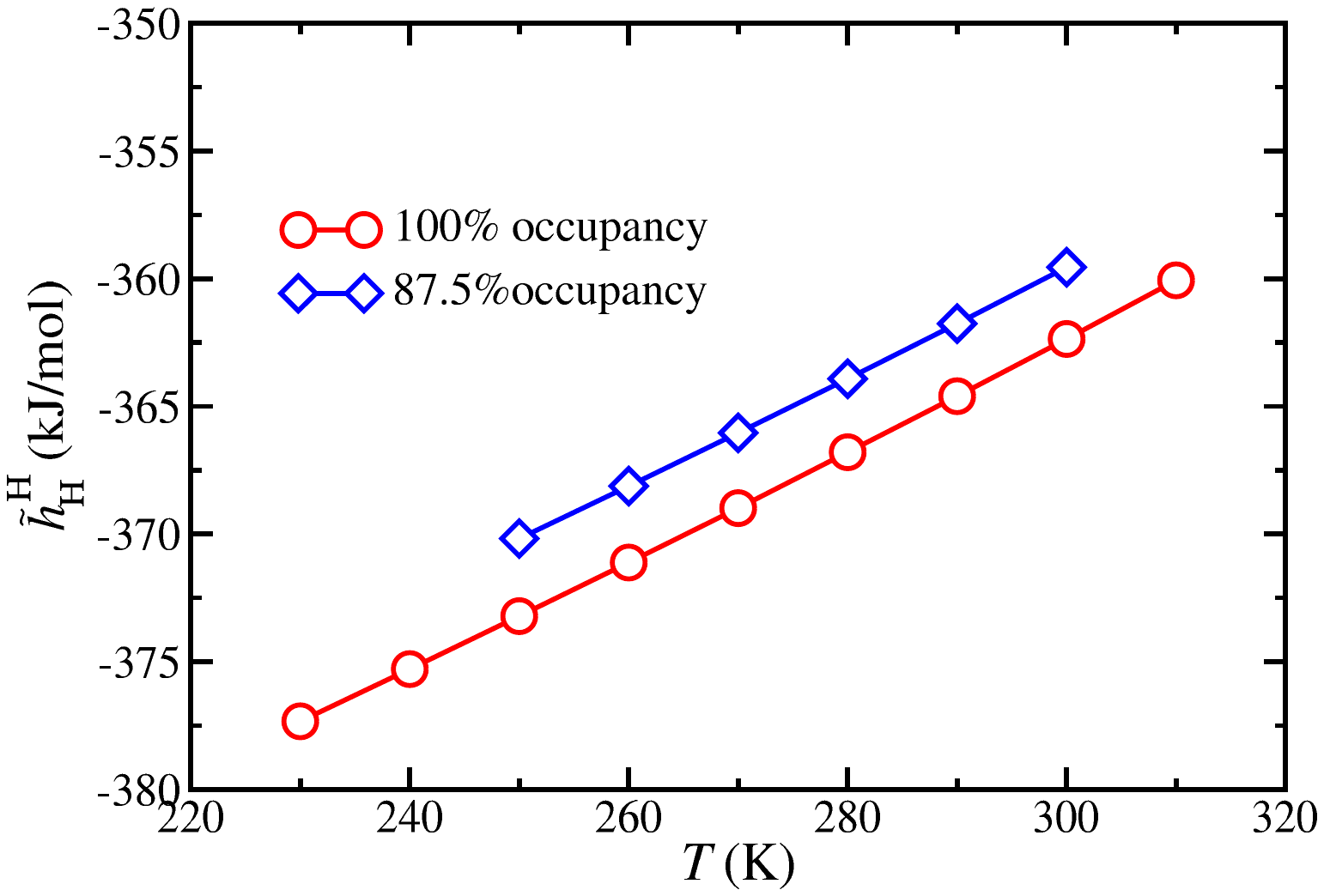}
\caption{Enthalpy per cage of the CO$_{2}$ hydrate at $400\,\text{bar}$, $\tilde{h}_{\text{H}}^{\text{H}}$, as a function of temperature with $100\%$ of occupancy (red circles) and with $87.5\%$ of occupancy (blue diamonds). The lines are guides to the eye. Note that for the case of $100\%$ occupancy the enthalpy per cage of the hydrate is identical to the enthalpy per CO$_{2}$ molecule but this is not the case when the occupancy is lower.
}
\label{figure18}
\end{figure}

\medskip

\noindent
In this particular case, since each unit cell of CO$_{2}$ hydrate is formed from $n_{\text{cg}}=8$ cages and $46$ water molecules, we only need $7/8=0.875$ CO$_{2}$ molecules (i.e., an occupancy $x_{\text{occ}}=7/8=0.875$) and $46/8=5.75$ water molecules in the solution to form $1$ cage of hydrate with the desired occupancy ($7$ CO$_{2}$ molecules per unit cell). The compound $[(\text{CO}_{2})_{x_{\text{occ}}}(\text{H}_{2}\text{O})_{5.75}]_{\text{H}}$ is simply a ``cage'' of hydrate. According to this, we call $[(\text{CO}_{2})_{x_{\text{occ}}}(\text{H}_{2}\text{O})_{5.75}]$ a ``molecule'' of one cage of the hydrate in the solid. Note that stoichiometry of reaction given by Eq.~\eqref{reaction2} is in agreement with an unit cell of this partially occupied hydrate, formed from 8 ``cages'' of hydrate with $8\times 0.875=7$ CO$_{2}$ molecules and $8\times 5.75=46$ water molecules.

\begin{figure}
\hspace*{-0.2cm}
\includegraphics[width=1.1\columnwidth]{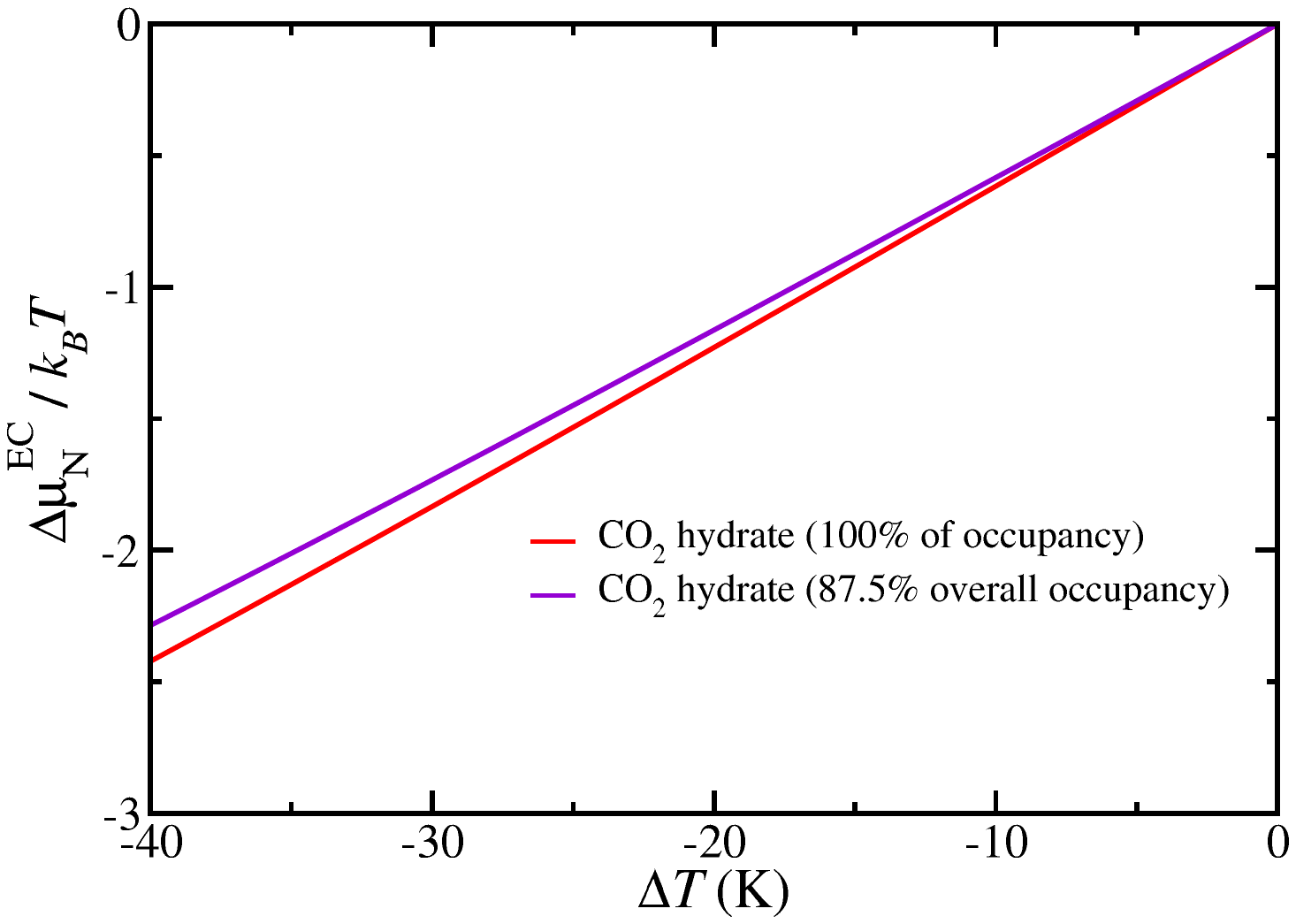}
\caption{Driving force for nucleation of the CO$_{2}$ hydrate at experimental conditions, as a function of the supercooling $\Delta T$ along the $400\,\text{bar}$ isobar, with $r_{c}=1.0\,\text{nm}$ when the hydrate is fully occupied by CO$_{2}$ molecules (red curve) and when the occupancy of the small or D cages is $50\%$ ($87.5\%$ overall occupancy) (magenta curve) obtained using the route 1 in both cases.
}
\label{figure19}
\end{figure}

We have used the route~1 described in Section III.E.1 with a cutoff distance for the dispersive interactions of $r_{c}=1.0\,\text{nm}$. We have followed the same procedure previously explained but instead of simulating a hydrate fully occupied by CO$_{2}$ molecules we have considered a hydrate with occupancy of the small or D cages of $50\%$ ($87.5\%$ overall occupancy of the hydrate). To be consistent with the description of the previous paragraph, we have used Eq.~\eqref{driving_force_route1} to evaluate the driving force for nucleation of the partially occupied hydrate per cage of hydrate instead of per CO$_{2}$ molecule. According to this, the corresponding molar enthalpy of the hydrate, $h_{\text{H}}^{\text{H}}$, as a function of the temperature, must be expressed as an enthalpy per cage of the hydrate, $\tilde{h}_{\text{H}}^{\text{H}}$,

\begin{equation}
h_{\text{H}}^{\text{H}}=\dfrac{H}{N_{\text{CO}_{2}}}=
\dfrac{H}{N_{\text{cg}}}\left(\frac{N_{\text{cg}}}{N_{\text{CO}_{2}}}\right)=
\tilde{h}_{\text{H}}^{\text{H}}\left(\frac{N_{\text{cg}}}{N_{\text{CO}_{2}}}\right)=
\dfrac{\tilde{h}_{\text{H}}^{\text{H}}}{x_{\text{occ}}}
\label{enthalpy_per_cage}    
\end{equation}

\noindent
Here $H$ is the enthalpy of the hydrate, and $N_{\text{CO}_{2}}$ and $N_{\text{cg}}$ are the total number of CO$_{2}$ molecules and cages used in the simulations, respectively.   
$N_{\text{CO}_{2}}=n_{\text{cells}}\times n_{\text{CO}_{2}}$ and $N_{\text{cg}}=n_{\text{cells}}\times n_{\text{cg}}$, with $n_{\text{cells}}=3\times 3\times 3=27$ the number of unit cells in the simulations. Note that the enthalpy per cage, $\tilde{h}_{\text{H}}^{\text{H}}$, is obtained dividing the enthalpy of the hydrate by the total number of cages, $N_{\text{cg}}=n_{\text{cells}}\times n_{\text{cg}}=27\times 8=216$. $H$ is calculated in the same way as in Section III.E.1 for the fully occupied hydrate but now using a simulation box with 189 CO$_{2}$ molecules and 27 replicas of the unit cell in a $3\times 3\times 3$ geometry used in Section III.E.1 for the $1242$ water molecules.

Fig.~\ref{figure18} shows the enthalpy per cage of the hydrate $\tilde{h}_{\text{H}}^{\text{H}}$ partially occupied by CO$_{2}$ molecules as a function of temperature (blue diamonds). We also present the enthalpy per cage of the hydrate with $100\%$ of occupancy (red circles). Note that 
$\tilde{h}_{\text{H}}^{\text{H}}$ is equal to the enthalpy of the hydrate per molecule of CO$_{2}$, $h_{\text{H}}^{\text{H}}$, in the case of full occupancy. As can be seen, the enthalpy per cage when occupancy is $87.5\%$ is systematically less negative than $\tilde{h}_{\text{H}}^{\text{H}}$ when the hydrate is fully occupied. The difference between both values is $\sim 3\,\text{kJ/mol}$. This is an expected result since there is one CO$_{2}$ molecule less per unit cell ($7$ instead of $8$) in the hydrate with occupancy of $87.5\%$. Although the difference is below $1\%$, there is less CO$_{2}$-water favorable (negative) dispersive interactions and this contributes to increase the energy and consequently the enthalpy of the system. Note that the lattice parameters of the unit cell (for a certain $P$ and $T$) of the hydrate depend on the occupancy. Particularly, it becomes about $0.16\%$ smaller when the occupancy changes from $100\%$ to $87.5\%$.

Once $\tilde{h}_{\text{H}}^{\text{H}}(T)$ is known, it is possible to use Eq.~\eqref{driving_force_route1} to evaluate the driving force for nucleation of the hydrate. However, Eq.~\eqref{driving_force_route1} is only valid for hydrates with $100\%$ occupancy. It is possible to reformulate route 1 for hydrates partially occupied taken into account that the enthalpy of the hydrate is expressed per cage of the hydrate, $\tilde{h}_{\text{H}}^{\text{H}}$, and using the appropriate stiochiometry when the hydrate has an occupancy of $x_{\text{occ}}$. According to this, the driving force for nucleation per cage of hydrate is given by,

\begin{widetext}
\begin{equation}
\dfrac{\Delta\mu^{\text{EC}}_{\text{N}}(T,x^{\text{eq}}_{\text{CO}_{2}})}
{k_{B}T}=
-\bigintss_{T_{3}}^{T} 
\dfrac{\tilde{h}_{\text{H}}^{\text{H}}(T')- \Big\{x_{\text{occ}}\,h_{\text{CO}_{2}}^{\text{pure}}(T')+5.75
h_{\text{H}_{2}\text{O}}^{\text{pure}}(T')\Big\}}{k_{B}T'^{2}}dT'
- \Big[k_{B}T\ln\{x^{\text{eq}}_{\text{H}_{2}\text{O}}(T)\}-k_{B}T_{3}\ln\{x^{\text{eq}}_{\text{H}_{2}\text{O}}(T_{3})\}\Big]
\label{driving_force_route1bis}
\end{equation}
\end{widetext}

\noindent
Note that Eq.~\eqref{driving_force_route1bis} is consistent with the view of Kashchiev and Firoozabadi~\cite{Kashchiev2002a} and with the reaction given by Eq.~\eqref{reaction2} in which the hydrate is a new compound formed from $x_{\text{occ}}$ molecules of CO$_{2}$ and $5.75$
 molecules of water when the hydrate with occupancy $x_{\text{occ}}$.

Fig.~\ref{figure19} shows the comparison between the driving force for nucleation obtained using the route 1 and a cutoff distance of $r_{c}=1.0\,\text{nm}$ when the hydrate is fully occupied and when only half of the small or D cages are occupied by CO$_{2}$ molecules. It is important to remark here that we are calculating driving forces for nucleation per cage of hydrate. This allows to compare $\Delta\mu_{\text{N}}^{\text{EC}}$ for both hydrates at the same conditions since the number of water molecules that form both solids is the same. Particularly, it is possible to know if a hydrate fully occupied is thermodynamically more stable than a hydrate partially occupied ($85.5\%$) when both are formed from an aqueous solution of CO$_{2}$ at fixed conditions of pressure and temperature.

As can be seen, $\Delta\mu_{\text{N}}^{\text{EC}}$ is similar in both cases for low supercoolings. However, as the supercooling increases, the differences between both values increases. Particularly, $\Delta\mu_{\text{N}}^{\text{EC}}$ becomes less negative (driving force for nucleation is lower) when the hydrate is partially occupied than when the hydrate is fully occupied. This means that the fully occupied hydrate is more stable, from the thermodynamic point of view, than the hydrate with occupancy of $87.5\%$ since the driving force for nucleation is higher. However it remains to be studied in the future if an occupancy between $0.875$ and $1$ could be more stable than the fully occupied hydrate.

\section{Conclusions}

In this work, we have studied the solubility of CO$_{2}$ in aqueous solutions when they are in contact via planar interfaces with a CO$_{2}$-rich liquid phase and with the hydrate phase at $400\,\text{bar}$ using molecular dynamics computer simulations. We have also estimated the driving force for the nucleation of the CO$_{2}$ hydrate using four different routes. These properties are key to understanding, from a thermodynamics point of view, the parameters that control the nucleation of CO$_{2}$ hydrates. Water is described using the TIP4P/Ice water model and CO$_{2}$ using the TraPPE model. The unlike dispersive interactions between water and CO$_{2}$ are taken into account using the approach proposed by us several years ago. This selection allows to describe very accurately, not only the dissociation temperature of the hydrate at the pressure considered in this work, but also the CO$_{2}$ hydrate--water interfacial free energy. Calculations of solubilities have been carried out using the direct coexistence technique between two phases. Additional simulations of the pure systems, at several temperatures, have also been performed to calculate the driving force for nucleation along the isobar considered in this work.

We have analyzed the aqueous solution of CO$_{2}$ when it is in contact with the liquid phase (pure CO$_{2}$) and with the hydrate using two different values of the cutoff associated with the dispersive interactions. From this information, we have obtained the solubility of CO$_{2}$ in water when the solution is in contact with the CO$_{2}$ liquid phase. The solubility of CO$_{2}$ decreases with temperature, in a similar way to that of methane. However, the solubility of CO$_{2}$ is one order of magnitude larger than that of methane. We also observed an important effect of the long-range dispersive interactions in the solubility curve along the isobar of $400\,\text{bar}$. The solubility of methane in water is also affected by these contributions but their effect is smaller. It is interesting to remark that corrections due to the long-range dispersive interactions affect in a different way both systems. Whereas the solubility of CO$_{2}$ increases with the cutoff distance, in the case of methane it decreases. This is probably an effect due to the CO$_{2}$-CO$_{2}$ and CO$_{2}$-water electrostatic interactions. We have also studied the solubility of CO$_{2}$ in the aqueous solution when it is in contact with the hydrate and analyzed its interfacial structure. This magnitude increases with the temperature, as it happens with the solubility of methane in water. Contrary to what happens when the aqueous solution is in contact with the CO$_{2}$ liquid phase, the variation of the cutoff distance due to the long-range dispersive interaction has no effect on the solubility. This behavior has been also observed in our previous study dealing with the solubility of methane in water.

The dissociation temperature of the CO$_{2}$ hydrate ($T_{3}$), at $400\,\text{bar}$, can be evaluated from the intersection of the two solubility curves obtained in this work. This intersection is possible because the formation of the hydrate phase, at $T<T_{3}$, and the formation of the CO$_{2}$ liquid phase, at $T>T_{3}$, are activated processes. This means that there exists metastability below and above the dissociation temperature of the hydrate at $400\,\text{bar}$, and because of this one can find the intersection between the two solubility curves. The temperature at which this occurs is the $T_{3}$ of the hydrate at the fixed pressure. From this analysis we find that the dissociation temperature of the hydrate is located at $290(2)\,\text{K}$ when the cutoff distance for dispersive interactions is equal to $1.0\,\text{nm}$. This is in good agreement (within the error bars) with our previous estimation of $T_{3}$ obtained from direct coexistence simulations and using the same cutoff distance, $287(2)\,\text{K}$. If the cutoff distance is larger ($1.9\,\text{nm}$), the $T_{3}$ is located at $292(2)\,\text{K}$. Although the value obtained in this work for a cutoff distance of $1.0\,\text{nm}$ compares well with our previous estimation, $287(2)\,\text{K}$ (within the error bars), it is possible that finite-size effects produce a shift of $T_{3}$ towards higher temperatures, as well as the value used to account for the long-range dispersive interactions.

We also estimate the driving force for nucleation of the CO$_{2}$ hydrate. Particularly, we have calculated $\Delta\mu_{N}$ using the three routes proposed in our previous paper (routes 1-3).~\cite{Grabowska2022a} Since the solubility of CO$_{2}$ in water is higher that that of methane by one order of magnitude, we have proposed a novel and alternative route based on the use of the solubility curve of CO$_{2}$ with the hydrate. This new route (which we refer to as route 4 in this paper) considers rigorously the non-ideality of the aqueous solution of CO$_{2}$ and provides reliable results for $\Delta\mu_{N}$. Routes 1, 3, and 4 provide similar values of the driving force for nucleation of the CO$_{2}$ hydrate in a wide range of supercoolings. Unfortunately, the route 2 can not be used for CO$_{2}$ hydrates due to the non-ideality of the water + CO$_{2}$ mixture at the conditions considered.

Finally, we have also analyzed the effect of the cutoff distance due to dispersive interactions and the occupancy of the cages on the driving force for nucleation of the CO$_{2}$ hydrate. In both cases, there is a non-negligible effect on the driving force for nucleation. Particularly, the driving force for nucleation increases when the cutoff distances increases and when the occupancy of the small or D cages of the hydrate increases from $87.5\%$ of occupancy to $100\%$.

\section*{Conflicts of interest}
The authors have no conflicts to disclose.

\section*{Acknowledgements}
This work was finnanced by Ministerio de Ciencia e Innovaci\'on (Grant No.~PID2021-125081NB-I00), Junta de Andalucía (P20-00363), and Universidad de Huelva (P.O. FEDER UHU-1255522 and FEDER-UHU-202034), all four cofinanced by EU FEDER funds. We also acknowledge the Centro de Supercomputaci\'on de Galicia (CESGA, Santiago de Compostela, Spain) for providing access to computing facilities. The authors also acknowledge Project No. PID2019-105898GB-C21 of the Ministerio de Educaci\'on y Cultura. We also acknowledge access to supercomputer time from RES from project FI-2022-1-0019. J.~G.~acknowledges the nacional support from Gdansk University of Technology by the DEC-09/2021/IDUB/II.1/AMERICIUM/ZD grant under the AMERICIUM - ``Excellence Initiative - Research University'' program. Part of the computations were carried out at the Centre of Informatics Tricity Academic Supercomputer \& Network. The research was supported in part by PL-Grid Infrastructure.

\section*{Data availability}

The data that support the findings of this study are available within the article.

\bibliography{bibfjblas}

\end{document}